\documentclass[11pt,a4paper]{article}
\usepackage{amsmath}
\usepackage{graphicx}
\usepackage{xcolor}
\usepackage[caption=false]{subfig}
\usepackage{mathrsfs,mathtools}
\usepackage{physics,amssymb}
\usepackage{siunitx}
\usepackage{bm}
\usepackage{braket}
\usepackage{listings}

\usepackage{cases}
\usepackage{comment}
\usepackage{soul}
\usepackage{cancel}
\usepackage{cases}
\usepackage[utf8]{inputenc}
\usepackage{url}
\usepackage{float}
\usepackage{longtable}
\usepackage[normalem]{ulem}
\usepackage{xspace}
\usepackage{xcolor}
\usepackage{aas_macros}
\setstcolor{red}
\usepackage{jcappub}
\hypersetup{colorlinks=true
,urlcolor=DARKBLUE
,anchorcolor=DARKBLUE
,citecolor=DARKBLUE
,filecolor=DARKBLUE
,linkcolor=DARKBLUE
,menucolor=DARKBLUE
,linktocpage=true
,pdfproducer=medialab
}

\definecolor{MONZA}{HTML}{CF000F}
\definecolor{DARKBLUE}{HTML}{00008b}
\definecolor{DARKMAGENTA}{HTML}{8b008b}
\definecolor{DARKCYAN}{HTML}{008B8B}
\definecolor{DARKORANGE}{HTML}{FF8C00}
\definecolor{PURPLE}{HTML}{8800CC}

\begin{document}

\title{Gravitational wave emission from nonspherical collapse in an early matter-dominated era using N-body simulations}

\author[a,b,c,d]{Albert Escriv\`a}
\author[e]{,Tomohiro Harada}
\author[f,g,h,i]{,Kazunori Kohri}
\author[j]{,Takahiro Terada}
\author[c,j]{,Chul-Moon Yoo}

\affiliation[a]{Asia Pacific Center for Theoretical Physics, Pohang 37673, Republic of Korea}
\affiliation[b]{Department of Physics, Pohang University of Science and Technology, Pohang 37673, Republic of Korea}
\affiliation[c]{Department of Physics, Nagoya University, Furo-cho Chikusa-ku, Nagoya 464-8602, Japan}
\affiliation[d]{Institute for Advanced Research, Nagoya University, 
Furo-cho Chikusa-ku, Nagoya 464-8601, Japan}
\affiliation[e]{Department of Physics, Rikkyo University, Toshima, Tokyo 171-8501, Japan}
\affiliation[f]{Division of Science, NAOJ, and SOKENDAI, 2-21-1 Osawa, Mitaka, Tokyo 181-8588, Japan}
\affiliation[g]{Department of Astronomy, University of Tokyo, Bunkyo-ku, Hongo, Tokyo 113-0033, Japan}
\affiliation[h]{Theory Center, IPNS, KEK, 1-1 Oho, Tsukuba, Ibaraki 305-0801, Japan}
\affiliation[i]{Kavli IPMU (WPI), UTIAS, University of Tokyo, Kashiwa, Chiba 277-8583, Japan}
\affiliation[j]{Kobayashi-Maskawa Institute for the Origin of Particles and the Universe (KMI), Nagoya University, Furo-cho Chikusa-ku, Nagoya 464-8602, Japan}

\emailAdd{alberto.escriva@apctp.org}


\date{\today}
\abstract{We study the dynamics of the collapse of a nonspherical overdense patch during an early matter-dominated era and the associated production of gravitational waves (GWs) using a semirelativistic N-body framework that we develop. The collapsing patch is initialized through a Zel'dovich deformation of a homogeneous sphere and evolved in an Einstein--de Sitter background, while the emitted signal is computed directly from the numerical quadrupole evolution. We show that a reliable prediction of the signal requires a fully numerical treatment of the nonlinear collapse dynamics. In particular, fitting-based procedures and Zel'dovich-based estimates fail to capture the post-shell-crossing evolution and can over/under-estimate the emitted power of the GWs. After averaging over realizations weighted by the Doroshkevich and BBKS (peak theory) distributions, we find that the two spectra have similar shapes and remain within the same overall order of magnitude at the peak amplitude, although the BBKS result is systematically smaller. The dominant contribution arises from peaks of relatively modest height, around $\nu \simeq 3$, while a larger variance significantly enhances the signal. Finally, by varying the horizon mass and reheating temperature, we map the present-day GW spectra to the sensitivity bands of different classes of detectors. In this way, the signal can populate a broad range of frequencies, from pulsar timing arrays to very high-frequency experiments, showing that GWs from nonspherical collapse can provide a probe of the pre-BBN thermal history.}

\maketitle
\flushbottom

\section{Introduction}

The early Universe may have experienced non-standard thermal histories prior to big-bang nucleosynthesis (BBN), including a transient epoch effectively dominated by nonrelativistic matter \cite{Allahverdi2020}. Such a phase can have important consequences for the evolution of primordial density perturbations, since in a pressureless background subhorizon overdensities grow efficiently and may eventually collapse to form compact objects such as primordial black holes \cite{10.1093/mnras/168.2.399} (see Refs.~\cite{Carr:2020gox,Green:2020jor,EscrivaKuhnelTada2022} for a review) or virialized bound structures, such as early halos and minihalos \cite{Cooray:2002dia,ErickcekSigurdson2011}. In particular, an early matter-dominated era can significantly enhance the growth of subhorizon perturbations and promote the formation of compact substructure well before the standard onset of structure formation.

Gravitational-wave (GW) emission from collapsing matter during an early matter-dominated era provides a particularly interesting probe of pre-BBN cosmology. In contrast to the radiation-dominated case, where pressure gradients tend to smoothen anisotropies and significantly influence the collapse dynamics, a dust-like phase allows nonspherical features to persist and grow throughout the evolution \cite{10.1093/mnras/203.2.265}. Moreover, unlike perfectly spherical collapse, which does not produce GWs \cite{Thorne1980}, nonspherical collapse generically leads to a time-dependent quadrupole moment and thus to GW emission. As a result, in an early matter-dominated era the GW signal is expected to depend sensitively on both the geometry of the collapsing configuration and the details of the subsequent nonlinear dynamics. Related GW production from early structure formation has been discussed in Refs.~\cite{Jedamzik:2010dq, Nakama:2020kdc, Eggemeier:2022gyo, Flores:2022uzt, Eggemeier:2023nyu, Fernandez:2023ddy, Fernandez:2024nhg, Zeng:2025law}, while the specific problem of collapse of pressureless matter in the early Universe has recently been studied in Refs.~\cite{Dalianis:2020gup, DalianisKouvaris2024}.\footnote{
For perturbatively induced GWs with an early matter-domianted era, see Refs.~\cite{Assadullahi:2009nf, Kohri:2018awv, Inomata:2019zqy, Inomata:2019ivs, Pearce:2023kxp, Kumar:2024hsi, Inomata:2025wiv}.
}

At the same time, a pressureless perfect-fluid description ceases to be valid once shell crossing occurs. Beyond this point, the flow becomes multivalued and the density formally develops caustics, so that the single-stream fluid approximation breaks down and a kinetic description is required \cite{Rampf2021,Rampf2017}. In this sense, N-body simulations provide a natural framework to follow the subsequent collisionless evolution, since they approximate the underlying Vlasov--Poisson dynamics and can consistently describe the multistream regime beyond shell crossing \cite{Rampf2021,Adamek2014,PueblasScoccimarro2009}.

This motivates the main purpose of the present work: to study the GW emission from the collapse of isolated nonspherical overdense patches during an early matter-dominated era using a numerical framework with N-body simulations tailored to the relevant dynamical regime. We model the system as a set of collisionless particles evolving in an Einstein--de Sitter background, with the self-gravity treated in the Newtonian approximation and the particle kinematics upgraded to a semirelativistic form. To address this problem, we have developed a dedicated numerical methodology to the N-body simulations relevant for our study and accurately calculate the emission of GWs.

A further objective is to maintain close control over the construction of the initial data and over potential numerical artifacts. The particle realizations are generated from an initially homogeneous Lagrangian sphere and then deformed through the Zel'dovich approximation, with the anisotropy parameterized by the eigenvalues $(\alpha,\beta,\gamma)$ of the deformation tensor. This setup makes it possible to connect the dynamical evolution directly to the statistical description of primordial nonspherical perturbations, while at the same time suppressing spurious quadrupoles associated with particle discreteness. We then evolve each realization through nonlinear collapse, monitor the approach to virialization, and compute the GW spectrum from the numerical quadrupole and its time derivatives.

Our analysis is aimed at clarifying several related questions. First, we determine how the collapse dynamics and the emitted gravitational waves depend on the anisotropy of the initial configuration and on the numerical treatment of the system using the novel numerical N-body methodology that we develop, including the role of the softening scale and particle number. Second, we assess to what extent fitting-based and Zel'dovich-based descriptions remain reliable for calculating the generation of GWs once the system reaches the nonlinear post-shell-crossing stage. Third, by averaging over realizations weighted with the Doroshkevich and BBKS distributions \cite{Doroshkevich1970,BBKS1986}, we estimate the resulting GW spectrum produced by nonspherical collapse during an early matter-dominated era. 
This also allows us to compare the predictions obtained with the two statistical prescriptions and to analyze in detail how the GW amplitude depends on the nonsphericity parameters. Finally, we translate the resulting spectra to the present epoch and compare them with the sensitivity windows of current and future GW detectors. For a concrete implementation of the statistical averaging, we focus
on the monochromatic limit, in which the relevant Gaussian field has
support at a single comoving wavenumber $k_*$.

This paper is organized as follows. In Sec.~\ref{sec:theory} we present the theoretical framework, including the semirelativistic N-body equations, the virial relations, the Zel'dovich initial conditions, the statistical description of the deformation parameters, and the formalism to calculate the GW emission. In Sec.~\ref{ref:num_simul} we describe the numerical implementation and the construction of the particle realizations. In Sec.~\ref{sec:numerical_results_all} we present the numerical results for the collapse dynamics, the statistically averaged spectra, and the comparison with present-day observational prospects. We summarize our conclusions in Sec.~\ref{sec:conclusions}.

\section{Theoretical set-up}
\label{sec:theory}
In this section, we introduce the basic theoretical concepts used in our study. Specifically, we present the methodology and equations employed to describe the dynamics of the N-body simulations, the implementation of the initial conditions, the virial theorem, and the basic elements of the GW calculations performed.

\subsection{N-body dynamics and methodology}

We consider the collapse of a collisionless overdense patch during an early matter-dominated
epoch by evolving a set of N particles in an Einstein--de Sitter (EdS) background. The
gravitational interaction between particles is treated in the Newtonian
approximation, while the particle kinematics are upgraded to a semirelativistic form in
order to remain well defined when the peculiar velocities become moderately relativistic, since we consider roughly horizon-sized patches.
In this sense, the numerical scheme adopted here should be regarded as a
\emph{semirelativistic N-body model}.

We first define the physical and comoving positions of particle $n$ as
\begin{equation}
\bm r_n(t)=a(t)\,\bm x_n(t),
\label{eq:r_ax}
\end{equation}
where $\bm x_n$ is the comoving coordinate and $a(t)$ is the background scale factor.
Differentiating Eq.~\eqref{eq:r_ax}, the physical velocity is
\begin{equation}
\dot{\bm r}_n
=
H\bm r_n+a\dot{\bm x}_n,
\qquad
H\equiv \frac{\dot a}{a}.
\label{eq:r_dot}
\end{equation}
This naturally motivates the definition of the peculiar velocity,
\begin{equation}
\bm v_n^{\rm pec}
\equiv
\dot{\bm r}_n-H\bm r_n
=
a\dot{\bm x}_n.
\label{eq:vpec_def}
\end{equation}
In an EdS background, the scale factor and Hubble rate are
\begin{equation}
a(t)=a_0\left(\frac{t}{t_0}\right)^{2/3},
\qquad
H(t)=\frac{2}{3t},
\qquad
\frac{\ddot a}{a}=-\frac12H^2.
\label{eq:HH_thing}
\end{equation}

Differentiating Eq.~\eqref{eq:r_ax} once more gives
\begin{equation}
\ddot{\bm r}_n
=
\ddot a\,\bm x_n
+2\dot a\,\dot{\bm x}_n
+a\,\ddot{\bm x}_n.
\label{eq:r_ddot}
\end{equation}
To isolate the motion relative to the background expansion, we split the density and
gravitational potential into homogeneous and inhomogeneous parts,
\begin{equation}
\rho(\bm r,t)=\rho_b(t)+\delta\rho(\bm r,t),
\qquad
\Phi(\bm r,t)=\Phi_b(\bm r,t)+\phi(\bm x,t),
\label{eq:rho_phi_split}
\end{equation}
where $\rho_b(t)=3H^2/(8\pi G)$ is the background density and $\phi$ denotes the peculiar
gravitational potential. The homogeneous background contribution satisfies
\begin{equation}
\nabla_{\bm r}^2\Phi_b=4\pi G\rho_b.
\label{eq:poisson_background}
\end{equation}
A convenient solution is
\begin{equation}
\Phi_b(\bm r,t)=\frac{2\pi G}{3}\rho_b(t)\,r^2,
\label{eq:phi_background}
\end{equation}
for which
\begin{equation}
-\nabla_{\bm r}\Phi_b
=
-\frac{4\pi G}{3}\rho_b\,\bm r
=
\frac{\ddot a}{a}\,\bm r,
\label{eq:grad_phi_background}
\end{equation}
where in the last equality we used the Friedmann acceleration equation for pressureless matter. Taking into account that $\ddot{\bm r}_n \equiv -\nabla_{\bm r} \Phi=-\nabla_{\bm r} \Phi_b-\nabla_{\bm r} \phi$ and using Eq.~\eqref{eq:r_ddot}
one obtains the comoving equation of motion
\begin{equation}
\ddot{\bm x}_n+2H\dot{\bm x}_n
=
-\frac{1}{a^2}\nabla_{\bm x}\phi,
\label{eq:x_eom}
\end{equation}
where the background term $-\nabla_{r} \Phi_b = \ddot{a} {\bm x}_n$ cancels out and we have used the fact that $\nabla_{\bm r}  = \nabla_{\bm x}/a$. Equivalently, in terms of the peculiar velocity,
\begin{equation}
\dot{\bm v}^{\rm pec}_n+H\bm v^{\rm pec}_n
=
-\frac{1}{a}\nabla_{\bm x}\phi,
\label{eq:vpec_eom}
\end{equation}
where the peculiar potential is sourced by the density contrast,
\begin{equation}
\nabla_{\bm x}^2\phi
=
4\pi G a^2\,\delta\rho,
\qquad
\delta\rho\equiv \rho-\rho_b.
\label{eq:poisson_comoving}
\end{equation}
Hence only the inhomogeneous part of the density distribution drives the peculiar motion. In the purely Newtonian formulation one introduces the canonical momentum per unit mass
\begin{equation}
\bm p_n \equiv a\,\bm v_n^{\rm pec}=a^2\dot{\bm x}_n,
\label{eq:p_def}
\end{equation}
which removes the explicit Hubble-friction term from Eq.~\eqref{eq:vpec_eom}.


In this work, our aim is to construct a numerical framework that follows the nonlinear
collisionless collapse of an isolated overdense patch in an expanding
matter-dominated background, including the regime after shell
crossing. For this purpose, we keep the gravitational interaction in the
standard Newtonian Vlasov--Poisson form, as in cosmological $N$-body dynamics,
while treating the homogeneous expansion through the FLRW scale factor, as implemented in the previous equations. 

At the same time, because the collapse of horizon-scale configurations may
generate mildly relativistic peculiar velocities, we promote the particle
momentum--velocity relation to its exact special-relativistic form in the
unperturbed FLRW background. With this prescription, we can avoid superluminal particles that would otherwise appear. The resulting prescription is therefore not a
purely Newtonian calculation, but an effective semirelativistic extension of the
usual comoving N-body equations: the force law remains Newtonian, whereas the
kinematics of the particles are treated relativistically. This formulation is
defined by the effective single-particle Lagrangian\footnote{This prescription should be distinguished from a weak-field or post-Newtonian
expansion in which the Newtonian potential is promoted to a metric perturbation,
for example through
\(ds^2=-(1+2\Phi)dt^2+a^2(1-2\Psi)d{\bm x}^2\).
Such a treatment would generate additional velocity-dependent gravitational
terms and would also require a corresponding upgrade of the gravitational source
and of the radiation extraction. In the present implementation, the Poisson
equation is not modified: the peculiar gravitational potential is sourced by the
Newtonian rest-mass density contrast, while relativistic corrections are included
through the particle momentum--velocity relation. The prescription should
therefore be understood as an effective weak-field N-body model, designed to
incorporate the leading kinematic effect of mildly relativistic peculiar velocities,
while retaining the standard Newtonian treatment of the gravitational field. This
approximation is expected to be appropriate as long as relativistic corrections
to the gravitational source remain subdominant compared with the Newtonian
rest-mass contribution. A fully post-Newtonian or general-relativistic
extension is beyond the scope of the present work and is left for future investigation.}

\begin{equation}
L_n^{\rm eff}
=
-m\sqrt{1-a^2|\dot{\bm x}_n|^2}
-m\,\phi(\bm x_n,t).
\label{eq:particle_action}
\end{equation}
It leads to the canonical momentum
\begin{equation}
\bm q_n=a\gamma_n\bm v^{\rm pec}_n,
\end{equation}
where 
\begin{equation}
    \gamma_n =\frac{1}{\sqrt{1-a^2|\dot{\bm x}_n|^2}}
    \label{eq:gamma_extra}
\end{equation} is the Lorentz factor of particle $n$, and reduces smoothly to the standard Newtonian $N$-body system in the limit
$|\bm v^{\rm pec}_n|\ll 1$.


The canonical momentum
conjugate to the comoving coordinate \(\bm x_n\) is then
\begin{equation}
\bm P_n \equiv \frac{\partial L_n}{\partial \dot{\bm x}_n}
=
m\,\frac{a^2\dot{\bm x}_n}{\sqrt{1-a^2|\dot{\bm x}_n|^2}}
=
m\,a^2\gamma_n\dot{\bm x}_n
=
m\,a\,\gamma_n\,\bm v_n^{\rm pec}. 
\label{eq:Pcanonical_def}
\end{equation}
Therefore, the variable evolved in the code is naturally identified as the canonical momentum
per unit mass,
\begin{equation}
\bm q_n \equiv \frac{\bm P_n}{m}
=
a\,\gamma_n\,\bm v_n^{\rm pec}.
\label{eq:q_def}
\end{equation}
Using Eq.~\eqref{eq:vpec_def}, this can also be written as
\begin{equation}
\bm q_n = a^2\gamma_n \dot{\bm x}_n.
\label{eq:q_def_xdot}
\end{equation}
This expression shows that \(\bm q_n\) is the semirelativistic generalization of the usual
Newtonian comoving canonical momentum \( \bm p_n=a^2\dot{\bm x}_n=a\,\bm v_n^{\rm pec}\),
which is recovered in the limit \(\gamma_n\to 1\).

The inverse relations are particularly useful in practice. Combining Eqs.~\eqref{eq:gamma_extra} and \eqref{eq:q_def}, one finds
\begin{equation}
\gamma_n
=
\sqrt{1+\frac{|\bm q_n|^2}{a^2}},
\label{eq:gamma_of_q}
\end{equation}
and therefore
\begin{equation}
\bm v_n^{\rm pec}
=
\frac{\bm q_n}{\sqrt{a^2+|\bm q_n|^2}},
\qquad
\dot{\bm x}_n
=
\frac{\bm q_n}{a^2\gamma_n}.
\label{eq:vpec_xdot_q}
\end{equation}
These expressions reduce smoothly to the Newtonian ones when \( |\bm q_n|/a\ll 1 \).

Moreover, the Euler--Lagrange equation derived from Eq.~\eqref{eq:particle_action} gives
\begin{equation}
\frac{d}{dt}\left(a^2\gamma_n\dot{\bm x}_n\right)
=
-\nabla_x\phi(\bm x_n,t).
\label{eq:EL_q}
\end{equation}
Equivalently, in terms of \(\bm q_n\),
\begin{equation}
\dot{\bm q}_n = -\nabla_x\phi.
\label{eq:qdot_phi}
\end{equation}
Using the definition of the comoving force adopted below,
\begin{equation}
-\nabla_x\phi \equiv \frac{\bm F_n}{a},
\end{equation}
the equations of motion become

\begin{equation}
\dot{\bm x}_n=\frac{\bm q_n}{a^2\gamma_n},
\qquad
\dot{\bm q}_n=\frac{\bm F_n}{a}.
\label{eq:semi_rel_equations}
\end{equation}
In this sense, the scheme used here is semirelativistic: particle kinematics are described by the exact special-relativistic momentum--velocity relation on an FLRW background, while self-gravity is still treated within the Newtonian approximation\footnote{In this work we adopt a direct particle-particle (PP) method, in which the force on each particle is obtained by summing the softened pairwise interactions from all the others. In contrast, a particle-mesh (PM) scheme computes the force by solving the Poisson equation on a grid after assigning the particle density to the mesh \cite{HockneyEastwood1981}. While PM methods are more efficient for very large cosmological simulations, in our case the collapse leads to a strongly concentrated and highly nonlinear configuration, so that accurate short-range force resolution is essential. For the moderate particle numbers used here, the PP approach is therefore more appropriate and robust for our settings, since it avoids mesh-induced force smoothing and gives more direct control over the dynamics relevant for the GW calculation.}.

The force term accounts for the interaction between all particles. A simple possibility would
be the Plummer-softened force \cite{HockneyEastwood1981}
\begin{equation}
\bm F_n
=
-G\sum_{m\neq n}m\,
\frac{\bm x_n-\bm x_m}{\left(r_{nm}^2+\epsilon_{\rm com}^2\right)^{3/2}},
\label{eq:forcepp_cosa}
\end{equation}
where
\begin{equation}
r_{nm}\equiv |\bm x_n-\bm x_m|.
\end{equation}
Although this prescription regularizes close encounters with the softening $\epsilon_{\rm com}$, the corresponding potential does not
recover the exact Newtonian form immediately outside the softening scale, since it
corresponds to an extended Plummer sphere. For this reason, and following the common
practice in some cosmological N-body codes (see for instance Ref.~\cite{Springel:2000yr}), we instead adopt a
cubic spline softening kernel, as discussed for example in
Refs.~\cite{Price:2006iz,1985A&A...149..135M}.

The total force on particle $n$ is written as
\begin{equation}
\bm F_n
=
-G\sum_{m\neq n} m\,(\bm x_n-\bm x_m)\,\mathcal K(r_{nm};h),
\label{eq:force_pp}
\end{equation}
where the spline kernel is
\begin{equation}
\mathcal K(r;h)=
\begin{cases}
\dfrac{32}{h^3}
\left(
\dfrac13-\dfrac65u^2+u^3
\right), & 0\le u<\dfrac12,\\[12pt]
\dfrac{32}{3}\dfrac{1}{r^3}
\left(
2u^3-\dfrac92u^4+\dfrac{18}{5}u^5-u^6-\dfrac{1}{160}
\right), & \dfrac12\le u<1,\\[10pt]
\dfrac{1}{r^3}, & u\ge 1,
\end{cases}
\qquad
u\equiv\frac{r}{h}.
\label{eq:spline_kernel}
\end{equation}
The softening scale used in the code is $h=14/5 \, \epsilon_{\rm com}=2.8\,\epsilon_{\rm com}$. With this choice, the central potential of the spline-softened particle (see Eq.~\eqref{eq:potential_softening}) matches that of a Plummer softening with length $\epsilon_{\rm com}$, while recovering the exact Newtonian force outside the kernel
radius $r\ge h$.

A quantity of direct relevance for the GW calculation is the time derivative
of the force. Differentiating Eq.~\eqref{eq:force_pp}, one obtains
\begin{equation}
\dot{\bm F}_n
=
-Gm\sum_{m\neq n}
\left[
\dot{\mathcal K}(r_{nm};h)\,\bm x_{nm}
+
\mathcal K(r_{nm};h)\,\bm v_{nm}^{\rm(com)}
\right],
\label{eq:Fdot}
\end{equation}
where
\begin{equation}
\bm x_{nm}\equiv \bm x_n-\bm x_m,
\end{equation}
and the relative comoving velocity is
\begin{equation}
\bm v_{nm}^{\rm(com)}
\equiv
\dot{\bm x}_n-\dot{\bm x}_m
=
\frac{\bm q_n}{a^2\gamma_n}
-
\frac{\bm q_m}{a^2\gamma_m}.
\label{eq:vrel_com_semirel}
\end{equation}
For $r\ge h$, where the interaction becomes exactly Newtonian and
$\mathcal K(r;h)=r^{-3}$, one has
\begin{equation}
\dot{\mathcal K}(r;h)
=
-3\,\frac{\bm x_{nm}\cdot \bm v_{nm}^{\rm(com)}}{r^5}.
\label{eq:Kdot_newton}
\end{equation}
For $r<h$, the corresponding expressions are obtained by differentiating the spline kernel
piecewise with respect to $r$ and using $\dot r=(\bm x_{nm}\cdot\bm v_{nm}^{\rm(com)})/r$.
In practice, these expressions are implemented explicitly in the code and are used in the
analytic calculation of the third time derivative of the quadrupole tensor. 

The equations of motion \eqref{eq:semi_rel_equations} are integrated with a
kick--drift--kick (KDK) leapfrog scheme. Suppose the state is known at time\footnote{Here and in the following, the superscript \(\ell\) labels the discrete time level, while the subscript \(n\) labels the particle.} $t_\ell$. The first
half-kick updates the momentum from $t_\ell$ to $t_\ell+d t/2$ according to
\begin{equation}
\bm q_n^{\,\ell+1/2}
=
\bm q_n^{\,\ell}
+
\bm F_n^{\,\ell}\,
K_{\rm fac}(t_\ell,t_\ell+d t/2),
\label{eq:kick1}
\end{equation}
where
\begin{equation}
K_{\rm fac}(t_1,t_2)
\equiv
\int_{t_1}^{t_2}\frac{\tilde{dt}}{a(\tilde{t})}.
\label{eq:Kfac_def}
\end{equation}
Here $K_{\rm fac}$ is the effective kick factor associated with the time-dependent prefactor in the momentum equation. During each kick, the particle positions are held fixed, so that the force $\bm F_n$ is treated as constant over the substep. However, since the equation of motion is $\dot{\bm q}_n=\bm F_n/a(t)$ rather than $\dot{\bm q}_n=\bm F_n$, the update of $\bm q_n$ involves the exact integral of $1/a(t)$ over the corresponding time interval. In the limit in which the scale factor is approximately constant over the sub-step, one gets $d\bm q_n\simeq (d t/2)\bm F^{\ell}_n/a(t_\ell)$. For an EdS background,
\begin{equation}
K_{\rm fac}(t_1,t_2)
=
\frac{3\,t_0^{2/3}}{a_0}
\left(
t_2^{1/3}-t_1^{1/3}
\right).
\label{eq:Kfac_EdS}
\end{equation}

During the drift, the momentum is held fixed at the half-step value, and the position is
advanced according to
\begin{equation}
\dot{\bm x}
=
\frac{\bm q}{a^2\gamma}
=
\frac{\bm q}{a\sqrt{a^2+q^2}},
\qquad q\equiv |\bm q|.
\label{eq:xdot_drift}
\end{equation}
Hence the drift update is
\begin{equation}
\bm x_n^{\,\ell+1}
=
\bm x_n^{\,\ell}
+
\bm q_n^{\,\ell+1/2}\,
D_{\rm SR}(t_\ell,t_{\ell+1};|\bm q_n^{\,\ell+1/2}|),
\label{eq:drift_semirel}
\end{equation}
where
\begin{equation}
D_{\rm SR}(t_1,t_2;q)
\equiv
\int_{t_1}^{t_2}
\frac{\tilde{dt}}{a(\tilde{t})\sqrt{a^2(\tilde{t})+q^2}}.
\label{eq:Dsr_def}
\end{equation}
This is the semirelativistic analogue of the usual Newtonian drift factor. In the
nonrelativistic limit $q\ll a$, one recovers
\begin{equation}
D_{\rm SR}(t_1,t_2;q)\longrightarrow
D_{\rm fac}(t_1,t_2)
\equiv
\int_{t_1}^{t_2}\frac{dt}{a^2(t)}.
\label{eq:Dfac_def}
\end{equation}
For an EdS background,
\begin{equation}
D_{\rm fac}(t_1,t_2)
=
\frac{3\,t_0^{4/3}}{a_0^2}
\left(
t_1^{-1/3}-t_2^{-1/3}
\right).
\label{eq:Dfac_EdS}
\end{equation}

The integral in Eq.~\eqref{eq:Dsr_def} can also be evaluated analytically in an EdS
background. Writing
\begin{equation}
a(t)=a_0\left(\frac{t}{t_0}\right)^{2/3},
\end{equation}
and performing the change of variables
\begin{equation}
z=\sqrt{\frac{a_0}{q}}\left(\frac{t}{t_0}\right)^{1/3},
\end{equation}
one finds
\begin{equation}
D_{\rm SR}(t_1,t_2;q)
=
\frac{3t_0}{2a_0^{3/2}\sqrt{q}}
\left[
F\!\left(
2\arctan\!\left(
\sqrt{\frac{a_0}{q}}
\left(\frac{t}{t_0}\right)^{1/3}
\right)
\,\middle|\,\frac12
\right)
\right]_{t_1}^{t_2},
\label{eq:Dsr_exact}
\end{equation}
where $F(i|j)$ denotes the incomplete elliptic integral of the first kind. This is the
analytic drift factor implemented in the code.

After the force has been recomputed from the updated positions, the second half-kick
completes the time step,
\begin{equation}
\bm q_n^{\,\ell+1}
=
\bm q_n^{\,\ell+1/2}
+
\bm F_n^{\,\ell+1}\,
K_{\rm fac}(t_\ell+d t/2,t_{\ell+1}).
\label{eq:kick2}
\end{equation}
The resulting KDK scheme is therefore
\begin{equation}
\bm q_n^\ell
\;\xrightarrow{\text{Kick }1}\;
\bm q_n^{\ell+1/2}
\;\xrightarrow{\text{Drift}}\;
\bm x_n^{\ell+1}
\;\xrightarrow{\text{Force update}}\;
\bm F_n^{\ell+1}
\;\xrightarrow{\text{Kick }2}\;
\bm q_n^{\ell+1}.
\label{eq:KDK_scheme}
\end{equation}
This preserves the standard leapfrog time-centering while incorporating the exact
semirelativistic relation between momentum and velocity during the drift. 

In summary, the numerical methodology we have developed in this work is a semirelativistic extension
of the usual comoving Newtonian N-body formalism: the particles evolve in an FLRW
background with exact special-relativistic kinematics encoded in the momentum variable
$\bm q_n=a\gamma_n \bm v_n^{\rm pec}$, while the self-gravity is still modeled through the
Newtonian peculiar potential and a spline-softened particle--particle force law. In Appendix~\ref{subsec:rel_non_rel}, we compare the resulting GW signals for several specific configurations obtained in the semirelativistic framework with those derived using a purely Newtonian treatment. As shown there, the inclusion of these semi-relativistic corrections can be important for the final prediction of the generated GW signal.

Finally, let us remark that the particles evolved in the simulation represent only the finite inhomogeneous overdense patch. The surrounding Universe is not represented by additional particles, but is treated as an exactly homogeneous Einstein-de Sitter background whose evolution is prescribed by the scale factor $a(t)$. Accordingly, cosmic time $t$ is the independent time variable of the N-body evolution and is tracked through the background relations $a(t)$ and $H(t)$. The absence of particles outside the collapsing patch therefore does not affect the definition or numerical evolution of cosmic time. This treatment corresponds to an isolated-patch approximation and neglects environmental tidal fields and the backreaction of the patch on the global expansion.

\subsection{Semirelativistic virial theorem}
\label{sec:virial_theorem}
To diagnose when the collapsing configuration approaches a quasi-stationary bound state, it is useful to monitor the virial theorem in the semirelativistic formulation introduced above. The following derivation may be viewed as a semirelativistic extension of the standard tensor virial theorem to a particle system evolving in an expanding background~\cite{Lebovitz1961,Layzer1963}.

Since the system is evolved in terms of the comoving positions $\bm{x}_n$ and the semirelativistic canonical momenta $\bm{q}_n$, satisfying
\begin{equation}
\dot{\bm{x}}_n=\frac{\bm{q}_n}{a^2\gamma_n},
\qquad
\dot{\bm{q}}_n=\frac{\bm{F}_n}{a},
\qquad
\gamma_n=\sqrt{1+\frac{|\bm{q}_n|^2}{a^2}},
\end{equation}
the natural virial quantity is not the usual inertia tensor, but rather a mixed position--momentum tensor.

For an isolated system, it is convenient to describe the internal dynamics using positions relative to the instantaneous mass centroid and canonical momenta with the uniform bulk contribution removed.
We therefore define

\begin{equation}
\widetilde{\bm{x}}_n
=
\bm{x}_n-\bm{x}_{\rm CM},
\qquad
\widetilde{\bm{q}}_n
=
\bm{q}_n-\bm{q}_{\rm CM},
\qquad
\widetilde{\bm{F}}_n
=
\bm{F}_n-\bm{F}_{\rm CM}.
\end{equation}

These definitions correspond to two distinct centering operations and
do not assume that the instantaneous mass centroid coincides with the
center of momentum. The subtraction of $\bm{x}_{\rm CM}$ isolates the
internal particle positions, while imposing $\bm{q}_{\rm CM}=0$
removes the residual bulk canonical momentum.

It is important to note, however, that in the semirelativistic formulation the condition
\(\bm q_{\rm CM}=0\) does not in general imply a vanishing arithmetic mean comoving
velocity, since \(\dot{\bm x}_n\) depends nonlinearly on \(\bm q_n\) through the Lorentz factor. Thus, subtracting
the bulk canonical momentum defines the natural center-of-momentum frame for the
virial diagnostics, but it does not guarantee that the arithmetic mean of the comoving
velocities vanishes exactly. In the numerical implementation we subtract the residual
bulk canonical momentum of the initial configuration, so that \(\bm q_{\rm CM}=0\)
is satisfied up to numerical round-off. In what follows we use this center-of-momentum
notation, which is the relevant one for diagnosing the internal dynamics of the collapsing
patch.

We then introduce the symmetric tensor
\begin{equation}
S_{ij}\equiv \frac{1}{2}\sum_{n=1}^N m
\left(
\tilde{x}_{n,i}\tilde{q}_{n,j}
+
\tilde{x}_{n,j}\tilde{q}_{n,i}
\right),
\end{equation}
which plays the role of the virial generator in the present semirelativistic formulation. Differentiating it with respect to time and using the equations of motion in the center-of-momentum frame, \(\bm q_{\rm CM}=0\), gives
\begin{align}
\frac{dS_{ij}}{dt}
&=
\frac{1}{2}\sum_{n=1}^N m
\left(
\dot{\tilde{x}}_{n,i}\tilde{q}_{n,j}
+
\tilde{x}_{n,i}\dot{\tilde{q}}_{n,j}
+
\dot{\tilde{x}}_{n,j}\tilde{q}_{n,i}
+
\tilde{x}_{n,j}\dot{\tilde{q}}_{n,i}
\right) \nonumber\\[2mm]
&=
\sum_{n=1}^N m\,\frac{\tilde{q}_{n,i}\tilde{q}_{n,j}}{a^2\gamma_n}
+
\frac{1}{2a}\sum_{n=1}^N m
\left(
\tilde{x}_{n,i}\tilde{F}_{n,j}
+
\tilde{x}_{n,j}\tilde{F}_{n,i}
\right).
\end{align}
This naturally defines the tensor virial relation
\begin{equation}
\frac{dS_{ij}}{dt}=V^{\rm rel}_{ij}+W_{ij},
\end{equation}
with
\begin{equation}
V^{\rm rel}_{ij}
=
\sum_{n=1}^N m\,\frac{\tilde{q}_{n,i}\tilde{q}_{n,j}}{a^2\gamma_n},
\qquad
W_{ij}
=
\frac{1}{2a}\sum_{n=1}^N m
\left(
\tilde{x}_{n,i}\tilde{F}_{n,j}
+
\tilde{x}_{n,j}\tilde{F}_{n,i}
\right).
\end{equation}

Taking the trace, we obtain the scalar virial theorem
\begin{equation}
\frac{dS}{dt}=V_{\rm rel}+W,
\label{eq:virial_scalar_semirel}
\end{equation}
where
\begin{equation}
S\equiv {\rm Tr}\,S_{ij}
=
\sum_{n=1}^N m\,\tilde{\bm{x}}_n\cdot\tilde{\bm{q}}_n,
\end{equation}
\begin{equation}
V_{\rm rel}
=
\sum_{n=1}^N m\,\frac{|\tilde{\bm{q}}_n|^2}{a^2\gamma_n},
\qquad
W
=
\frac{1}{a}\sum_{n=1}^N m\,\tilde{\bm{x}}_n\cdot\tilde{\bm{F}}_n.
\end{equation}

The quantity \(V_{\rm rel}\) is the kinetic combination that enters the
semirelativistic virial theorem. In the center-of-momentum frame used in the
simulations, where \(\bm q_{\rm CM}=0\) and therefore
\(\tilde{\bm q}_n=\bm q_n\), one may also write

\begin{equation}
V_{\rm rel}
=
\sum_{n=1}^N m\,\gamma_n |\bm{v}^{\rm pec}_n|^2
=
\sum_{n=1}^N m\left(\gamma_n-\frac{1}{\gamma_n}\right),
\end{equation}
which should be contrasted with the relativistic kinetic energy,
\begin{equation}
K=\sum_{n=1}^N m(\gamma_n-1).
\end{equation}
Therefore, in the semirelativistic formulation the virial theorem is controlled by the combination $V_{\rm rel}+W$, not by $2K+W$.

In the nonrelativistic limit, $|\bm{v}^{\rm pec}_n|\ll 1$, one has
\begin{equation}
\gamma_n \simeq 1+\frac{|\bm{v}^{\rm pec}_n|^2}{2},
\end{equation}
and hence
\begin{equation}
V_{\rm rel}
\simeq
\sum_{n=1}^N m\,|\bm{v}^{\rm pec}_n|^2
=
2K,
\end{equation}
so that Eq.~\eqref{eq:virial_scalar_semirel} reduces smoothly to the standard Newtonian result.

Virial equilibrium in the present framework therefore corresponds to
\begin{equation}
\left\langle V_{\rm rel}+W \right\rangle \simeq 0,
\end{equation}
where the brackets denote either an exact equality in a stationary state or, more generally, a time average over a relaxed bound configuration. Equivalently, once the object has virialized, one expects $dS/dt$ to become small.

For numerical diagnostics, it is convenient to monitor
\begin{equation}
K=\sum_n m(\gamma_n-1),
\qquad
V_{\rm rel}=\sum_n m\,\frac{|\tilde{\bm{q}}_n|^2}{a^2\gamma_n},
\qquad
W=\frac{1}{a}\sum_n m\,\tilde{\bm{x}}_n\cdot\tilde{\bm{F}}_n,
\qquad
S=\sum_n m\,\tilde{\bm{x}}_n\cdot\tilde{\bm{q}}_n.
\end{equation}
A useful virial ratio is then
\begin{equation}
\eta_{\rm Virial}\equiv \frac{V_{\rm rel}}{|W|},
\end{equation}
so that a relaxed halo is expected to satisfy
\begin{equation}
\eta_{\rm Virial}\simeq 1,
\end{equation}
together with a small value of $|V_{\rm rel}+W|$.

For a pairwise force law of the form
\begin{equation}
\bm{F}_n=-Gm\sum_{m\neq n}\bm{x}_{nm}\,\mathcal{K}(r_{nm};h),
\qquad
\bm{x}_{nm}\equiv \bm{x}_n-\bm{x}_m,
\qquad
r_{nm}\equiv |\bm{x}_{nm}|,
\end{equation}
the virial term may also be written as
\begin{equation}
W_{\rm pair}
=
-\frac{Gm^2}{a}\sum_{n<m} r_{nm}^2 \mathcal{K}(r_{nm};h).
\end{equation}
This is the pairwise representation of the same virial contribution and is therefore useful as an internal consistency check of the force calculation and of the softening implementation.

For completeness, one may also define the softened potential-energy diagnostic
\begin{equation}
U
=
-\frac{Gm^2}{a}\sum_{n<m}\psi(r_{nm};h),
\end{equation}
where the softened pair potential $\psi(r;h)$ is defined by
\begin{equation}
\frac{d\psi}{dr}=-\,r\,\mathcal{K}(r;h),
\qquad
\psi(r;h)=\frac{1}{r}\qquad (r\ge h).
\end{equation}
Writing $u\equiv r/h$, the explicit form corresponding to the cubic spline kernel used in the code is
\begin{equation}
\psi(r;h)=
\begin{cases}
\displaystyle
\frac{1}{h}\left(
\frac{14}{5}
-\frac{16}{3}u^2
+\frac{48}{5}u^4
-\frac{32}{5}u^5
\right),
& 0\le u<\frac{1}{2},
\\[10pt]
\displaystyle
\frac{1}{h}\left(
\frac{16}{5}
-\frac{32}{3}u^2
+16u^3
-\frac{48}{5}u^4
+\frac{32}{15}u^5
-\frac{1}{15u}
\right),
& \frac{1}{2}\le u<1,
\\[10pt]
\displaystyle
\frac{1}{r},
& u\ge 1.
\end{cases}
\label{eq:potential_softening}
\end{equation}
This definition guarantees consistency between the softened force and the potential-energy diagnostic. From that we define the parameter $\tilde{\eta}_{\rm Virial} \equiv V_{\rm rel}/\abs{U}$. However, it is important to emphasize that, once semirelativistic corrections are included, the virial theorem is governed by $V_{\rm rel}+W$, whereas $K$ and $U$ should be regarded only as auxiliary diagnostics for the energy budget and for the numerical evolution.

In summary, the semirelativistic virial theorem in the present FLRW framework takes the form
\begin{equation}
\frac{dS}{dt}=V_{\rm rel}+W,
\end{equation}
with $V_{\rm rel}$ replacing the usual Newtonian combination $2K$. This is therefore the appropriate relation to use when diagnosing the onset of virialization in the numerical simulations. For an unsoftened inverse-square interaction, one has $W=U$, and the two virial ratios coincide. For the spline-softened interaction,
however, $W$ and $U$ differ for particle pairs with separations
$r<h$. As the collapsing core becomes denser, an increasing fraction of particle pairs enters this softened regime. The softened potential remains finite at small separation, whereas the factor
$r^2K(r;h)$ entering $W$ tends to zero. Consequently,
$|W|/|U|$ decreases as the compact core forms. Since virialization is controlled by $V_{\rm rel}\simeq |W|$, the auxiliary ratio
$\widetilde\eta_{\rm Virial}=V_{\rm rel}/|U|$ can decrease below
unity. It is therefore not used as the virialization criterion.

\subsection{Zel'dovich approximation for the initial conditions}
\label{sec:zeldovich}

Let us now discuss the initial conditions adopted for the N-body simulation, which are constructed using the Zel'dovich approximation \cite{Zeldovich1970}. This provides a convenient description of the early evolution of a mildly nonlinear overdense patch, where particle trajectories can still be treated as a linear displacement from an initially homogeneous Lagrangian configuration. 

In this approach, the Eulerian comoving position $\bm x$ of each particle is written as
\begin{equation}
\bm{x}(\bm{\xi},t)=\bm{\xi}-D(t)\,\bm{\nabla}_{\bm \xi}\Psi(\bm \xi),
\label{eq:x_zd_thing}
\end{equation}
where $D(t)$ is the linear growth factor, $\bm{\xi}$ is the Lagrangian coordinate and $\Psi(\bm \xi)$ is the displacement potential. Equivalently, the deformation is governed by the tensor
\begin{equation}
d_{ij}(\bm \xi)\equiv \frac{\partial^2 \Psi}{\partial \xi_i \partial \xi_j},
\end{equation}
whose eigenvalues determine the local anisotropic compression or expansion along the principal axes. Working in the basis in which this tensor is diagonal, we identify the three eigenvalues as
\begin{equation}
\lambda_1=\alpha,\qquad
\lambda_2=\beta,\qquad
\lambda_3=\gamma,
\label{eq:lambdas_abg}
\end{equation}
subject to the ordering
\begin{equation}
\alpha>0,
\qquad
\gamma\le \beta\le \alpha,
\qquad
\alpha+\beta+\gamma>0.
\label{eq:abg_domain}
\end{equation}
The last condition ensures that the configuration corresponds to an overdense region, since the linear density contrast is proportional to the trace of the deformation tensor,
\begin{equation}
\delta_{\rm L}(t)=D(t)\left(\lambda_1+\lambda_2+\lambda_3\right)
= D(t)\left(\alpha+\beta+\gamma\right).
\end{equation}

We consider a nonspherical patch of comoving Lagrangian size $R_L$ associated with the scale $k_*^{-1}$. At horizon reentry $t_k$, the horizon mass is
\begin{equation}
M_k=\frac{c^3}{2GH_k},
\end{equation}
where $H_{k}=H(t_k)$. The total patch mass is then taken to be $M_{\rm patch}=M_k$. Here \(M_{\rm patch}\) denotes the Lagrangian mass associated with the undeformed patch. Since the particle load is specified in Lagrangian
coordinate space, this mass is fixed before the Zel'dovich deformation is applied. We therefore identify it with the horizon mass \(M_k\) associated
with the comoving scale \(k_*^{-1}\). Then, each of the N particles is assigned equal mass $m=M_k/N$.

The initial particle load is first generated inside a homogeneous Lagrangian sphere of radius $R_L$, representing the undeformed patch. Denoting the Lagrangian particle coordinates by $\bm \xi_n=(\xi_{n,1},\xi_{n,2},\xi_{n,3})$, the Zel'dovich map gives the initial comoving positions
\begin{equation}
x_{n,i}(t_0)=\left(1-D_0\lambda_i\right)\xi_{n,i},
\qquad i=1,2,3,
\label{eq:ZA_positions}
\end{equation}
where $D_0\equiv D(t_0)$.

The corresponding initial comoving velocities follow from differentiating the Zel'dovich map,
\begin{equation}
\dot x_{n,i}(t_0)=-\dot D_0\,\lambda_i\,\xi_{n,i}.
\label{eq:ZA_velocities_general}
\end{equation}
For a matter-dominated background, the growing mode satisfies $D(t)\propto a(t)$. With the normalization $D=a$, this becomes
\begin{equation}
\dot x_{n,i}(t_0)=-H_0 a_0 \lambda_i \xi_{n,i}.
\label{eq:ZA_velocities}
\end{equation}

Since the code evolves the semirelativistic canonical momentum per unit mass, the initial momentum variable is
\begin{equation}
q_{n,i}(t_0)=a_0^2\gamma_{n,0}\,\dot x_{n,i}(t_0),
\label{eq:ZA_momenta_semirel}
\end{equation}
where
\begin{equation}
\gamma_{n,0}=\frac{1}{\sqrt{1-a_0^2|\dot{\bm x}_n(t_0)|^2}}.
\end{equation}
In the nonrelativistic limit at the initial time, this reduces to the familiar Newtonian expression $q_{n,i}(t_0)\simeq a_0^2\dot x_{n,i}(t_0)$.

If the simulation is started at the turnaround time of the shortest axis, then using Eq.~\eqref{eq:ZA_positions} with $D=a$ gives
\begin{equation}
r_i(t)=a(t)\,x_i(t)=a(t)\bigl(1-a(t)\lambda_i\bigr)\xi_i,
\end{equation}
and therefore
\begin{equation}
\dot r_i(t)=\dot a(t)\,\bigl(1-2a(t)\lambda_i\bigr)\xi_i.
\end{equation}
The turnaround condition $\dot r_i(t_{\rm turn}^{(i)})=0$ then implies
\begin{equation}
a_{\rm turn}^{(i)}=\frac{1}{2\lambda_i},
\qquad \lambda_i>0.
\end{equation}
Because $\alpha=\lambda_1$ is the largest eigenvalue, the first axis to reach maximum expansion is the shortest one, and the corresponding starting time fulfils that
\begin{equation}
a_{\rm turn}=\frac{1}{2\alpha}.
\end{equation}
Assuming an Einstein--de Sitter background and choosing $a_k=1$ at horizon entry, the corresponding time is
\begin{equation}
t_{\rm turn}=t_k\left(\frac{a_{\rm turn}}{a_k}\right)^{3/2},
\qquad
t_k=\frac{2}{3H_k},
\label{eq:turn_time}
\end{equation}
where $a_{k}=a(t_k)$. This construction ensures that the initial particle positions and velocities are aligned with the Zel'dovich growing mode. Since the simulation is started at first-axis turnaround rather than at an asymptotically early linear time, one should not impose $|D_0\lambda_i|\ll 1$; instead, the relevant requirement is that shell crossing has not yet occurred at the initial time, so that the Zel'dovich map still provides a meaningful description of the pre-collapse configuration.

We emphasize that the Zel'dovich approximation is used only to construct the particle positions and momenta at the initial time $t_0=t_{\rm turn}$. After the initial data have been generated, the subsequent evolution is obtained entirely by integrating the N-body equations of motion. In particular, after shell crossing the Zel'dovich approximation is no longer used, while cosmic time continues to evolve according to the prescribed Einstein-de Sitter background.

Finally, after generating the initial configuration, we shift the
spatial origin so that $\bm{x}_{\rm CM}(t_0)=0$ and subtract the
uniform bulk canonical momentum so that $\bm{q}_{\rm CM}(t_0)=0$.
These are two distinct centering operations: the first fixes the
initial origin at the mass centroid, while the second removes the
residual bulk canonical momentum of the finite particle realization.
Since the total internal force vanishes, $\bm{q}_{\rm CM}=0$ is
preserved throughout the evolution up to numerical round-off, whereas
$\bm{x}_{\rm CM}$ need not remain zero because of the nonlinear
momentum--velocity relation.

\subsection{Statistics of the initial conditions}
\label{subsec:statistics_initial_conditions}

In order to estimate the total GW contribution from a population of
nonspherical collapsing patches, one needs a prescription for the
statistical weight associated with each initial triaxial configuration. In
this work we compare two related, but conceptually distinct, statistical
descriptions of the initial conditions: the Doroshkevich distribution
\cite{Doroshkevich1970} and the BBKS peak formalism
\cite{BBKS1986}.

Both descriptions can be formulated starting from the same smoothed Gaussian
linear field, evaluated at the reference epoch \(t_k\). In the Zel'dovich
approximation (see section \ref{sec:zeldovich}), the local deformation is described by the Hessian of the
displacement potential. We absorb the linear growth factor at the reference
epoch into the definition of this potential and write
\begin{equation}
    d_{ij}
    \equiv
    \partial_i\partial_j\Psi_k ,
    \qquad
    \Psi_k \equiv D(t_k)\Psi .
\end{equation}
Equivalently, with the normalization \(D(t_k)=a(t_k)=1\), \(\Psi_k\) coincides
with \(\Psi\). Since \(d_{ij}\) is symmetric, it can be diagonalized, and we
denote its ordered eigenvalues by
\begin{equation}
    (\lambda_1,\lambda_2,\lambda_3)
    =
    (\alpha,\beta,\gamma),
    \qquad
    \alpha\geq\beta\geq\gamma .
\end{equation}
With this convention, the smoothed linear density contrast at the reference
epoch is
\begin{equation}
    \delta_L(\mathbf r,t_k)
    =
    \nabla^2\Psi_k
    =
    {\rm Tr}\,d_{ij}
    =
    \alpha+\beta+\gamma ,
\end{equation}
and we require it to be positive $\alpha+\beta+\gamma >0$. The corresponding peak-height variable is
\begin{equation}
    \nu\equiv\frac{\delta_L(t_k)}{\sigma},
\end{equation}
where
\[
\sigma\equiv\sigma_\delta\equiv
\left\langle \delta_L^2(t_k)\right\rangle^{1/2}
\]
is the root-mean-square amplitude of the smoothed linear density contrast at
the reference epoch.

The Doroshkevich distribution describes the statistics of the ordered
eigenvalues \((\alpha,\beta,\gamma)\) at a random point of the Gaussian
field. It therefore provides a probability distribution for the local
triaxial deformation, but it does not impose that the point is a true density
maximum. By contrast, the BBKS formalism describes maxima of a Gaussian field. In this case, the point is additionally required to satisfy
the peak condition, and the statistical weight is expressed as a number
density of peaks labelled by their height and shape parameters. Thus the
two prescriptions are based on the same underlying Gaussian field, but
correspond to different statistical conditionings.

\paragraph{Doroshkevich distribution}
Following Ref.~\cite{Doroshkevich1970}, for a Gaussian random field, the joint distribution of the ordered eigenvalues \((\alpha,\beta,\gamma)\) is given by the Doroshkevich probability distribution. In terms of the invariants
\begin{equation}
I_1 = \alpha+\beta+\gamma,
\qquad
I_2 = \alpha\beta+\alpha\gamma+\beta\gamma,
\end{equation}
it can be written as
\begin{equation}
P_{\rm D}(\alpha,\beta,\gamma)\,d\alpha\,d\beta\,d\gamma
=
\frac{3375}{8\sqrt{5}\,\pi\,\sigma^6}
\exp\!\left[
-\frac{3 I_1^2}{\sigma^2}
+\frac{15 I_2}{2\sigma^2}
\right]
(\alpha-\beta)(\alpha-\gamma)(\beta-\gamma)\,
d\alpha\,d\beta\,d\gamma,
\label{eq:Doroshkevich_abg}
\end{equation}
with support only on the ordered domain \(\alpha\ge\beta\ge\gamma\). This distribution quantifies the likelihood of different local triaxial deformations in a Gaussian field, independently of whether the point is a true density peak.

It is often more convenient to replace the three eigenvalues by one variable controlling the overall amplitude of the perturbation and two variables describing its shape. For this purpose we define the ellipticity \(e\) and prolateness \(p\) as
\begin{equation}
e \equiv \frac{\alpha-\gamma}{2(\alpha+\beta+\gamma)},
\qquad
p \equiv \frac{\alpha-2\beta+\gamma}{2(\alpha+\beta+\gamma)}.
\label{eq:ep_def_abg}
\end{equation}
These variables have a simple geometrical interpretation: \(e\) measures the degree of anisotropy, while \(p\) distinguishes between prolate and oblate configurations. Inverting these relations, one obtains
\begin{equation}
\alpha = \frac{\delta_L}{3}(1+3e+p),
\qquad
\beta  = \frac{\delta_L}{3}(1-2p),
\qquad
\gamma = \frac{\delta_L}{3}(1-3e+p),
\label{eq:abg_inverse_ep}
\end{equation}
or equivalently,
\begin{equation}
\alpha = \frac{\sigma\nu}{3}(1+3e+p),
\qquad
\beta  = \frac{\sigma\nu}{3}(1-2p),
\qquad
\gamma = \frac{\sigma\nu}{3}(1-3e+p).
\label{eq:abg_inverse_nuep}
\end{equation}

The ordering \(\alpha\ge\beta\ge\gamma\) immediately implies the allowed domain
\begin{equation}
e\ge 0,
\qquad
-e \le p \le e.
\label{eq:doro_ep_domain}
\end{equation}
Thus, in the Doroshkevich description the \((e,p)\)-plane is the full triangular region bounded by \(p=e\), \(p=-e\), and \(e=0\).

Using the Jacobian
\begin{equation}
\left|
\frac{\partial(\alpha,\beta,\gamma)}{\partial(\nu,e,p)}
\right|
=
\frac{2}{3}\,\sigma^3 \nu^2,
\label{eq:jacobian_nuep}
\end{equation}
the Doroshkevich distribution can be rewritten in terms of \((\nu,e,p)\). Up to the overall normalization, one finds
\begin{equation}
P_{\rm D}(\nu,e,p)\,d\nu\,de\,dp
=
\frac{225\sqrt{5}}{2\pi}
\nu^5\,e(e^2-p^2)
\exp\!\left[
-\frac{\nu^2}{2}\left(1+15e^2+5p^2\right)
\right]
d\nu\,de\,dp,
\label{eq:Doro_nuep}
\end{equation}
again with support on the domain Eq.~\eqref{eq:doro_ep_domain}. The Doroshkevich distribution therefore provides a natural way to sample triaxial initial conditions from a Gaussian field, but it does not enforce that the point be a true local maximum.

\paragraph{BBKS peak statistics}

To impose the peak condition following the BBKS peak-theory approach
\cite{BBKS1986}, we apply the formalism to the Gaussian field
\begin{equation}
F(\mathbf r)\equiv -\Psi_k(\mathbf r).
\end{equation}
This sign convention is convenient because, with the Zel'dovich convention
\(\delta_L=\nabla^2\Psi_k\), an overdensity corresponds to
\begin{equation}
\delta_L=-\nabla^2 F>0 .
\end{equation}

Around a maximum of \(F\), we write
\begin{equation}
F(\mathbf r)
\simeq
F(0)
-\frac{1}{2}
\sum_{i=1}^{3}\lambda_i r_i^2 ,
\end{equation}
where \(\lambda_i\) are the ordered eigenvalues of
\(-\partial_i\partial_jF\). Since
\[
-\partial_i\partial_jF=\partial_i\partial_j\Psi_k=d_{ij},
\]
these eigenvalues coincide with the Zel'dovich deformation eigenvalues,
\begin{equation}
\lambda_1=\alpha,\qquad
\lambda_2=\beta,\qquad
\lambda_3=\gamma,
\qquad
\alpha\geq\beta\geq\gamma .
\end{equation}

The BBKS peak height and curvature variables are then defined by
\begin{equation}
\nu_F
\equiv
\frac{F(0)}{\sigma_{F,0}},
\qquad
x
\equiv
-\frac{\nabla^2F}{\sigma_{F,2}}
=
\frac{\delta_L}{\sigma_{F,2}} ,
\end{equation}
where the spectral moments of \(F\) are
\begin{equation}
\sigma_{F,j}^2
=
\int \frac{dk}{2\pi^2}
k^{2(j+1)}P_F(k),
\end{equation}
where $P_F(k)$ is the power spectrum of $F$, \textit{i.e.}, $\langle F^*(\bm{k})F(\bm{k}')\rangle = (2\pi)^3 \delta(\bm{k}-\bm{k}') P_F(k)$. 

The corresponding ellipticity and prolateness are
\begin{equation}
e=
\frac{\lambda_1-\lambda_3}
{2(\lambda_1+\lambda_2+\lambda_3)},
\qquad
p=
\frac{\lambda_1-2\lambda_2+\lambda_3}
{2(\lambda_1+\lambda_2+\lambda_3)}.
\end{equation}

Inverting these relations and using
\(\lambda_1+\lambda_2+\lambda_3=\sigma_{F,2}x\), one obtains
\begin{equation}
\alpha
=
\frac{\sigma_{F,2}x}{3}(1+3e+p),
\qquad
\beta
=
\frac{\sigma_{F,2}x}{3}(1-2p),
\qquad
\gamma
=
\frac{\sigma_{F,2}x}{3}(1-3e+p).
\label{eq:relation_ep}
\end{equation}

The BBKS differential peak number density may then be written as
\begin{align}
n_{\rm pk}(\nu_F,x,e,p)\,d\nu_F\,dx\,de\,dp
&=
\frac{3^2\,5^{5/2}}
{(2\pi)^3 R_*^3 \sqrt{1-\gamma_{\rm BBKS}^2}}\,
x^8 W(e,p)
\nonumber\\[1mm]
&\quad\times
\exp\!\left(-\frac{\nu_F^2}{2}\right)
\exp\!\left[
-\frac{(x-\gamma_{\rm BBKS}\nu_F)^2}
{2(1-\gamma_{\rm BBKS}^2)}
-\frac{5}{2}x^2(3e^2+p^2)
\right]
\,d\nu_F\,dx\,de\,dp .
\label{eq:d4npk_nuxep}
\end{align}
Here,
\begin{equation}
\gamma_{\rm BBKS}
=
\frac{\sigma_{F,1}^2}{\sigma_{F,0}\sigma_{F,2}},
\qquad
R_*=
\sqrt{3}\frac{\sigma_{F,1}}{\sigma_{F,2}} ,
\end{equation}
and
\begin{equation}
W(e,p)
=
e\,(e^2-p^2)\,(1-2p)\,
\big[(1+p)^2-9e^2\big]\,
\chi(e,p).
\label{eq:Wep}
\end{equation}
The support function \(\chi(e,p)\) is
\begin{equation}
\chi(e,p)=
\begin{cases}
1, & 0<e<\frac14,\quad -e<p<e, \\[2mm]
1, & \frac14<e<\frac12,\quad 3e-1<p<e, \\[2mm]
0, & \text{otherwise}.
\end{cases}
\label{eq:chi_ep}
\end{equation}

The origin of the difference with respect to the Doroshkevich domain is
transparent. The ordering
\({\lambda}_1\geq {\lambda}_2\geq {\lambda}_3\) implies the
same bounds \(-e\leq p\leq e\) as before. However, the additional requirement that the point be a true maximum of \(F\) imposes
\({\lambda}_3>0\). Using Eq.~\eqref{eq:relation_ep}, this gives
\begin{equation}
1-3e+p>0
\qquad\Longrightarrow\qquad
p>3e-1 ,
\end{equation}
which removes part of the Doroshkevich triangle. Therefore, the BBKS
support is smaller because it retains only true maxima of the smoothed gravitational potential.

\paragraph{Monochromatic limit}

In the monochromatic case, where the field has support only at a single
wavenumber \(k_*\),
\begin{equation}
P_F(k)\propto \delta_{\rm D}(k-k_*),
\end{equation}
one has
\begin{equation}
\nabla^2 F \rightarrow -k_*^2 F .
\end{equation}
Consequently,
\begin{equation}
\delta_L=-\nabla^2F \rightarrow k_*^2 F .
\end{equation}
The spectral moments satisfy
\begin{equation}
\sigma_{F,1}=k_*\sigma_{F,0},
\qquad
\sigma_{F,2}=k_*^2\sigma_{F,0}.
\end{equation}
Therefore
\begin{equation}
\gamma_{\rm BBKS}
=
\frac{\sigma_{F,1}^2}{\sigma_{F,0}\sigma_{F,2}}
\rightarrow 1,
\qquad
R_*=\frac{\sqrt{3}}{k_*}.
\end{equation}
Moreover, the curvature variable reduces to the peak height,
\begin{equation}
x
=
-\frac{\nabla^2F}{\sigma_{F,2}}
=
\frac{k_*^2F}{k_*^2\sigma_{F,0}}
=
\frac{F}{\sigma_{F,0}}
=
\nu_F .
\end{equation}

Equivalently, the linear density contrast $\delta_L$ and the field \(F\) have related root-mean-square 
amplitudes,
\begin{equation}
\sigma = k_*^2\sigma_{F,0}=\sigma_{F,2},
\end{equation}
so that
\begin{equation}
\frac{\delta_L}{\sigma}
=
\frac{k_*^2F}{k_*^2\sigma_{F,0}}
=
\nu_F .
\end{equation}
Thus, in the monochromatic limit, the normalized height of the \(F\)-peak coincides with the normalized linear density contrast $\delta_L$. In what follows we therefore
write simply
\begin{equation}
\nu \equiv \nu_F = \frac{\delta_L}{\sigma},
\qquad
\sigma=\sigma_{F,2}.
\end{equation}

Equivalently, the Gaussian factor in Eq.~\eqref{eq:d4npk_nuxep},
\begin{equation}
\frac{1}{\sqrt{1-\gamma_{\rm BBKS}^2}}
\exp\!\left[
-\frac{(x-\gamma_{\rm BBKS}\nu)^2}
{2(1-\gamma_{\rm BBKS}^2)}
\right],
\end{equation}
collapses to a Dirac delta enforcing \(x=\nu\).

After integrating over \(x\), one obtains the monochromatic BBKS expression
\begin{equation}
\frac{d^3 n_{\rm pk}}{d\nu\,de\,dp}
=
\frac{3^2\,5^{5/2}}
{(2\pi)^2\sqrt{2\pi}\,R_*^3}\,
\nu^8
\exp\!\left[
-\frac{\nu^2}{2}
\left(1+15e^2+5p^2\right)
\right]
W(e,p).
\label{eq:npk_mono_nuep}
\end{equation}
The corresponding conditional probability density at fixed \(\nu\) can be
written as
\begin{equation}
P_{\rm BBKS}(e,p\mid \nu)\,de\,dp
=
\frac{1}{f(\nu)}
\frac{3^2\,5^{5/2}}{\sqrt{2\pi}}\,
\nu^8
\exp\!\left[
-\frac{5}{2}\nu^2(3e^2+p^2)
\right]
W(e,p)\,de\,dp ,
\label{eq:Pepnu_BBKS}
\end{equation}
where \(f(\nu)\) is the function which plays the role of the normalization factor for the conditional
distribution at fixed peak height. In the monochromatic limit, \(x=\nu\), and
therefore
\begin{equation}
\begin{aligned}
f(\nu)
={}&
\frac{\nu^3-3\nu}{2}
\left[
\operatorname{erf}\!\left(\sqrt{\frac{5}{2}}\,\nu\right)
+
\operatorname{erf}\!\left(\sqrt{\frac{5}{8}}\,\nu\right)
\right]
\\
&+
\sqrt{\frac{2}{5\pi}}
\left[
\left(\frac{31}{4}\nu^2+\frac{8}{5}\right)
e^{-5\nu^2/8}
+
\left(\frac{\nu^2}{2}-\frac{8}{5}\right)
e^{-5\nu^2/2}
\right],
\end{aligned}
\end{equation}

\begin{equation}
\int de\,dp\,P_{\rm BBKS}(e,p\mid \nu)=1
\end{equation}

over the allowed BBKS domain.

In particular, within the BBKS support \eqref{eq:chi_ep}, the combination
entering the largest deformation eigenvalue satisfies
\begin{equation}
1<1+3e+p<3,
\end{equation}
so that
\begin{equation}
\frac{\sigma\nu}{3}<\alpha<\sigma\nu .
\end{equation}

The additional cutoff \(\alpha\leq 1/2\) (which comes from requiring that the first-axis turnaround does not occur
before the reference epoch \(a_k=1\), since \(a_{\rm turn}=1/(2\alpha)\)) then imposes a restriction on the
allowed BBKS shapes at fixed \(\nu\). Since
\(\alpha=(\sigma\nu/3)(1+3e+p)\) and
\(1<1+3e+p<3\) within the BBKS support, all shapes automatically satisfy
\(\alpha\leq 1/2\) if \(\nu\leq 1/(2\sigma)\). For
\(1/(2\sigma)<\nu<3/(2\sigma)\), only the subset satisfying
\begin{equation}
    1+3e+p \leq \frac{3}{2\sigma\nu}
\end{equation}
remains allowed. Finally, for \(\nu\geq 3/(2\sigma)\), the cutoff excludes
the whole BBKS shape domain.

Equation~\eqref{eq:npk_mono_nuep} should be interpreted as a number density
of peaks, whereas Eq.~\eqref{eq:Pepnu_BBKS} is a conditional probability
density in the \((e,p)\)-plane at fixed \(\nu\). This distinction is
important for the construction of the GW background: normalized
distributions such as \(P_{\rm BBKS}(e,p\mid\nu)\) determine the relative
statistical weight of different shapes, whereas
\(d^3n_{\rm pk}/d\nu\,de\,dp\) additionally includes the abundance of peaks
per unit volume.

In summary, the Doroshkevich and BBKS descriptions differ in two important
respects. First, the BBKS domain is more restrictive, since it retains only true maxima
of the smoothed field \(F=-\Psi_k\), which correspond to overdense
configurations in the monochromatic limit. Second, even within the common
region of the \((e,p)\)-plane, the statistical weighting differs. The two
approaches therefore correspond to different ensembles of initial
conditions, and the resulting GW predictions need not coincide, even in the
monochromatic limit.

\subsection{Calculation of the gravitational waves}
\label{sec:theory_gws_calculation}

In the present implementation, the GW emission is estimated using the standard
leading-order formula based on the Newtonian mass quadrupole
\cite{PhysRev.131.435}. We construct the trace-free rest-mass quadrupole tensor
with respect to the ordinary center of mass, using the particle trajectories
obtained from the semirelativistic N-body evolution introduced above. This mass-centroid subtraction should be distinguished from the center-of-momentum condition used for the virial analysis. The quadrupole is constructed from particle positions relative to the instantaneous mass centroid, whereas the internal dynamical frame is fixed by subtracting the bulk canonical momentum. Thus, the
particle dynamics retain the relativistic momentum--velocity relation, while
the radiation extraction is performed at the leading order in the weak-field
multipolar expansion and slow-motion approximation.

This prescription provides the leading weak-field and slow-motion contribution
to the emitted radiation. Corrections beyond this approximation would arise
from higher-order terms in the radiative source moments, including
velocity-dependent contributions to the mass multipoles, stress terms, current
multipoles, and post-Newtonian or fully relativistic radiation effects. The inclusion of such corrections is left for future work. For the cases in which
particle velocities become moderately relativistic, these higher-order effects may introduce a factor-of-order-unity uncertainty. In Sec.~\ref{sec:numerical_results_all}, we
assess the slow-motion approximation by analyzing the particle velocity
distributions.

The basic momentum variable evolved in the simulation is
\begin{equation}
\bm q_n = a\,\gamma_n\,\bm v^{\rm pec}_n,
\qquad
\gamma_n = \sqrt{1+\frac{|\bm q_n|^2}{a^2}},
\qquad
\bm v^{\rm pec}_n
=
\frac{\bm q_n}{\sqrt{a^2+|\bm q_n|^2}}.
\label{eq:q_sr_gw}
\end{equation}
Here and in the dynamical equations below we use units with $c=1$, while explicit factors
of $c$ are restored in the final GW observables.

The physical position of particle $n$ is
\begin{equation}
\bm r_n = a\,\bm x_n,
\label{eq:r_phys_gw}
\end{equation}
and the center of mass is
\begin{equation}
\bm R \equiv \bm r_{\rm CM}
=
\frac{\sum_{n=1}^N m_n \bm r_n}{\sum_{n=1}^N m_n}.
\label{eq:Rcm_gw}
\end{equation}
We then define the relative position
\begin{equation}
\bm y_n \equiv \bm r_n - \bm R.
\label{eq:yalpha_def}
\end{equation}
The trace-free mass quadrupole tensor is

\begin{equation}
Q_{ij}(t)
=
\sum_{n=1}^{N} m_n
\left[
y_{n,i}y_{n,j}
-\frac{1}{3}\delta_{ij}|\bm y_n|^2 
\right],
\qquad
|\bm y_n|^2
\equiv
\bm y_n\cdot\bm y_n
=
\sum_{k=1}^{3} y_{n,k}^2 .
\label{eq:Qij_srnewt}
\end{equation}

By construction, \(\sum_{i=1}^{3} Q_{ii}=0\). A direct numerical differentiation of $Q_{ij}(t)$ is in general very noisy, especially for the
higher derivatives entering the quadrupole formula. For this reason, instead of differentiating
the sampled quadrupole time series, we compute the required derivatives analytically from the
particle positions and from the force quantities already available during the evolution.

It is convenient to define
\begin{equation}
\bm u_n \equiv \bm v^{\rm pec}_n
=
\frac{\bm q_n}{s_n},
\qquad
s_n \equiv \sqrt{a^2+|\bm q_n|^2},
\label{eq:ualpha_def}
\end{equation}
so that the total physical velocity of particle $n$ is
\begin{equation}
\bm v_n \equiv \dot{\bm r}_n
=
H\,\bm r_n + \bm u_n.
\label{eq:valpha_def}
\end{equation}
The corresponding center-of-mass velocity is
\begin{equation}
\dot{\bm R}
=
\frac{\sum_{n=1}^N m_n \bm v_n}{\sum_{n=1}^N m_n},
\label{eq:Rdot_def}
\end{equation}
and therefore
\begin{equation}
\dot{\bm y}_n
=
\bm v_n-\dot{\bm R}.
\label{eq:ydot_def}
\end{equation}

The momentum equation of motion is
\begin{equation}
\dot{\bm q}_n = \frac{\bm F_n}{a},
\label{eq:qdot_force_gw}
\end{equation}
where $\bm F_n$ is the softened comoving gravitational force. Using 
\begin{equation}
\dot s_n
=
\frac{a^2H+\bm q_n\!\cdot\!\dot{\bm q}_n}{s_n},
\label{eq:sdot_def}
\end{equation}
the time derivative of the peculiar velocity is
\begin{equation}
\dot{\bm u}_n
=
\frac{\dot{\bm q}_n}{s_n}
-
\bm q_n \frac{\dot s_n}{s_n^2}.
\label{eq:udot_def}
\end{equation}

For an EdS 
background, one has
\begin{equation}
H=\frac{2}{3t},
\qquad
\dot H = -\frac{3}{2}H^2,
\qquad
A\equiv \frac{\ddot a}{a}=-\frac{1}{2}H^2,
\qquad
\dot A = \frac{3}{2}H^3.
\label{eq:EdS_relations_srnewt}
\end{equation}
Using these relations, the physical acceleration of particle $n$ is
\begin{equation}
\bm a_n \equiv \ddot{\bm r}_n
=
A\,\bm r_n + H\,\bm u_n + \dot{\bm u}_n,
\label{eq:acc_srnewt}
\end{equation}
and the center-of-mass acceleration is
\begin{equation}
\ddot{\bm R}
=
\frac{\sum_{n=1}^N m_n \bm a_n}{\sum_{n=1}^N m_n}.
\label{eq:Rddot_def}
\end{equation}
Hence the relative acceleration is
\begin{equation}
\ddot{\bm y}_n
=
\bm a_n-\ddot{\bm R}.
\label{eq:yddot_def}
\end{equation}

To obtain the jerk, which is defined as the third time derivative of the physical position of the $n$-th particle $\bm j_n \equiv \dddot{\bm r}_n$, we first note that
\begin{equation}
\ddot{\bm q}_n
=
\frac{\dot{\bm F}_n - H \bm F_n}{a}.
\label{eq:qddot_force_gw}
\end{equation}
Then
\begin{equation}
\ddot s_n
=
\frac{
a^2(2H^2+\dot H)
+\dot{\bm q}_n^{\,2}
+\bm q_n\!\cdot\!\ddot{\bm q}_n
-\dot s_n^{\,2}
}{s_n},
\label{eq:sddot_def}
\end{equation}
and therefore
\begin{equation}
\ddot{\bm u}_n
=
\frac{\ddot{\bm q}_n}{s_n}
-
2\dot{\bm q}_n\,\frac{\dot s_n}{s_n^2}
-
\bm q_n\,\frac{\ddot s_n}{s_n^2}
+
2\bm q_n\,\frac{\dot s_n^2}{s_n^3}.
\label{eq:uddot_def}
\end{equation}
The physical jerk of particle $n$ is then
\begin{equation}
\bm j_n 
=
\dot A\,\bm r_n
+
A\,\bm v_n
+
\dot H\,\bm u_n
+
H\,\dot{\bm u}_n
+
\ddot{\bm u}_n.
\label{eq:jerk_srnewt}
\end{equation}
Similarly, the center-of-mass jerk is
\begin{equation}
\dddot{\bm R}
=
\frac{\sum_{n=1}^N m_n \bm j_n}{\sum_{n=1}^N m_n},
\label{eq:Rdddot_def}
\end{equation}
so that the relative jerk is
\begin{equation}
\dddot{\bm y}_n
=
\bm j_n-\dddot{\bm R}.
\label{eq:ydddot_def}
\end{equation}

With these definitions, the first time derivative of the quadrupole tensor is
\begin{equation}
\dot Q_{ij}
=
\sum_{n=1}^{N}
m_n
\left[
\dot y_{n,i} y_{n,j}
+
y_{n,i} \dot y_{n,j}
-
\frac{2}{3}\delta_{ij}
\left(
\bm y_n\cdot\dot{\bm y}_n
\right)
\right].
\label{eq:dQij_srnewt}
\end{equation}
The second derivative is
\begin{equation}
\ddot Q_{ij}
=
\sum_{n=1}^{N}
m_n
\left[
\ddot y_{n,i} y_{n,j}
+
y_{n,i} \ddot y_{n,j}
+
2\dot y_{n,i}\dot y_{n,j}
-
\frac{2}{3}\delta_{ij}
\left(
|\dot{\bm y}_n|^2
+
\bm y_n\cdot\ddot{\bm y}_n
\right)
\right].
\label{eq:ddQij_srnewt}
\end{equation}
Finally, the third derivative is
\begin{equation}
\dddot Q_{ij}
=
\sum_{n=1}^{N}
m_n
\left[
y_{n,i}\dddot y_{n,j}
+
\dddot y_{n,i}y_{n,j}
+
3\ddot y_{n,i}\dot y_{n,j}
+
3\dot y_{n,i}\ddot y_{n,j}
-
\frac{2}{3}\delta_{ij}
\left(
\bm y_n\cdot\dddot{\bm y}_n
+
3\,\dot{\bm y}_n\cdot\ddot{\bm y}_n
\right)
\right].
\label{eq:dddQij_srnewt}
\end{equation}

In the numerical implementation, the quadrupole and its third derivative are evaluated at the
synchronized integer-time state of the kick--drift--kick leapfrog scheme. Namely, after the
second kick one has the updated variables $(\bm x^{\,\ell+1}_n,\bm q^{\,\ell+1}_n)$, from which the code
recomputes the force and its first time derivative at the same time slice. The quantities
$Q_{ij}(t_{\ell+1})$ and $\dddot Q_{ij}(t_{\ell+1})$ are then built directly from that synchronized
state, avoiding any mismatch between the leapfrog time-centering and the GW
source evaluation.

The instantaneous GW luminosity is obtained from the quadrupole formula,
\begin{equation}
P_{\rm GW}(t)
\equiv
\frac{dE_{\rm GW}}{dt}
=
\frac{G}{5c^5}
\sum_{i,j}
\dddot Q_{ij}(t)\,\dddot Q_{ij}(t).
\label{eq:GW_power_srnewt}
\end{equation}
For each configuration, we first compute the instantaneous GW power and the corresponding cumulative emitted energy,
\begin{equation}
E_{\rm GW}(<t)
=
\int_{t_{\rm ini}}^{t} dt'\,P_{\rm GW}(t') \, .
\label{eq:EGW_cumulative_srnewt}
\end{equation}
We then introduce an effective emission time, $t_e$, defined as the time beyond which GW production can be regarded as negligible. In practice, $t_e$ is determined as the time at which $99\%$ of the total GW energy has already been emitted. This condition reads
\begin{equation}
E_{\rm GW}(<t_e)=0.99\,E_{\rm GW}^{\rm tot},
\qquad
E_{\rm GW}^{\rm tot}=E_{\rm GW}(<t_{\rm end}) \, ,
\label{eq:teff_def_srnewt}
\end{equation}
where $t_{\rm end}$ is the final time of the simulation.

To characterize the signal in frequency space, we introduce the Fourier transform
\begin{equation}
\widetilde X(f)
\equiv
\int_{-\infty}^{+\infty} dt\, e^{-2\pi i f t}X(t).
\end{equation}
For the quadrupole derivative this implies
\begin{equation}
\widetilde{\dddot Q}_{ij}(f)
=
(2\pi i f)^3 \widetilde Q_{ij}(f),
\end{equation}
up to the usual boundary terms, which are negligible for the finite-duration windowed
signals used in the numerical implementation. Since \(\dddot Q_{ij}(t)\) is real,
\begin{equation}
\widetilde{\dddot Q}_{ij}(-f)
=
\widetilde{\dddot Q}_{ij}(f)^* .
\end{equation}
Thus the negative-frequency part of the two-sided transform gives the same contribution
as the positive-frequency part. In the numerical implementation we therefore use the
positive-frequency, one-sided spectrum. For \(f_e>0\), the emitted spectral energy
distribution is
\begin{equation}
\frac{dE_{\rm GW}}{df_e}
=
\frac{2G}{5c^5}
\sum_{i,j}
\left|
\widetilde{\dddot Q}_{ij}(f_e)
\right|^2 .
\end{equation}

To translate the emitted signal to the present epoch, we assume an abrupt transition from the
early matter-dominated era to radiation domination at reheating. Entropy conservation gives\footnote{We emphasize that the sudden-transition approximation adopted
here is used only to specify the cosmological redshift of the
collapse-generated GW signal. For this purpose, approximating a
finite-duration transition as instantaneous is adequate at the level
of accuracy considered here, provided that the transition does not
substantially modify the background expansion over an extended period.
By contrast, if the transition from matter domination to radiation
domination is sufficiently rapid, the scalar-induced GW contribution
can be strongly enhanced just after the transition through the
Poltergeist mechanism
\cite{Inomata:2019ivs}.
This contribution is physically distinct from the direct quadrupole
radiation sourced by the nonspherical motion and redistribution of
matter during the nonlinear gravitational collapse studied here, and
is not included in the spectra presented below. In a complete
cosmological scenario, the collapse-generated and scalar-induced
contributions may coexist, with their relative importance depending
on the primordial scalar spectrum, the duration of the transition,
and the timing of the nonlinear collapse relative to reheating.}
\begin{equation}
1 + z_{\rm rh}
=
\frac{T_{\rm rh}}{T_0}
\left(
\frac{g_{*s}(T_{\rm rh})}{g_{*s}(T_0)}
\right)^{1/3},
\label{eq:zRH_def}
\end{equation}
while the radiation energy density at reheating is
\begin{equation}
\rho_{\rm rh}
=
\frac{\pi^2}{30}\,
g_{*\rho}(T_{\rm rh})\,
\frac{(k_\text{B} T_{\rm rh})^4}{\hbar^3 c^3}.
\label{eq:rho_rad_RH}
\end{equation}
Here $T_{\rm rh}$ denotes the reheating temperature, $T_0$ the temperature today and \(g_{*\rho}(T)\) and \(g_{*s}(T)\) denote the effective relativistic degrees of freedom contributing to the energy density and entropy density, respectively. In the numerical implementation we use spline interpolations
of the tabulated Standard Model results of Ref.~\cite{Saikawa:2018rcs} (see also Ref.~\cite{Drees:2015exa}), where both quantities are given as functions of the temperature. Then the corresponding mass density is
\begin{equation}
\rho_{{\rm M},\,{\rm rh}}
=
\frac{\rho_{\rm rh}}{c^2},
\label{eq:rho_mass_RH}
\end{equation}
and the Hubble rate and reheating time are
\begin{equation}
H_{\rm rh}
=
\sqrt{\frac{8\pi G}{3c^2}\,\rho_{\rm rh}},
\qquad
t_{\rm rh}
=
\frac{2}{3H_{\rm rh}}.
\label{eq:tRH_code}
\end{equation}

Before reheating, the background is assumed to be matter-dominated, $a\propto t^{2/3}$.
Therefore, for an emission time $t_e<t_{\rm rh}$, the redshift at emission is
\begin{equation}
1+z_e
=
(1+z_{\rm rh})
\left(\frac{t_{\rm rh}}{t_e}\right)^{2/3}.
\label{eq:ze_code}
\end{equation}
The observed frequency today is then
\begin{equation}
f_0
=
\frac{f_e}{1+z_e}.
\label{eq:f0_code}
\end{equation}

In our framework and numerical methodology, we associate the horizon-mass scale $M_k$ to a comoving wavenumber
\begin{equation}
k_{\rm com}(M_k,T_{\rm rh})
=
\left[
\frac{3\,M_k\,(1+z_{\rm rh})^3}
{4\pi\,\rho_{{\rm M},\,{\rm rh}}}
\right]^{-1/3},
\label{eq:kcom_code}
\end{equation}
from which the corresponding comoving volume is
\begin{equation}
V_k
=
\frac{4\pi}{3}\,k_{\rm com}^{-3}.
\label{eq:Vk_code}
\end{equation}
Since the physical emission volume is $V_e=a_e^3V_k=V_k/(1+z_e)^3$, the present-day energy
density follows from the usual radiation redshifting, yielding one power of $(1+z_e)^{-1}$ in
the final spectrum per logarithmic frequency interval.

The present critical energy density is
\begin{equation}
\rho_{c,0}
=
\frac{3H_0^2c^2}{8\pi G},
\qquad
h\equiv
\frac{H_0}{100\,{\rm km\,s^{-1}\,Mpc^{-1}}}.
\label{eq:rho_c0_code}
\end{equation}
With these definitions, the present-day GW spectrum for a given single realization is
\begin{equation}
h^2\Omega_{{\rm GW},0}(f_0)
=
\frac{h^2}{\rho_{c,0}}\,
\frac{1}{V_k(1+z_e)}
\left.
\frac{dE_{\rm GW}}{d\ln f_e}
\right|_{f_e=(1+z_e)f_0}.
\label{eq:OmegaGW_today_code_single}
\end{equation}


Before performing the statistical average, it is useful to clarify the
meaning of the spectrum assigned to a single realization in
Eq.~\eqref{eq:OmegaGW_today_code_single}. For a given configuration labelled
by \((\nu,e,p)\), or equivalently by \((\alpha,\beta,\gamma)\), the simulation
directly gives the GW energy spectrum emitted by one collapsing patch,
\(dE_{\rm GW}/d\ln f_e\). To express this single-event result as an
energy-density parameter, we associate the patch with the representative
comoving volume \(V_k\) corresponding to the smoothing scale. Therefore
\(h^2\Omega^{(\nu,e,p)}_{\rm GW,0}\) should be understood as the contribution
per smoothing volume from a patch with fixed initial parameters, rather than
as the final stochastic background.

The statistically weighted spectrum is then obtained by integrating the
present-day spectrum of each individual realization over the space of
initial deformation parameters. For the BBKS peak ensemble, the
corresponding contribution is
\begin{equation}
h^2\Omega^{\rm BBKS}_{\rm GW,0}(f_0)
=
\int_{\mathcal{D}_{\rm BBKS}}
d\nu\,de\,dp\,
V_k\,
\frac{d^3 n_{\rm pk}}{d\nu\,de\,dp}\,
h^2\Omega^{(\nu,e,p)}_{\rm GW,0}(f_0) ,
\label{eq:Omega_BBKS_average}
\end{equation}
where \(\mathcal{D}_{\rm BBKS}\) denotes the BBKS support defined by
Eq.~\eqref{eq:chi_ep}. For the Doroshkevich ensemble, the weighted spectrum
is obtained by averaging over the probability distribution of the deformation
parameters \((\alpha,\beta,\gamma)\),
\begin{equation}
h^2\Omega^{\rm D}_{\rm GW,0}(f_0)
=
\int_{\mathcal{D}_{\rm D}}
d\alpha\,d\beta\,d\gamma\,
P_{\rm D}(\alpha,\beta,\gamma)\,
h^2\Omega^{(\alpha,\beta,\gamma)}_{\rm GW,0}(f_0) ,
\label{eq:Omega_D_average_alphabetagamma}
\end{equation}
where \(\mathcal{D}_{\rm D}\) is the domain satisfying
\(\alpha \geq \beta \geq \gamma\) and \(\alpha+\beta+\gamma>0\).

The interpretation of the weighting differs in the two cases. In
the BBKS case, \(d^3n_{\rm pk}/d\nu\,de\,dp\) is a comoving number density
of peaks, so that \(V_k\,d^3n_{\rm pk}/d\nu\,de\,dp\) gives the expected
number of peaks in one smoothing volume. In the Doroshkevich case, the
result corresponds instead to a probability-weighted estimate over the
deformation parameters, effectively assigning one representative region to
each smoothing volume.

Finally, in the numerical implementation a high-frequency cut is imposed from the observed Nyquist frequency. If $d t_e$ is the physical timestep of the emitted signal, then
\begin{equation}
f_{{\rm Nyq},e}
=
\frac{1}{2d t_e},
\qquad
f_{{\rm Nyq},0}
=
\frac{f_{{\rm Nyq},e}}{1+z_e},
\qquad
f_0 \le \eta_{\rm Nyq}\,f_{{\rm Nyq},0},
\label{eq:Nyquist_cut_code}
\end{equation}
with $\eta_{\rm Nyq}=0.25$ in the present analysis.

In summary, our numerical procedure is the following: we evolve the system with
semirelativistic particle kinematics, construct the standard rest-mass quadrupole
$Q_{ij}(t)$ from the synchronized particle data, compute its analytic third derivative
$\dddot Q_{ij}(t)$ from the same time slice, obtain the emitted spectrum through a Fourier
transform, and finally redshift the result to the present epoch using the reheating history of
the model.

\section{Numerical implementation and initial conditions}
\label{ref:num_simul}

In this section we describe the numerical methodology used to perform the calculations presented in this work. For this purpose, we have developed a dedicated C++/Python N-body code from scratch in order to evolve the dynamics of the collapsing patch and compute the corresponding observables. Since the force evaluation is performed through direct particle--particle summation, the computational cost scales as $\mathcal{O}(N^2)$. For this reason, the code includes parallelized routines specifically designed to accelerate the force calculation. We refer the reader to Ref.~\cite{escriva_github} for details of the simulations, along with some representative animations.

The simulation is performed in dimensionless code units normalized to the equivalent patch radius at horizon entry. For a characteristic horizon-mass scale \(M_k\), this fixes the time $t_k$. Then the corresponding equivalent physical radius at horizon entry is
\begin{equation}
R_{{\rm eq},k}
=
\left[
\frac{3M_{k}}{4\pi \rho_{\rm b}(t_k)}\right]^{1/3} = c \, H^{-1}_{k},
\label{eq:Reqk}
\end{equation}

The fundamental code units are then chosen as
\begin{equation}
L_{\rm unit}=R_{{\rm eq},k},
\qquad
T_{\rm unit}=\frac{L_{\rm unit}}{c},
\qquad
M_{\rm unit}=\frac{L_{\rm unit}^3}{G\,T_{\rm unit}^2},
\qquad
E_{\rm unit}=\frac{M_{\rm unit} L^2_{\rm unit}}{T_{\rm unit}^2},
\label{eq:code_units}
\end{equation}
so that $G=c=1$ in code units for the simulations. For equal-mass particles, the particle mass is $m=M_k/N$. On the other hand, in our setup we typically use a global constant time-step $dt$, although we have also tested an adaptive prescription based on the jerk of the acceleration. We refer the reader to Appendix~\ref{sec:time_step_convergence} for a discussion.

An important aspect of the numerical setup is the construction of the initial conditions. The initial particle load is generated in two conceptually distinct stages. First, we construct an approximately homogeneous and isotropic particle distribution inside a unit Lagrangian sphere. Second, we impose the desired triaxial perturbation through the Zel'dovich affine deformation associated with the three parameters $(\alpha,\beta,\gamma)$.

The motivation for the particular initialization procedure is the following. A naive random realization of particles inside a sphere inevitably introduces discreteness-induced anisotropies and inhomogeneities. In particular, for a finite number of particles N, the resulting particle cloud has a non-vanishing trace-free quadrupole moment already at the initial time. Since the GW signal is extracted from the time evolution of the quadrupole tensor, such residual discreteness effects can seed spurious radiation. For this reason, we construct the initial particle load so as to be simultaneously as homogeneous and isotropic as possible, while driving the initial quadrupole to a negligible level. The use of a glass-like particle load is motivated by the standard cosmological N-body literature (see for instance Ref.~\cite{YoshidaSugiyamaHernquist2003}), where such pre-initial conditions are employed to suppress spurious small-scale clumping and discreteness-induced correlations. The methodology developed in this work is divided into the following steps.

\paragraph{Step 1: glass-like homogeneous spherical load.}
Let \(N\) be the number of particles and let \(\bm{\xi}_a\in\mathbb{R}^3\) (\(a=1,\dots,N\)) denote their Lagrangian positions. We begin from a uniform random realization\footnote{For all non-spherical configurations, we use the same random seed generator.} inside a sphere of radius \(R_L=1\),
\[
|\bm{\xi}_a|\le R_L,
\qquad
P(<r)\propto r^3,
\]
where $P(<r)$ is the cumulative radial probability, and then iteratively relax the particle distribution through a short-range repulsive update. The characteristic mean inter-particle spacing is
\[
\ell = \left(\frac{V_L}{N}\right)^{1/3},
\qquad
V_L=\frac{4\pi}{3}R_L^3.
\]
The repulsive interaction is applied only for separations smaller than a cutoff
\[
r_{\rm cut}=\max\!\left(c_{\rm cut}\,\ell,\;0.9\,h_{\rm soft}\right),
\qquad
h_{\rm soft}=2.8\,\epsilon_{\rm com},
\]
where \(\epsilon_{\rm com}\) is the softening length used in the simulation and \(c_{\rm cut}\) is a tunable constant that we take $c_{\rm cut}=1.25$. For each pair with separation \(r<r_{\rm cut}\), we apply a compact repulsive displacement proportional to
\[
\Delta \xi \propto \frac{(1-r/r_{\rm cut})^2}{r}.
\]
After each relaxation sweep, particles that moved outside the sphere are projected back onto the surface, the center of mass is removed, and the radial distribution is remapped so as to preserve the cumulative profile of a homogeneous ball,
\[
\frac{N(<r)}{N}\simeq \left(\frac{r}{R_L}\right)^3,
\]
where $N(<r)$ is the number of particles within the radius $r$, and $N(\le R_L)= N$.

In the implementation used in this work, the main relaxation stage is followed by a weaker anisotropy-reduction step, a short post-relaxation stage, and finally an exact quadrupole-cleanup step.

\paragraph{Step 2: exact zero-quadrupole cleanup.}
To suppress the residual low-order anisotropy of the discrete particle realization, we compute the second-moment tensor of the Lagrangian cloud,
\[
C_{ij}=\frac{1}{N}\sum_{a=1}^{N}\xi_{a,i}\xi_{a,j}.
\]
For a perfectly isotropic cloud one has \(C_{ij}\propto \delta_{ij}\), which is equivalent to a vanishing trace-free quadrupole. We therefore apply the volume-preserving whitening transformation
\[
\bm{\xi}_a \;\mapsto\; M\,\bm{\xi}_a,
\qquad
M=(\det C)^{1/6}C^{-1/2},
\]
so that the transformed second moment satisfies
\[
C' = M C M^{\mathsf T} = (\det C)^{1/3} I.
\]
Hence the initial trace-free quadrupole of the Lagrangian cloud vanishes up to machine precision. After this transformation, the cloud is isotropically rescaled to fit again inside the sphere \(R_L=1\), and the center of mass is subtracted once more.

Once the isotropic Lagrangian cloud has been generated, we impose the desired triaxial perturbation through the Zel'dovich approximation described in Sec.~\ref{sec:zeldovich}. In this way, the deformation associated with the eigenvalues \((\alpha,\beta,\gamma)\) is introduced on top of a particle realization whose intrinsic discreteness quadrupole has already been strongly suppressed.

This construction is designed to reduce artificial GW emission seeded purely by shot noise. Compared with a purely random sampling of the initial sphere, the present method simultaneously suppresses close particle pairs, preserves an approximately homogeneous radial profile, enforces vanishing center of mass, enforces a vanishing initial trace-free quadrupole up to machine precision, and still allows one to realize arbitrary triaxial configurations through the affine Zel'dovich map. For a spherical benchmark, \(\alpha=\beta=\gamma\), the remaining spurious signal is therefore expected to be dominated by finite-\(N\) evolution effects rather than by the initialization itself.

\begin{figure}[!htbp]
\centering
\includegraphics[width=0.8\linewidth]{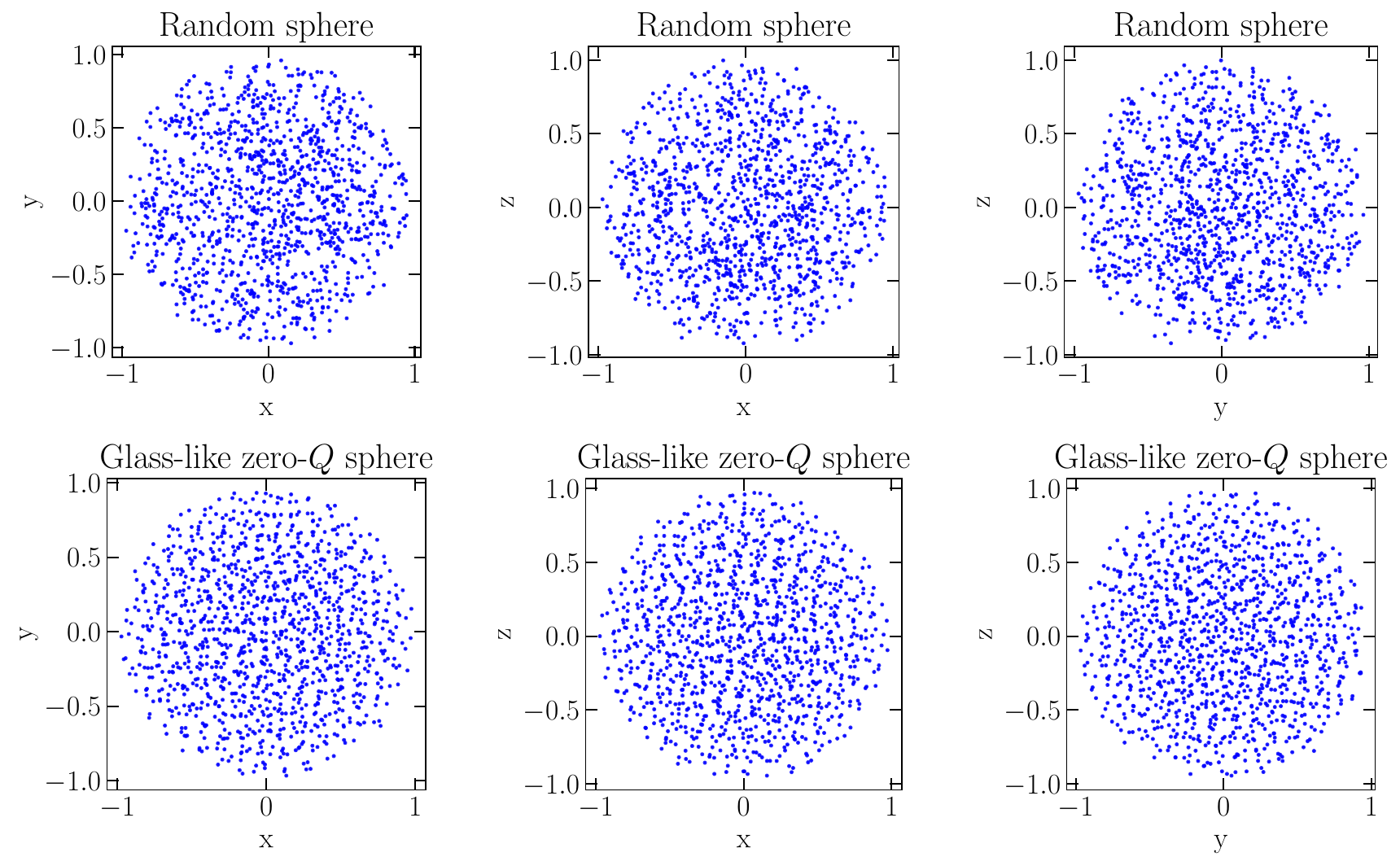}\hfill
\includegraphics[width=0.8\linewidth]{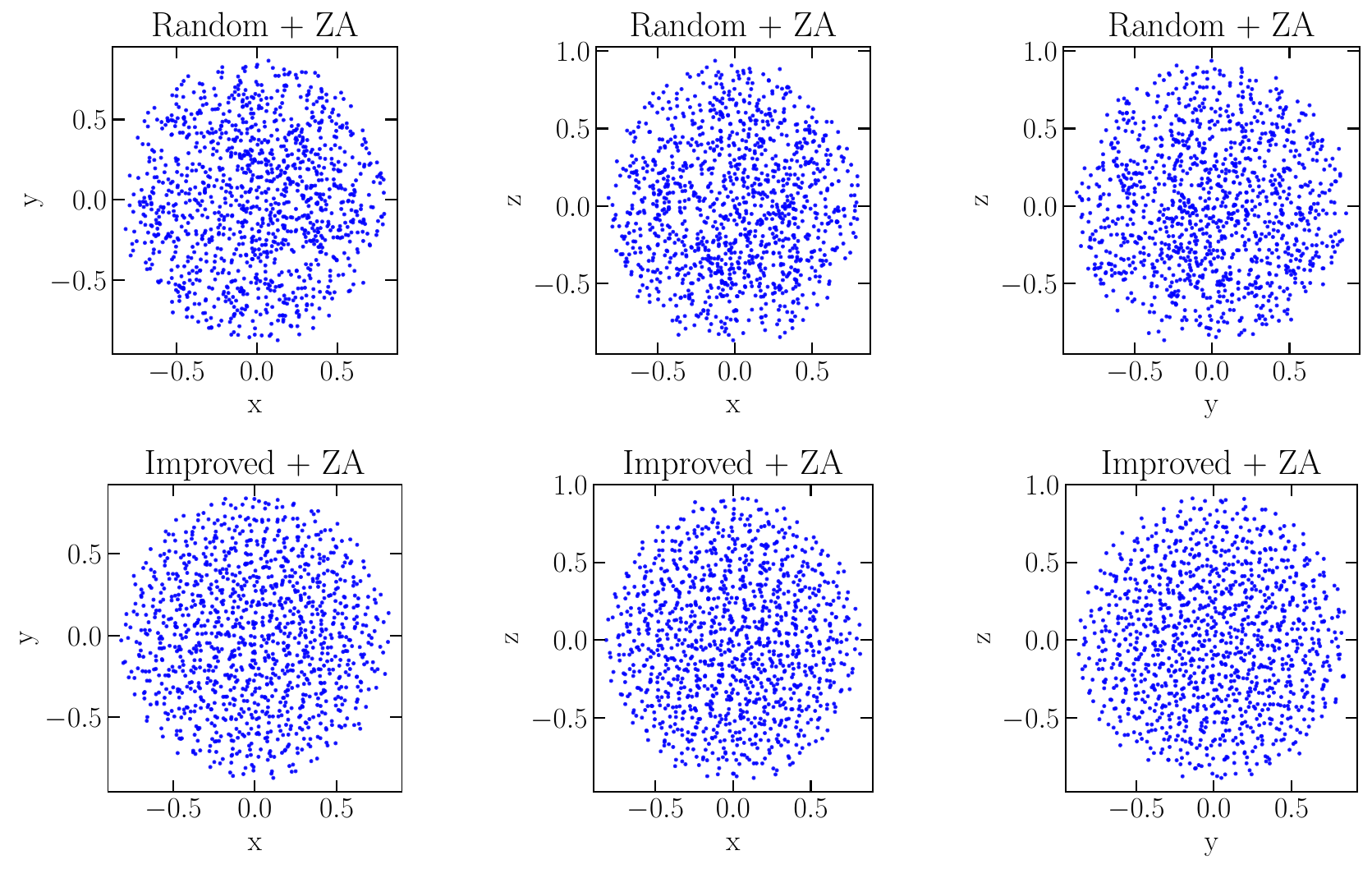}\hfill
\includegraphics[width=0.8\linewidth]{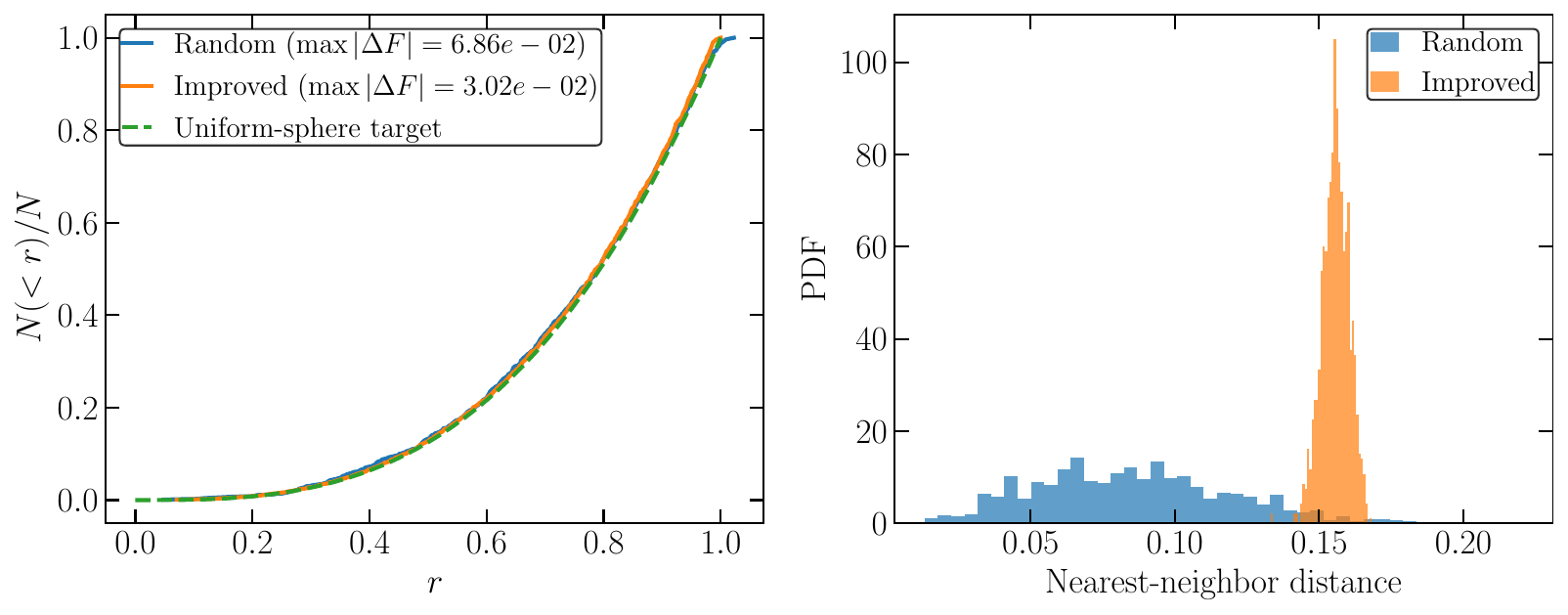}\hfill
\caption{Comparison between a purely random particle realization and the glass-like zero-quadrupole initialization adopted in this work. The first two rows show the initial particle distributions in the three coordinate projections for the random sphere and for the glass-like zero-$Q$ sphere (the improved approach), respectively. The next two rows show the corresponding configurations after applying the Zel'dovich affine deformation. The bottom-left panel displays the cumulative radial profile, where the empirical distribution $F_N(r)=N(<r)/N$ is compared with the homogeneous-sphere target $F_{\rm sph}(r)=(r/R_L)^3$; the quantity $\max|\Delta F|$, with $\Delta F(r)=F_N(r)-F_{\rm sph}(r)$, measures the maximum deviation from this target profile. The bottom-right panel shows the nearest-neighbour distance distribution.}
\label{snapshots_particles_Nbody_IC}
\end{figure}

A comparison between this methodology and the simpler case of a purely random particle realization is shown in Fig.~\ref{snapshots_particles_Nbody_IC}. In the scatter panels one can see the difference between the random initial distribution and the glass-like realization with vanishing quadrupole. The latter is visibly more homogeneous, while still preserving the global spherical profile. This improvement is especially evident in the lower-right panel, where the nearest-neighbour spacing distribution is shown: the random realization exhibits a much broader distribution, whereas the improved construction yields a significantly narrower one. At the same time, the lower-left panel shows that in both cases the cumulative radial profile remains close to that of an ideal homogeneous sphere.

Most importantly, the residual initial quadrupole is drastically reduced. For the purely random realization one typically finds values of order \(|Q_{ij}|\sim 10^{-1}\) in code units, whereas after the glass-like relaxation and quadrupole-cleanup procedure the residual value becomes as small as \(|Q_{ij}|\sim 10^{-14}\). This demonstrates the effectiveness of the method in suppressing spurious anisotropy in the quadrupole components and, consequently, in reducing artificial GW emission seeded by the initial particle load.

The nearest-neighbour distance distribution of the improved particle realization is shifted toward larger separations compared with a purely random realization. This is an expected consequence of the glass-like relaxation procedure. In a random (Poisson-like) sampling of a uniform sphere, there is a non-negligible probability of finding very close particle pairs, which produces a nearest-neighbour distribution with substantial support at small distances. By contrast, the present initialization includes a short-range repulsive relaxation that suppresses close pairs and introduces an effective exclusion scale. As a result, the nearest-neighbour distribution becomes narrower and its peak moves to larger distances. Importantly, this shift does not imply a lower mean density, since the cumulative radial profile remains close to that of a homogeneous sphere,
\[
\frac{N(<r)}{N}\simeq \left(\frac{r}{R_L}\right)^3.
\]
In the diagnostic plot, we quantify the residual deviation from this ideal radial profile by defining
\[
F_N(r)\equiv \frac{N(<r)}{N},
\qquad
F_{\rm sph}(r)\equiv \left(\frac{r}{R_L}\right)^3,
\]
and
\[
\Delta F(r)\equiv F_N(r)-F_{\rm sph}(r),
\qquad
\max|\Delta F|\equiv \max_r|\Delta F(r)| .
\]
Rather, it indicates that the particle load is locally more uniform and less affected by artificial shot-noise clustering.

\section{Numerical Results}
\label{sec:numerical_results_all}
In this section, we present the numerical results of our study. We first focus on the collapse dynamics, then consider averaging over different realizations in order to obtain statistically meaningful results for the GW emission, and finally compare these results with observational prospects by varying the model parameters.

\subsection{Dynamics of the collapse}
\label{sec:collapse_dynamics}

\begin{figure}[!htbp]
\centering
\includegraphics[width=0.3\linewidth]{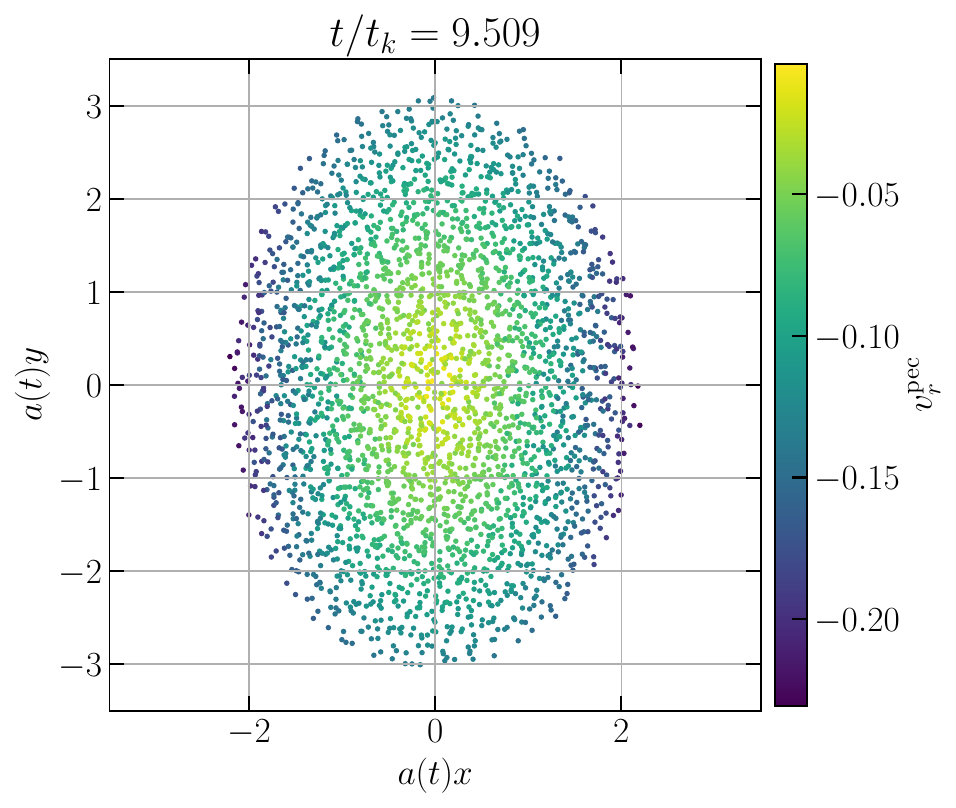}\hfill
\includegraphics[width=0.3\linewidth]{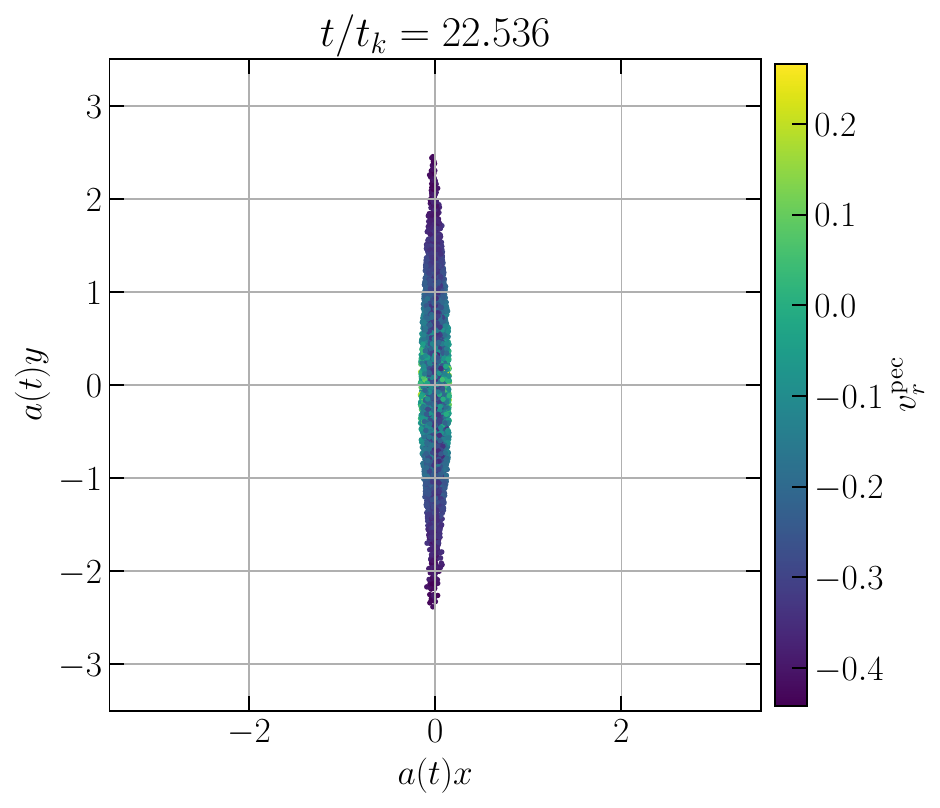}\hfill
\includegraphics[width=0.3\linewidth]{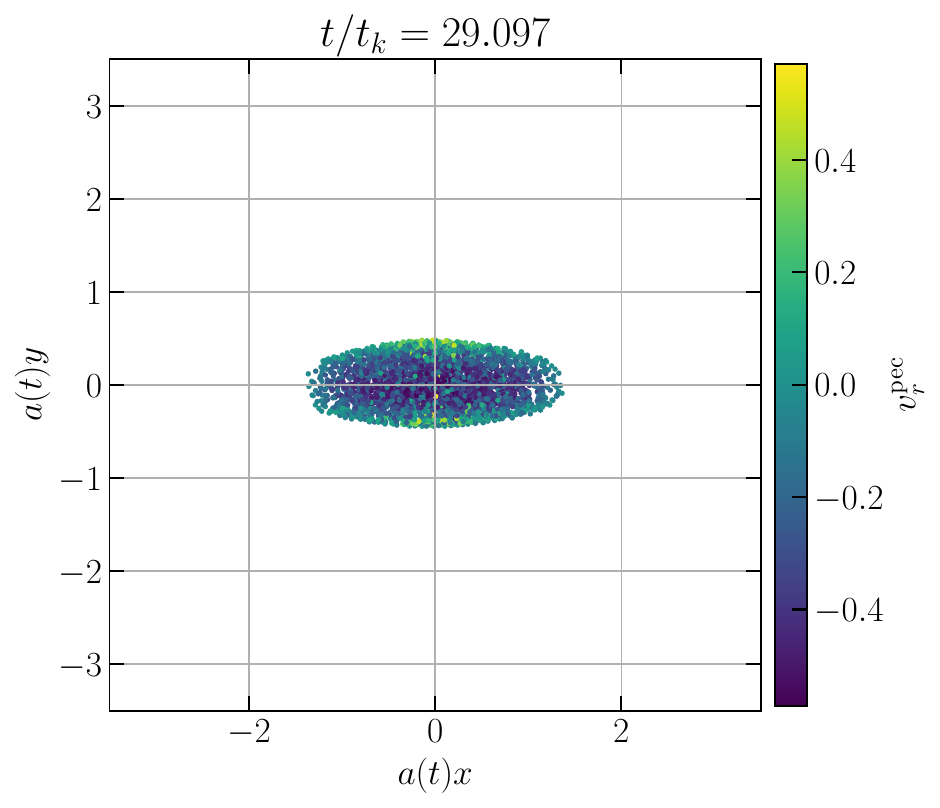}\hfill
\includegraphics[width=0.3\linewidth]{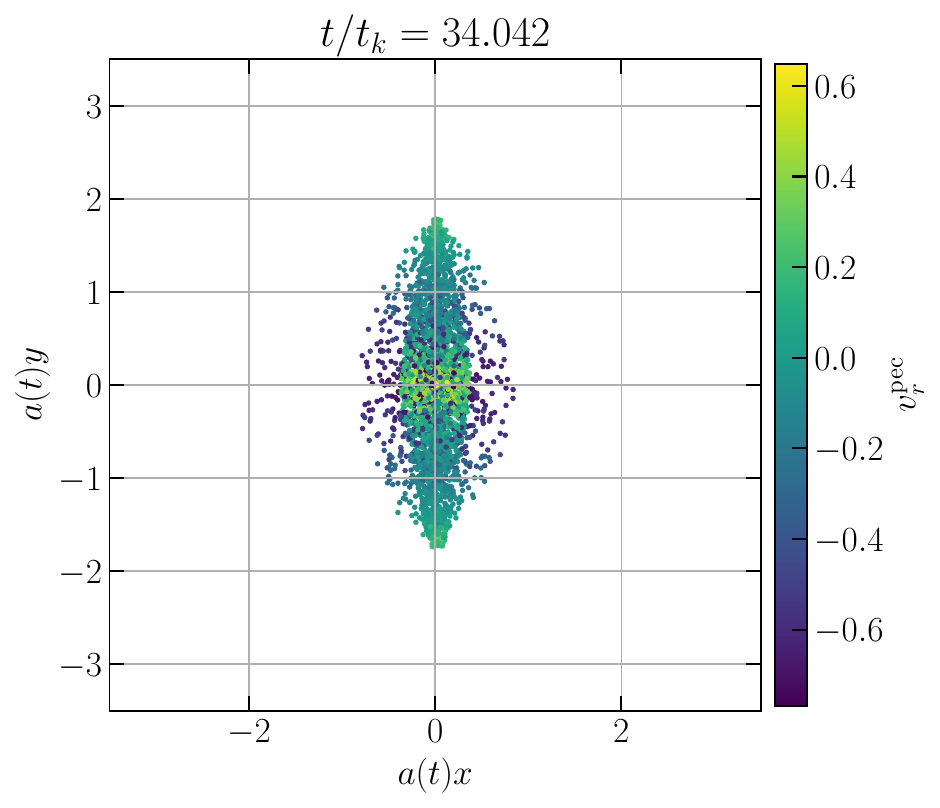}\hfill
\includegraphics[width=0.3\linewidth]{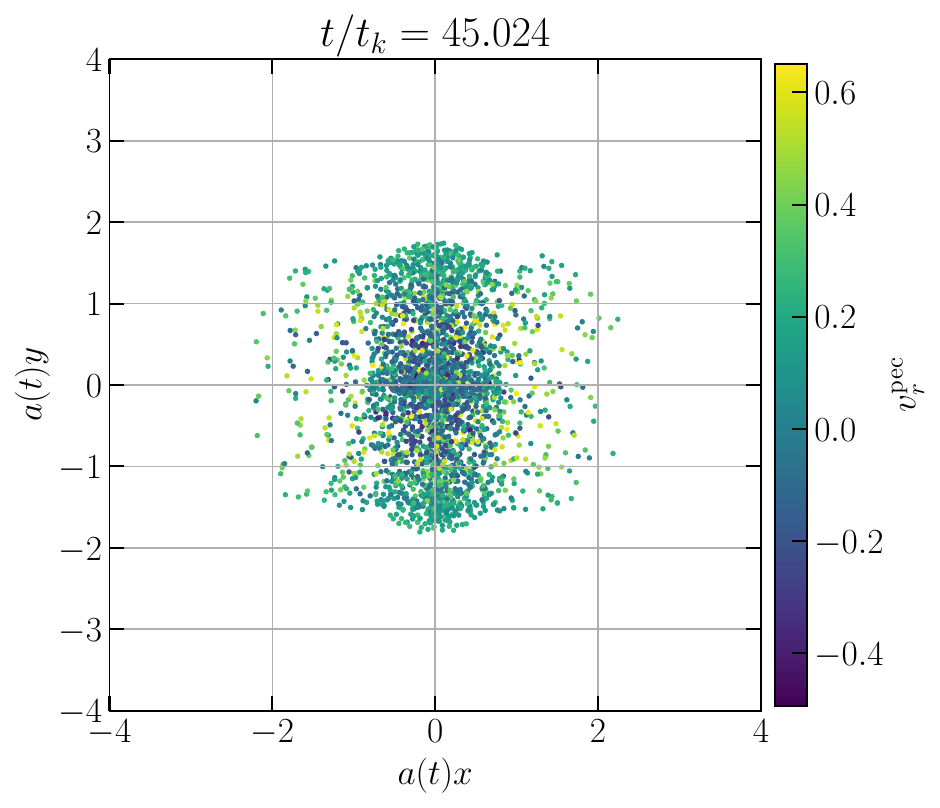}\hfill
\includegraphics[width=0.3\linewidth]{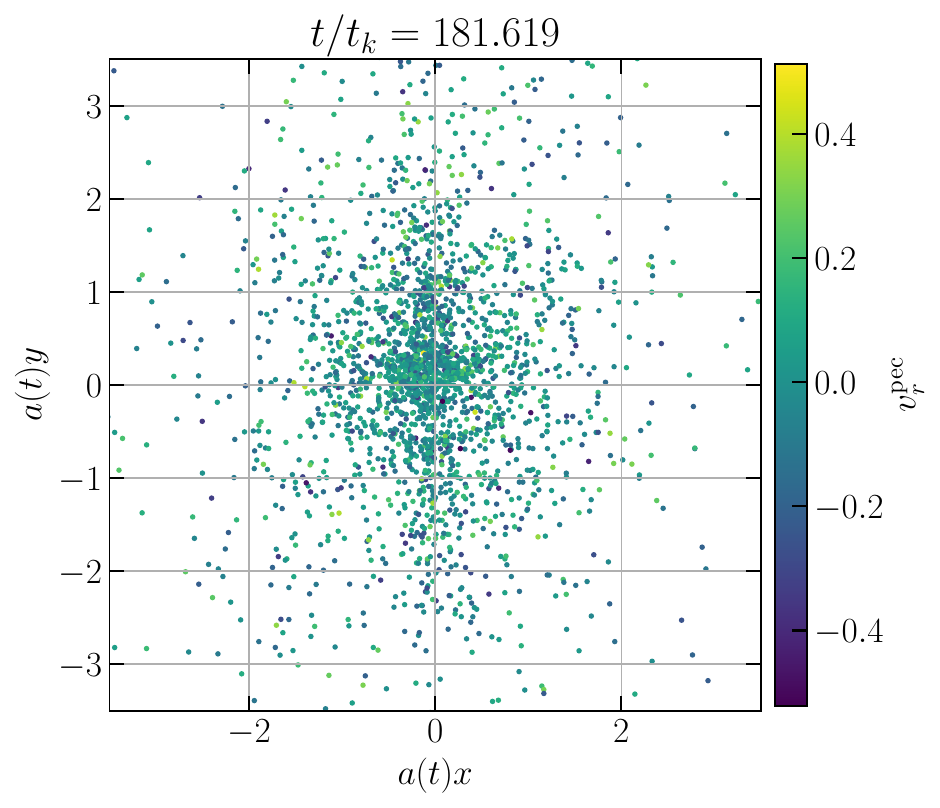}\hfill
\caption{Snapshots of the spatial particle distribution at different times, expressed in units of the horizon-crossing time of the overdense patch. The colour legend shows the radial components of the peculiar velocity $v^{\rm pec}_{r}$ (in units of $c$).}
\label{snapshots_particles_Nbody}
\end{figure}

Let us first illustrate the collapse dynamics of the N-body particle system. In Fig.~\ref{snapshots_particles_Nbody}, we present several snapshots of the collapse for a representative configuration. As an example, we consider the values $\sigma=0.1$, $\alpha \approx 0.111$, $\beta \approx 0.0668$, and $\gamma \approx 0.0219$, which correspond to $\nu \approx 2$, $e \approx 0.223$, and $p \approx -0.00125$ according to Eq.~\eqref{eq:ep_def_abg}. We also fix $N=3000$ particles, $\epsilon_{\rm com}=0.05$ and $M_{k} = 10^{-10}M_{\odot}$, we will discuss later the dependence of the results in terms of these two parameters. These snapshots highlight the main stages of the collapse.

\begin{figure}[!htbp]
\centering
\includegraphics[width=0.55\linewidth]{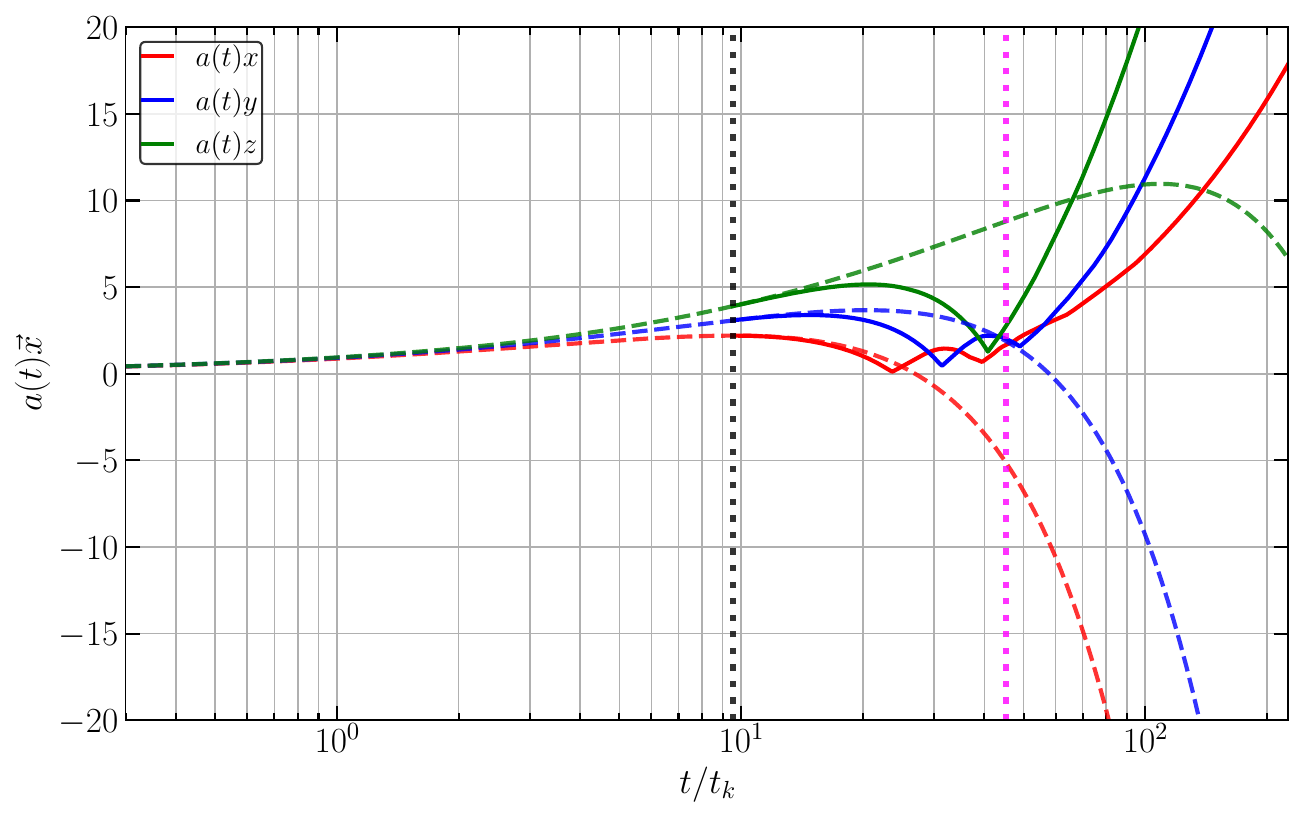}\hfill
\includegraphics[width=0.55\linewidth]{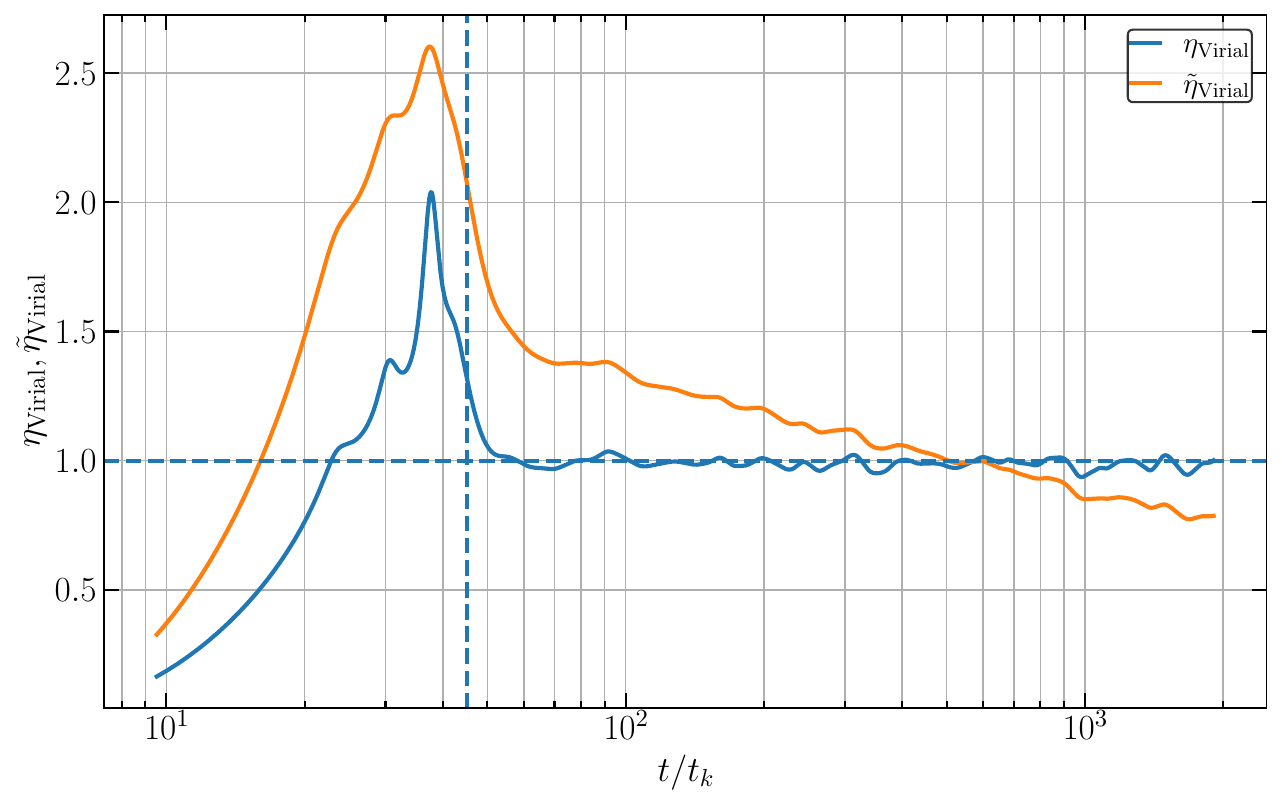}\hfill
\caption{Top panel: Evolution of the principal axes from maximum expansion in the N-body simulation (solid lines) and in the Zel'dovich approximation (dashed lines). The vertical black line denotes the initial time of the simulation, while the magenta line marks the time by which $99\%$ of the gravitational waves have been emitted. Bottom panel: Time evolution of the quantities $\eta_{\rm Virial}$ and $\tilde{\eta}_{\rm Virial}$. The horizontal line indicates unity, and the vertical line marks the time by which $99\%$ of the gravitational waves have been emitted. We note that, while $\eta_{\rm Virial}$ approaches unity and therefore signals virialization, $\tilde{\eta}_{\rm Virial}$ decreases with time after the peak value and does not behave as a direct virialization indicator, as discussed in section \ref{sec:virial_theorem}.}
\label{fig:zeldovich}
\end{figure}

The system is initialized at the time of maximum expansion, $t_{\rm turn}$, following the Zel'dovich approximation described in Section~\ref{sec:zeldovich}. Subsequently, the shortest semi-axis, associated with $\alpha$, is the first to collapse, in agreement with the qualitative analytical behaviour described in Ref.~\cite{1965ApJ...142.1431L}\footnote{This behavior is qualitatively different from that observed in Ref.~\cite{Escriva:2024ellipsoidal} within a fluid description. In that work, ellipsoidal fluctuations collapsing in the presence of pressure exhibit an oscillatory damping of the ellipticity, so that deviations from sphericity progressively decay with time. In this sense, pressure plays a crucial role in the collapse dynamics.}. In particular, shell crossing along the shortest axis occurs at $t/t_k \approx 22.536$, when the configuration develops a thin pancake-like geometry along the $x$-direction. At this stage, the dynamical evolution begins to depart from the Zel'dovich solution, as can be appreciated in the top panel of Fig.~\ref{fig:zeldovich}. This figure shows the evolution of the principal semi-axes of the ellipsoid, with the solid lines representing the N-body simulation and the dashed lines the Zel'dovich approximation extrapolated from times earlier than turnaround. Initially, the collapse is well described by the Zel'dovich approximation; however, significant deviations develop at later times, and the approximation breaks down for $t/t_k \gtrsim 20$.

After the first shell crossing, the remaining axes also collapse. In particular, collapse along the $y$-axis occurs at $t/t_k \approx 29.097$. The system then undergoes further compression until the axes reach their minimum extent, after which the particles begin to spread out, breaking the original ellipsoidal symmetry of the initial configuration. The evolution subsequently enters a strongly nonlinear stage characterized by increasing particle velocities, until the virial parameter $\eta_{\rm Virial}$ reaches its maximum value at $t/t_k \approx 34.271$. This quantity measures the relative importance of the kinetic energy with respect to the softened gravitational interaction term.

\begin{figure}[!htbp]
\centering
\includegraphics[width=0.5\linewidth]{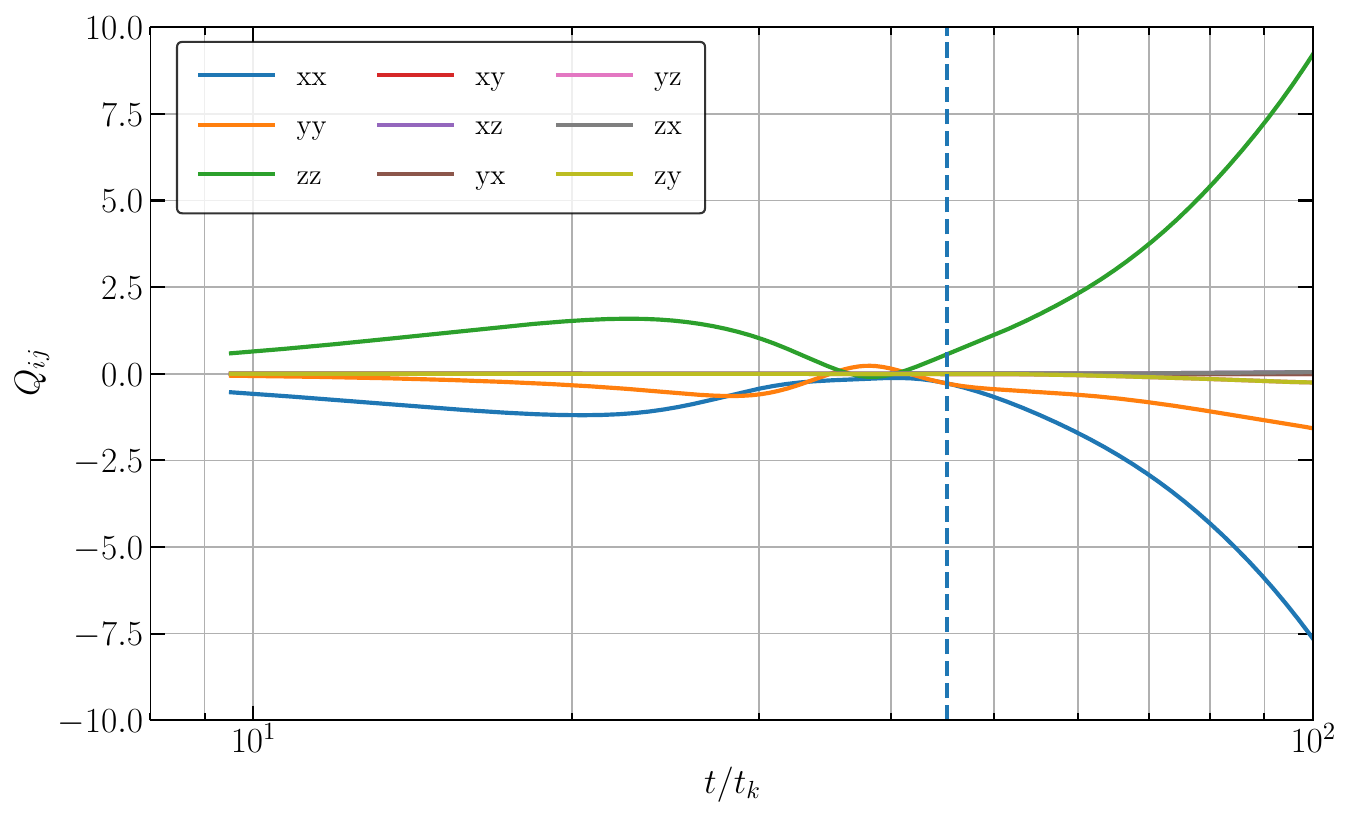}\hfill
\includegraphics[width=0.5\linewidth]{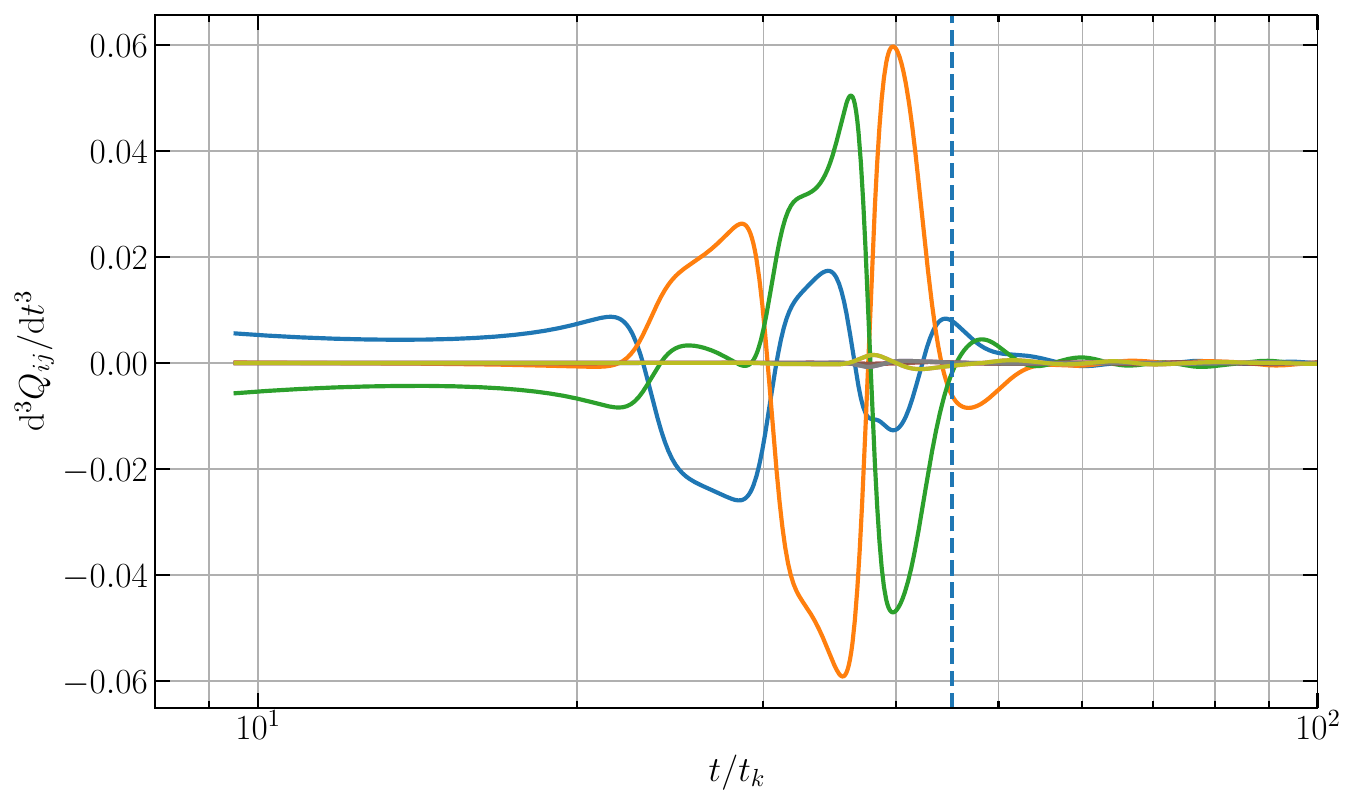}\hfill
\includegraphics[width=0.5\linewidth]{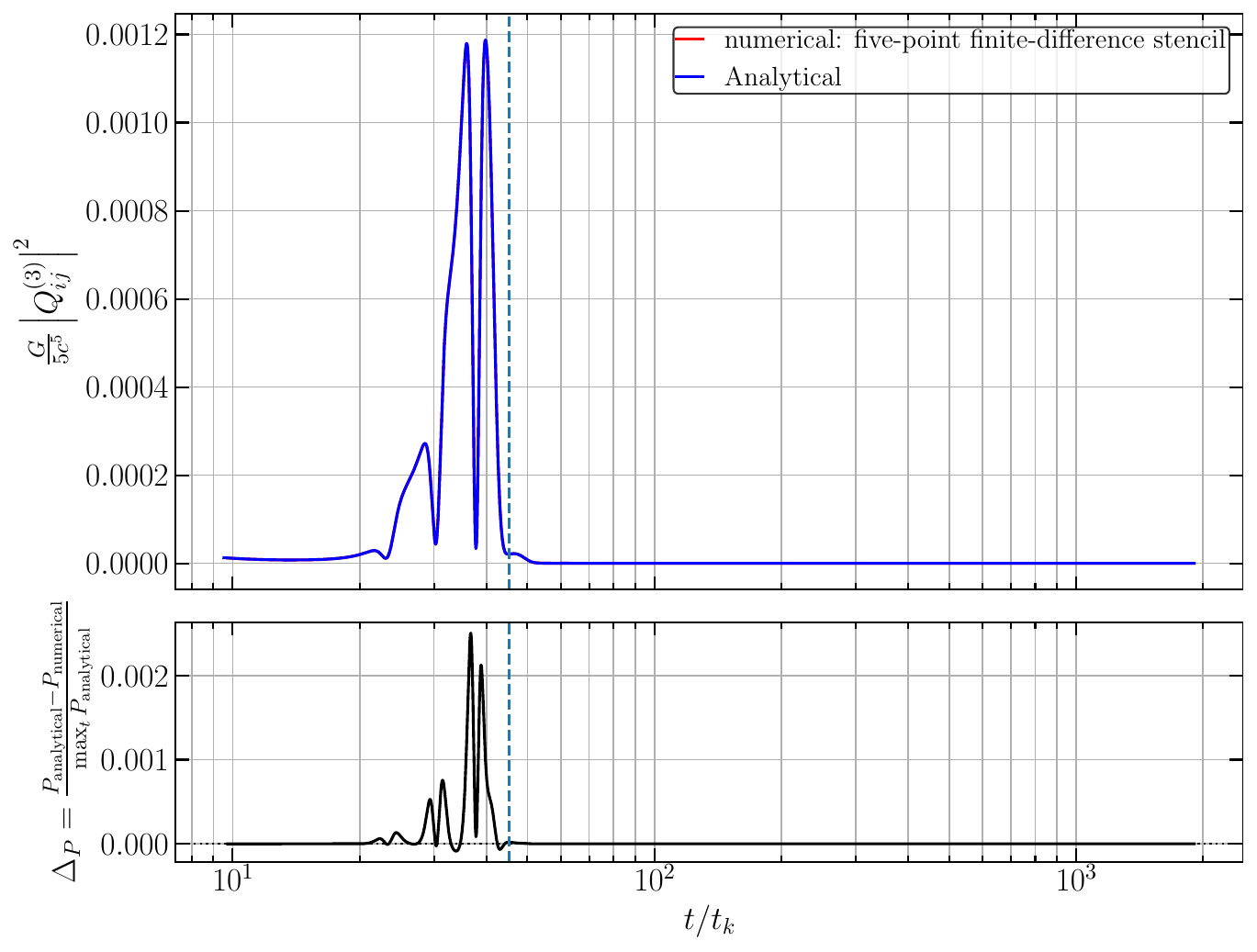}\hfill
\includegraphics[width=1.0\linewidth]{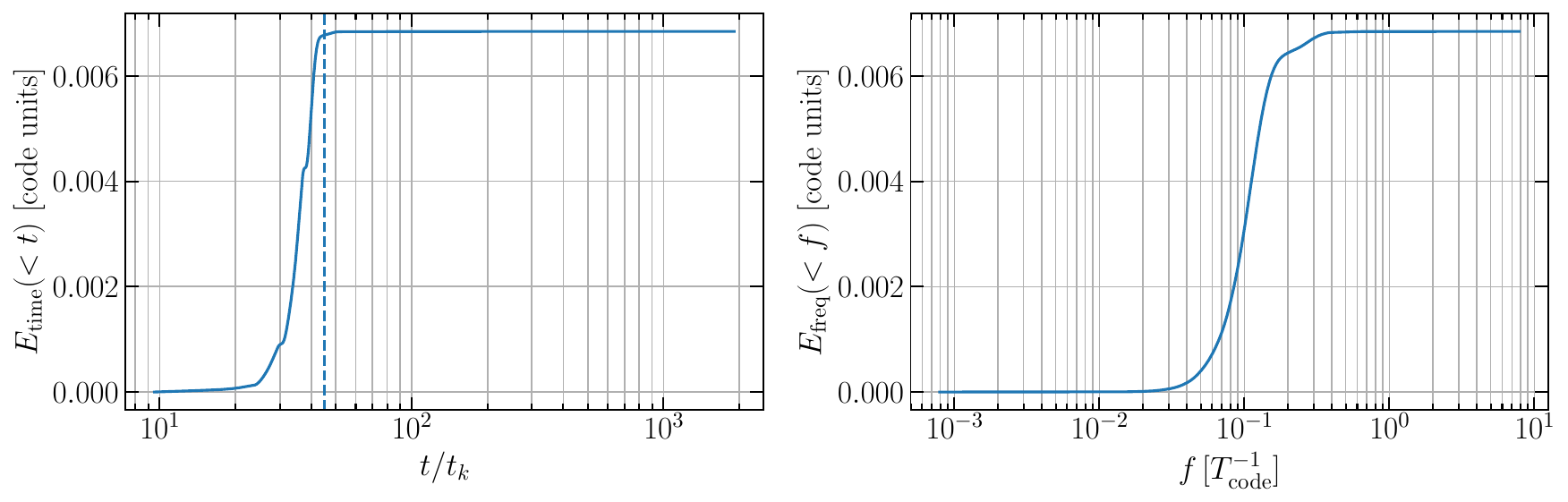}\hfill
\caption{Top panels: Different quadrupole components and their corresponding third time derivatives, computed using the analytical expression in Eq.~\eqref{eq:dddQij_srnewt}. Middle panel: GW power emitted as a function of time, computed using Eq.~\eqref{eq:GW_power_srnewt} with the analytical expression in Eq.~\eqref{eq:dddQij_srnewt} (blue line), and compared with the result obtained using a five-point stencil finite-difference method (red line). The lower panel shows the relative comparison between the two quantities. Bottom panel: Integrated GW energy emitted in time space (left) and frequency space (right). In all plots, the vertical dashed line denotes $t_{\rm e}$.}
\label{fig:quadrupole_components}
\end{figure}

This stage also corresponds to the main epoch of GW emission. In Fig.~\ref{fig:quadrupole_components}, we show the evolution of the quadrupole components and their third time derivatives, which develop pronounced peaks during the collapse, well after the first shell crossing. Integrating the emitted power using Eq.~\eqref{eq:GW_power_srnewt}, we find, as shown in the bottom panel of Fig.~\ref{fig:quadrupole_components}, that by $t/t_k \approx 45.024$ about $99\%$ of the total GW energy has already been emitted. It is worth emphasizing that, throughout this work, our default numerical calculation of the GW signal uses the analytical expression in Eq.~\eqref{eq:dddQij_srnewt} for the third time derivative of the quadrupole tensor. When compared with a five-point stencil finite-difference evaluation, we find excellent agreement in most cases, indicating that the numerical evolution is robust and affected by only a small level of numerical noise. In Ref.~\cite{DalianisKouvaris2024}, a phenomenological fitting formula was employed to estimate the third time derivative of the quadrupole, thus avoiding a direct numerical differentiation, which can be affected by noise in their numerical data. By comparing our numerical computation of $\dddot{Q}_{ij}$ with their fitting formula, we find differences that can lead to either an overestimation or an underestimation of the GW signal, depending on the non-spherical configuration considered. This comparison is presented in Appendix~\ref{subsec:fitting_comparison}, which shows the importance of the numerical calculation.

At later times, the system evolves toward a virialized state. According to the virial theorem discussed in Section~\ref{sec:virial_theorem}, this corresponds to $\langle V_{\rm rel} + W \rangle \approx 0$. This behavior is visible in the final snapshot of Fig.~\ref{snapshots_particles_Nbody}, where a halo has formed and is surrounded by particles. The virialization process can also be seen in the bottom panel of Fig.~\ref{fig:zeldovich}. During the collapse, the system converts gravitational potential energy into kinetic energy, leading to an increase in particle velocities up to a maximum value. This is followed by a relaxation phase, during which the particle velocities decrease and the system approaches a virialized configuration with $\eta_{\rm Virial} \approx 1$ when a stable halo is then established. The onset of virialization occurs approximately in the range $t/t_k \approx 50$--$60$, after which the system remains virialized throughout the rest of the numerical evolution with $\eta_{\rm Virial}$ oscillating around the mean value $1$, further supporting the robustness of the simulations.

\begin{figure}[!htbp]
\centering
\includegraphics[width=0.5\linewidth]{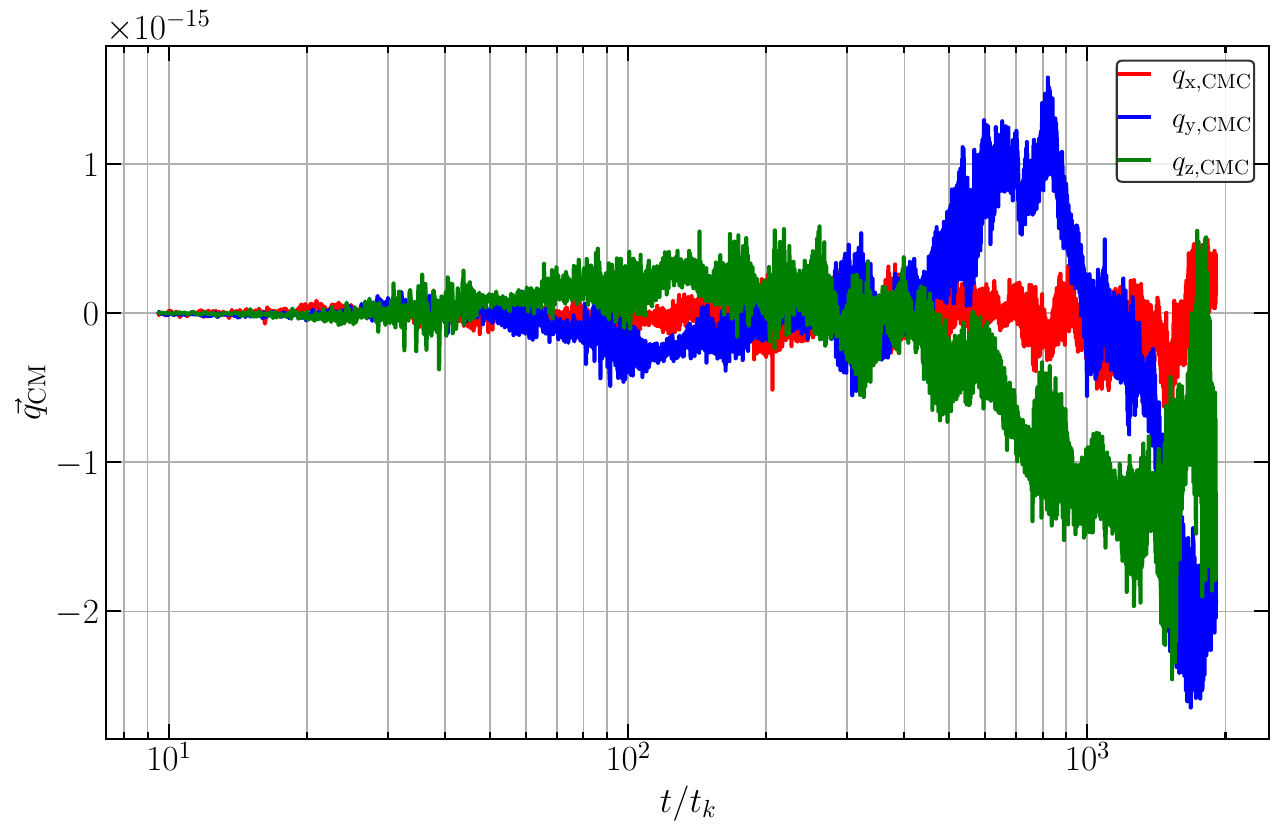}\hfill
\caption{Evolution of the mean canonical momentum along the three spatial directions.}
\label{fig:num_constraints}
\end{figure}

On the other hand, to test the robustness of the simulations, we monitor the mean value of the canonical momentum, whose total sum should vanish in the absence of external forces. The result is shown in Fig.~\ref{fig:num_constraints}, demonstrating excellent numerical accuracy. This behavior is observed in all the numerical evolutions considered in this work.

Finally, we study the dependence of the collapse dynamics and the GW signal on the softening parameter $\epsilon_{\rm com}$ and on the particle number $N$. It is important to stress that $\epsilon_{\rm com}$ should not be interpreted as a physical particle size, but rather as the numerical scale below which the pairwise gravitational interaction is smoothed. In our implementation, the force is computed with a cubic-spline softening kernel, so that the interaction is exactly Newtonian at separations larger than the kernel radius, while it is regularized at shorter distances. Therefore, $\epsilon_{\rm com}$ controls how strongly the simulation resolves the dense inner stage of the collapse and how much it suppresses spurious two-body effects.

\begin{figure}[!htbp]
\centering
\includegraphics[width=0.6\linewidth]{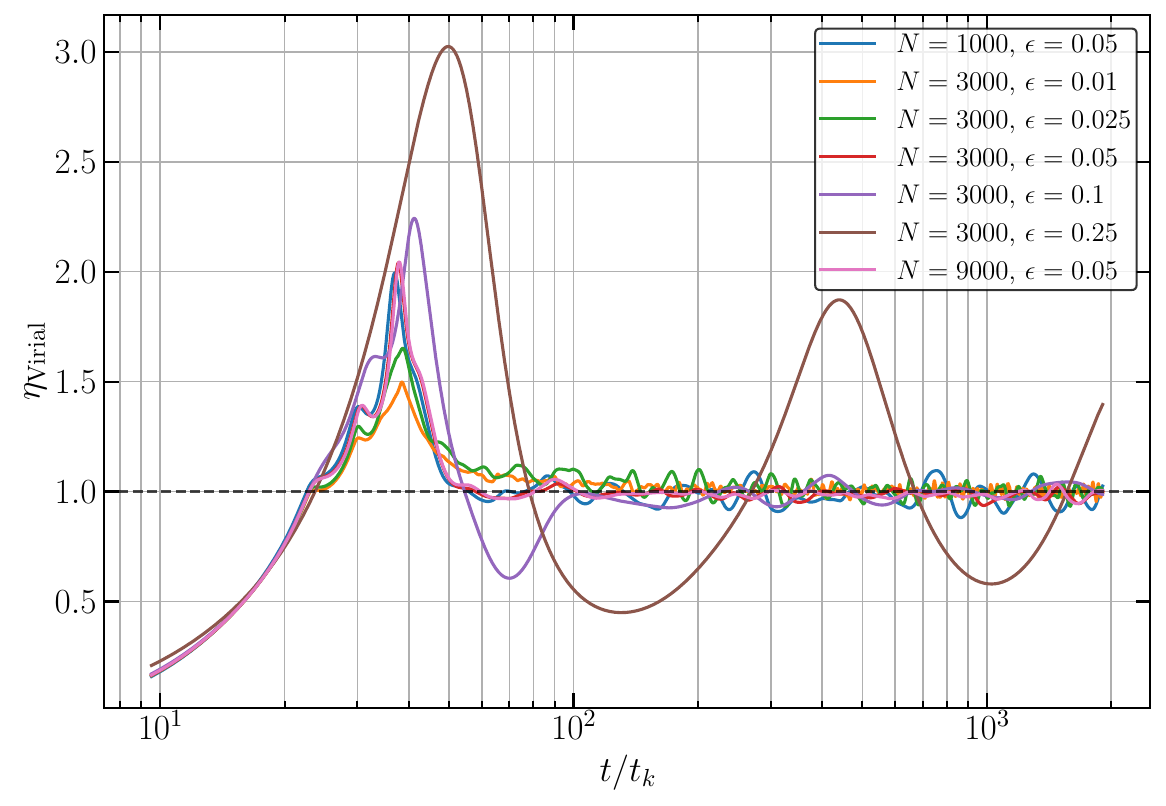}
\vspace{0.5cm}
\begin{minipage}{0.48\linewidth}
    \centering
    \includegraphics[width=\linewidth]{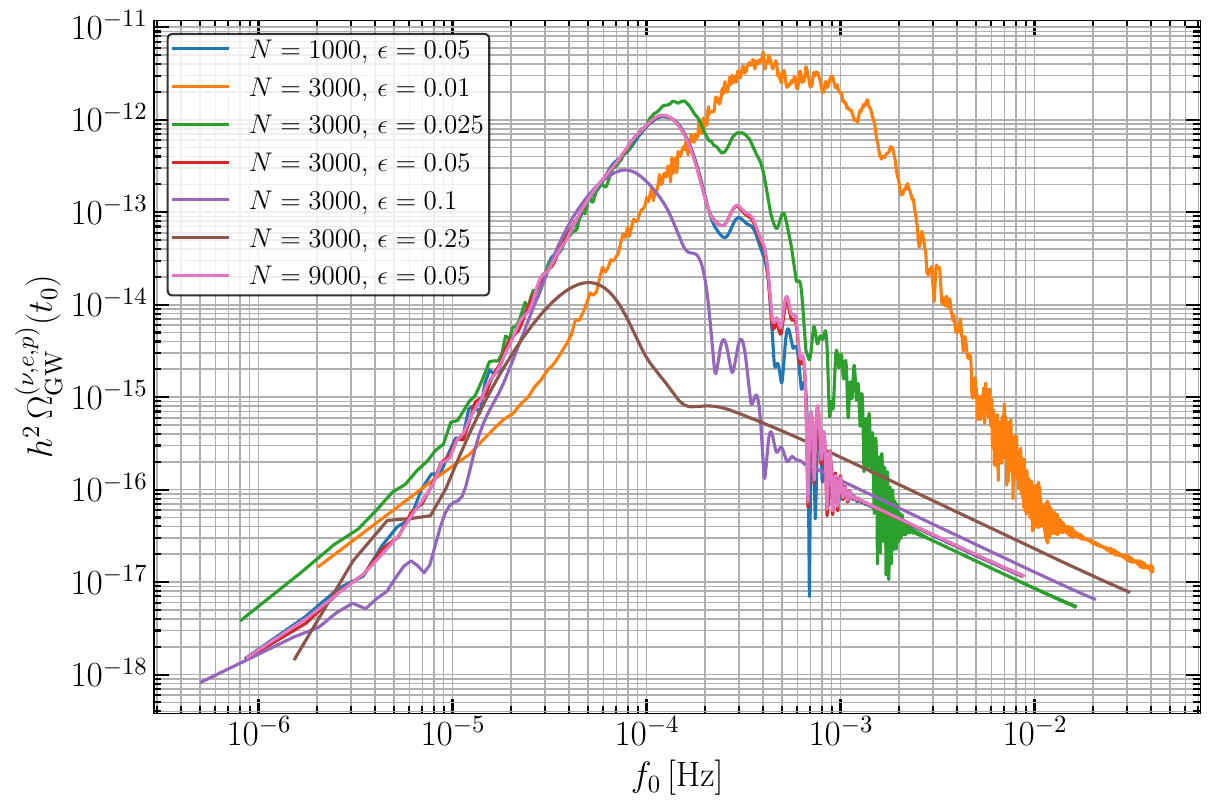}
\end{minipage}
\hfill
\begin{minipage}{0.48\linewidth}
    \centering
    \includegraphics[width=\linewidth]{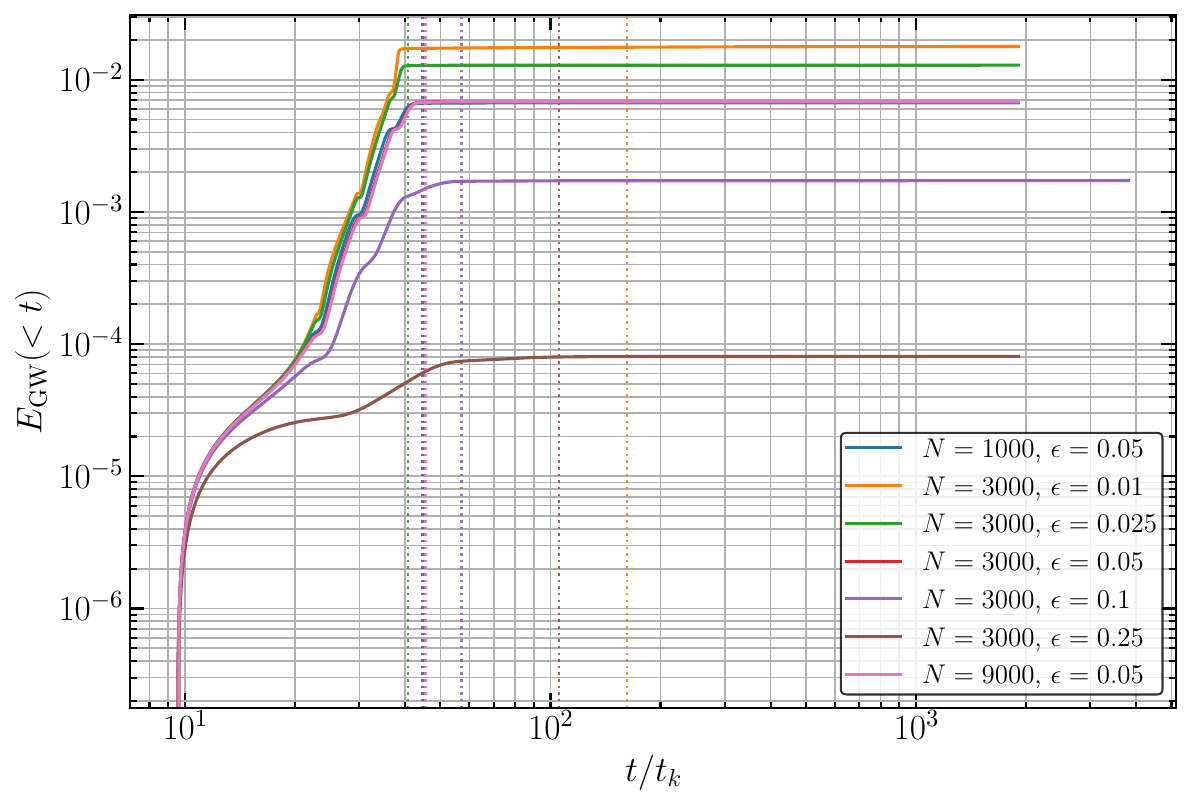}
\end{minipage}
\caption{Dependence of the collapse and GW signal on the particle number $N$ and the comoving softening length $\epsilon_{\rm com}$. Top panel: evolution of the virial ratio $\eta_{\rm Virial}$. Bottom-left panel: present-day GW spectrum. Bottom-right panel: cumulative emitted GW energy.}
\label{fig:comparison_parametters}
\end{figure}

The effect of varying $\epsilon_{\rm com}$ is visible in Fig.~\ref{fig:comparison_parametters}. In the top panel, smaller values of the softening lead to a faster approach to virialization, whereas larger values produce a more extended oscillatory phase before the system relaxes. Physically, when the softening is small, the particles experience stronger short-distance gravitational interactions during the post-shell-crossing stage, which enhances the violent redistribution of energy and accelerates the formation of a virialized halo. On the other hand, if the softening is too large, the inner gravitational field is excessively smoothed, close encounters are strongly suppressed, and the collapse becomes less efficient at converting the bulk infall into random motions. As a consequence, the system exhibits a slower relaxation and a more pronounced breathing behavior, especially for the largest value considered, $\epsilon_{\rm com}=0.25$.

This behaviour has a direct impact on the GW signal shown in the middle and bottom panels of Fig.~\ref{fig:comparison_parametters}. Since gravitational waves are sourced by the time variation of the nonspherical quadrupole moment, a more violent and rapidly changing collapse produces a stronger signal. Accordingly, smaller softening tends to increase the emitted GW energy and enhances the high-frequency part of the spectrum. At the same time, however, very small $\epsilon_{\rm com}$ makes the dynamics more sensitive to discrete close encounters between particles, which introduces a noisier and less smooth behaviour in the quadrupole derivatives. In contrast, for large softening the collapse is smoother and less abrupt, which suppresses the quadrupole variations and leads to a smaller GW amplitude.

Overall, Fig.~\ref{fig:comparison_parametters} shows that the results depend nontrivially on the choice of the softening length. If $\epsilon_{\rm com}$ is too large, the collapse is over-smoothed and virialization is artificially delayed. If it is too small, the simulation becomes increasingly sensitive to close-pair interactions and particle-discreteness effects, which can contaminate the GW signal. We therefore adopt $\epsilon_{\rm com}=0.05$ as a fiducial compromise between resolving the inner collapse and suppressing discreteness noise. The detailed slope, cutoff, and fine structure of the high-frequency tail are particularly sensitive to $\epsilon_{\rm com}$, since this part of the spectrum is generated by the shortest-scale post-shell-crossing dynamics and is therefore most affected by the regularization of close particle encounters. We consequently do not regard the far high-frequency tail as a quantitatively converged prediction independent of the adopted force regularization. Our qualitative conclusions concern instead the importance of the fully nonlinear post-shell-crossing evolution and the resulting enhancement of the GW emission relative to the Zel'dovich-based estimate. Nevertheless, the precise peak amplitude and high-frequency behavior retain a systematic dependence on the adopted softening scale. A precision determination of the high-frequency tail would require a dedicated continuum-limit analysis in which $N$ and $\epsilon_{\rm com}$ are varied jointly.


We also compare the dependence on the particle number $N$. For fixed $\epsilon_{\rm com}=0.05$, the difference between the results for $N=3000$ and $N=9000$ is relatively small, while the case $N=1000$ shows somewhat larger deviations. This indicates that the simulations are already reasonably converged for $N\sim 3000$. We therefore adopt $N=3000$ as our fiducial choice in the statistical analysis, since it provides a good compromise between numerical accuracy and computational efficiency, noting that increasing the number of particles significantly raises the cost and runtime of the N-body simulations which scales as $\sim N^2$.

\begin{figure}[!htbp]
\centering
\includegraphics[width=0.5\linewidth]{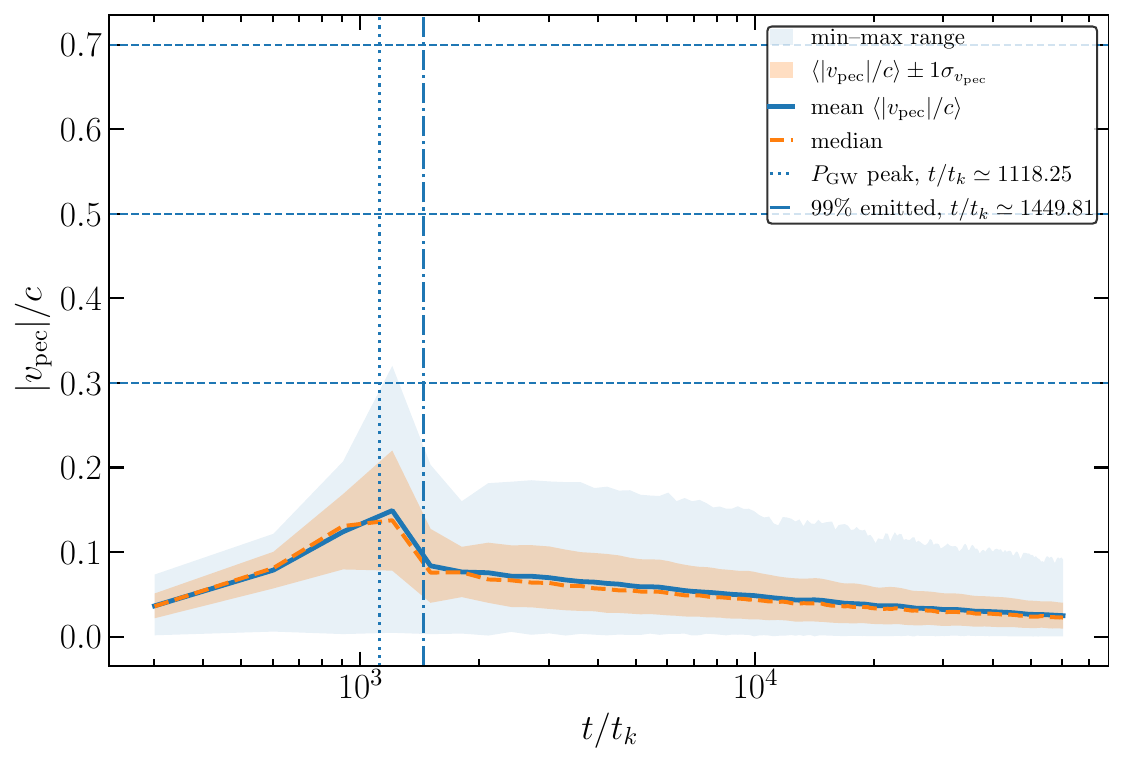}\hfill
\includegraphics[width=0.5\linewidth]{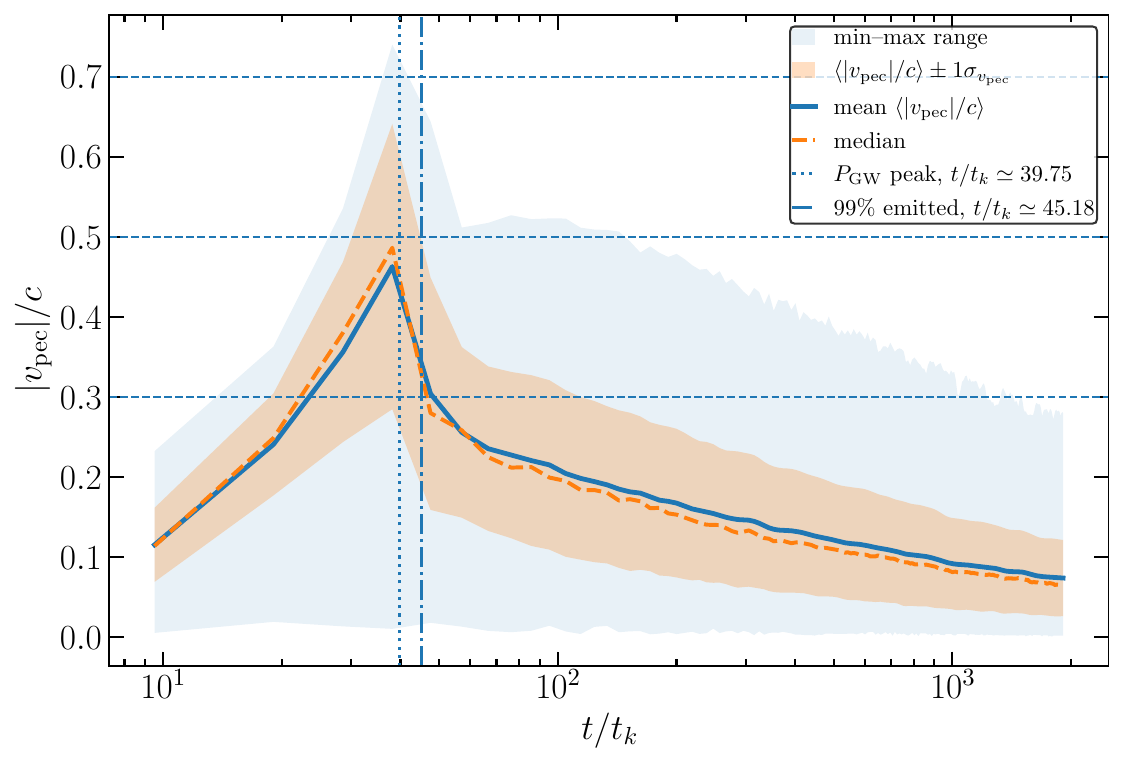}\hfill
 \caption{Time evolution of the particle peculiar-speed $v_{\rm pec}$ distribution for the
    representative nonspherical configuration discussed in the main text. The left and right panels correspond to
    \(\sigma=0.01\) and \(\sigma=0.1\) for the same $\nu,e,p$, respectively. The solid line shows the
    mean value of \(|v_{\rm pec}|/c\), the dashed line shows the median, the
    shaded orange region denotes the interval
    \(\langle |v_{\rm pec}|/c\rangle \pm 1\sigma_{v_{\rm pec}}\), and the light-blue
    region shows the full min--max range. The vertical dotted and dash-dotted
    lines indicate, respectively, the time at which the GW power reaches its
    peak and the time by which \(99\%\) of the total emitted GW energy has been
    accumulated.}
    \label{fig:velocity_diagnostics}
\end{figure}

To assess the slow-motion assumption entering the leading quadrupole estimate,
we examine the particle velocity distribution during the evolution. Figure
\ref{fig:velocity_diagnostics} shows the time dependence of the mean and median
values of \(|v_{\rm pec}|/c\), together with the $1\sigma_{v_{\rm pec}}$ interval and
the full min--max range, for the representative nonspherical configuration $\nu \approx 2 , e \approx 0.223$ and $p \approx -0.00125$, with
\(\epsilon_{\rm com}=0.05\). For \(\sigma=0.01\), the bulk of the particle
distribution remains nonrelativistic throughout the relevant emission
epoch. Around the peak of the GW power and by the time \(99\%\) of the total GW
energy has been emitted, the mean and median velocities are well below the speed
of light, while only a small tail reaches larger values. On the other hand, for the larger value \(\sigma=0.1\), the system becomes moderately relativistic
near the peak of the emission, with the mean and median values of
\(|v_{\rm pec}|/c\) reaching values of order \(0.4\)--\(0.5\). This indicates
that the leading quadrupole estimate may receive relativistic corrections in
this case. However, the velocities remain at most moderately relativistic for
the bulk of the particles, and the subsequent evolution rapidly moves back
towards smaller velocities. We therefore expect these corrections to affect
mainly the overall normalization of the GW spectrum, at the level of
a factor of order unity $\mathcal{O}(1)$, rather than changing the qualitative behaviour of the signal. This interpretation is also supported by the fact that the qualitative time evolution and spectral features remain similar between the small- and large-\(\sigma\) cases, see for instance the appendix~\ref{sec:extra_simulations} and Fig.~\ref{fig:average_spectra}. It is useful to compare the two values of \(\sigma\). For smaller \(\sigma\),
the corresponding value of \(\alpha\) is also smaller. Then, according to
Eq.~\eqref{eq:turn_time}, the turnaround condition places the initial time of
the collapse at a later epoch after horizon crossing. The system is therefore
already well inside the cosmological horizon when the simulation starts, which
explains why the peculiar velocities remain milder in the smaller-\(\sigma\)
case.

\subsection{Statistically weighted average over the different realizations}
\label{sec:statistics_weighted}
So far, we have analyzed in detail the collapse dynamics and the GW emission for individual representative realizations. However, in order to estimate the total stochastic signal produced during an early matter-dominated era, it is necessary to average over the ensemble of nonspherical initial configurations, weighted by their corresponding probability distributions. In this section, we perform such a statistically weighted average by combining the numerical N-body results for different realizations with the distributions introduced in Sec.~\ref{subsec:statistics_initial_conditions}. Throughout this analysis, we adopt as numerical parameters the values selected in the previous subsection, namely $N=3000$ and $\epsilon_{\rm com}=0.05$.

Our goal is to compute the GW contribution from the full set of relevant triaxial configurations. For this purpose, we consider two different prescriptions for the statistical weighting of the initial conditions introduced in Sec.~\ref{subsec:statistics_initial_conditions}: the Doroshkevich distribution in terms of the ordered deformation eigenvalues $(\alpha,\beta,\gamma)$, and the BBKS peak-theory description in terms of the variables $(\nu,e,p)$. In practice, the corresponding parameter spaces must be discretized in order to perform the ensemble average numerically. We therefore construct a finite set of realizations, $N_{\rm conf}$, chosen so as to cover efficiently the physically relevant region of each probability distribution while avoiding configurations whose statistical weight is negligibly small.

\begin{figure}[!htbp]
\centering
\includegraphics[width=0.5\linewidth]{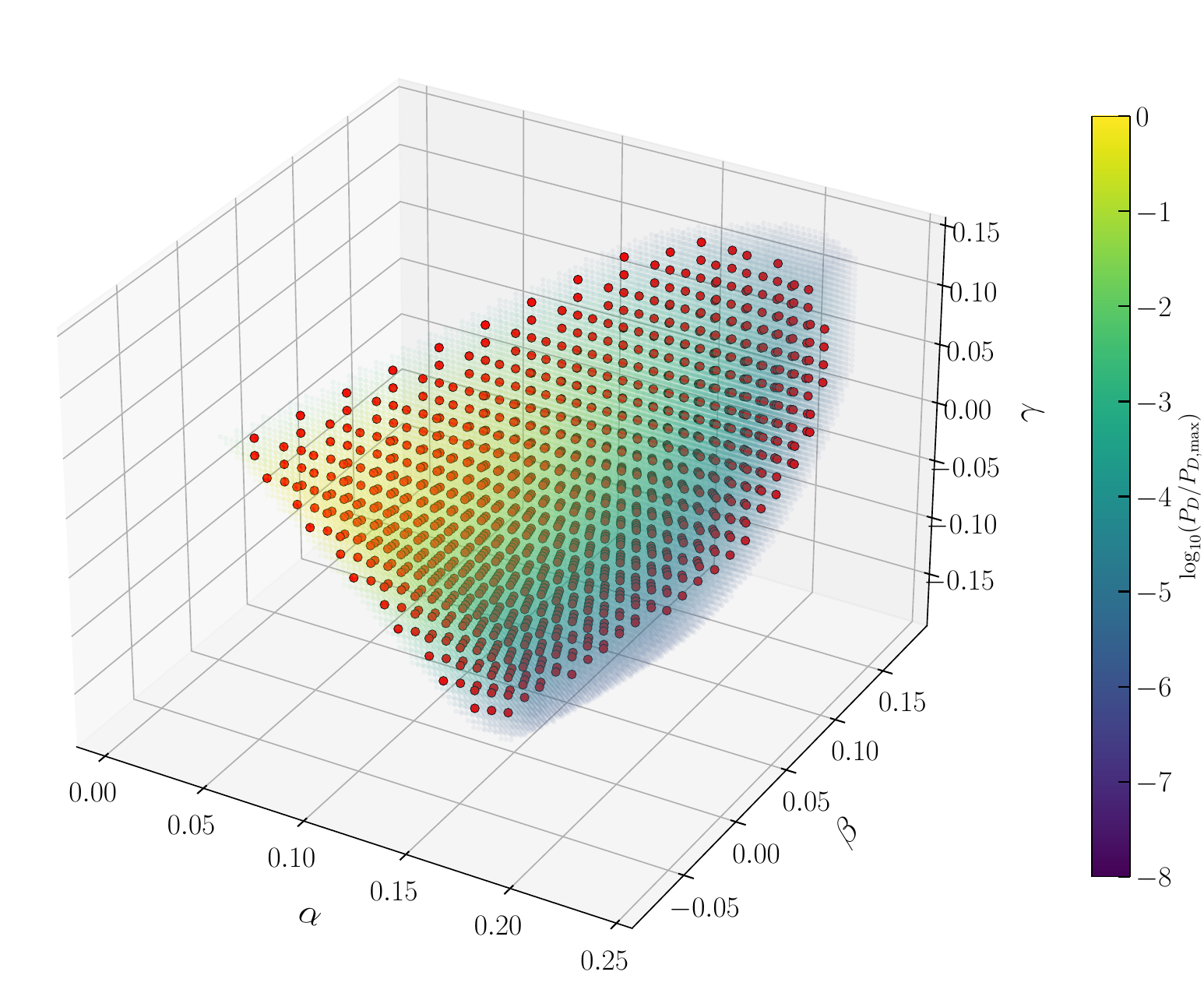}\hfill
\includegraphics[width=0.5\linewidth]{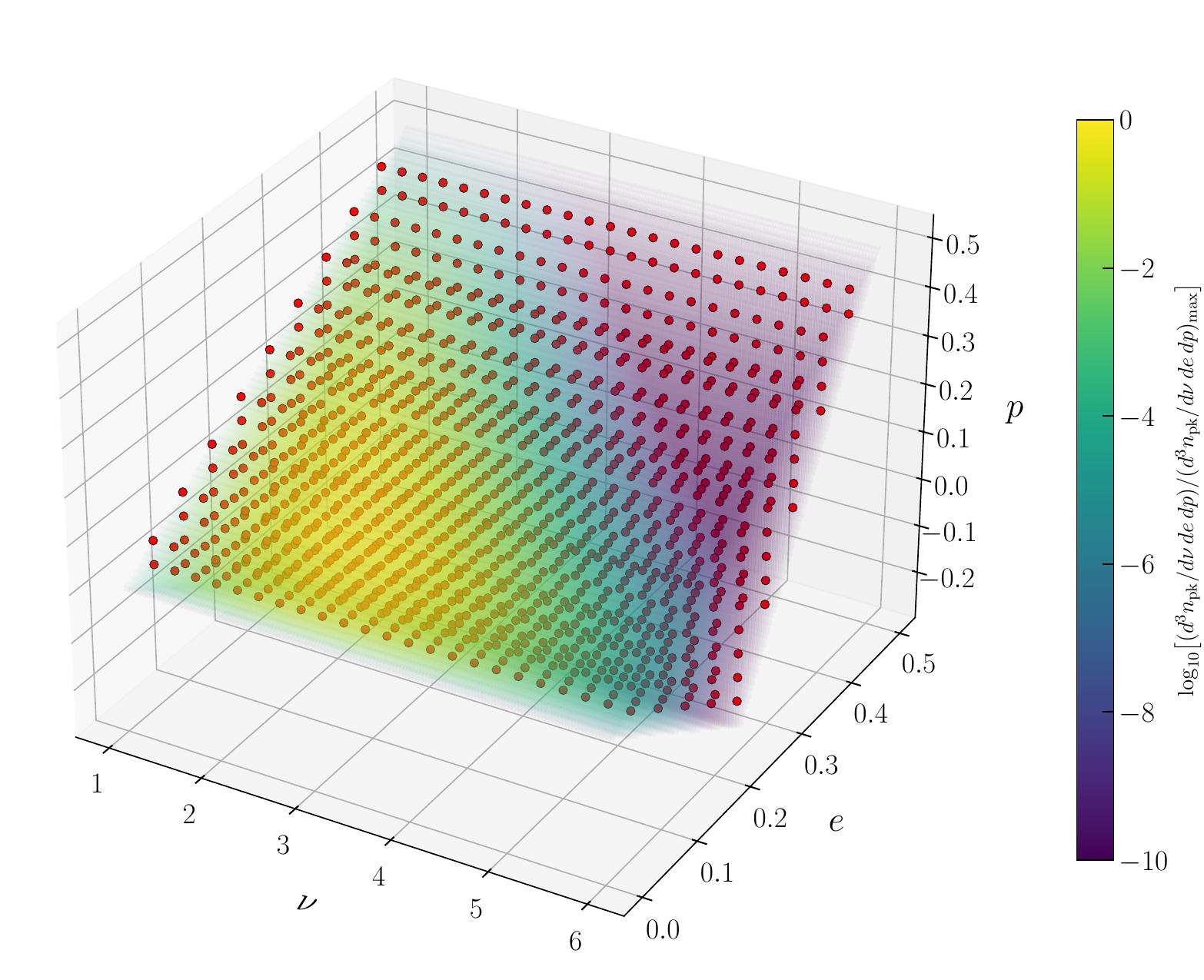}\hfill
\caption{Parameter-space sampling adopted for the statistically weighted average over realizations. Left panel: Doroshkevich probability density for the deformation parameters, displayed in the relevant eigenvalue domain and normalized to its maximum value. Right panel: BBKS peak-theory probability density in the ellipticity--prolateness $(e,p)$ plane, likewise normalized to its maximum value. The colour scale in both panels shows the logarithm of the corresponding probability weight. In both cases $\sigma=0.1$.}
\label{fig:doros_bbks_figure}
\end{figure}

The parameter-space domains adopted in this work are shown in Fig.~\ref{fig:doros_bbks_figure}, with $N_{\rm conf}\sim 10^3$. In the Doroshkevich case, we sample the ordered eigenvalue space $(\alpha,\beta,\gamma)$ on a uniform grid, restricting the domain to the region where the probability density remains sufficiently close to its peak value to contribute non-negligibly to the final average. Similarly, in the BBKS case we sample the space $(\nu,e,p)$ by fixing a set of peak heights $\nu$ and discretizing the allowed triangular domain in the $(e,p)$ plane; see Eq.~\eqref{eq:chi_ep}. In both cases, the sampling is chosen to retain the bulk of the statistically relevant support of the distribution while excluding the far tails, which are exponentially suppressed and would have a negligible impact on the final result. This procedure substantially reduces the computational cost of the averaging without affecting the dominant contribution to the final spectrum.

Once the individual realizations have been evolved numerically, we compute the statistically weighted average of the GW spectrum by assigning to each configuration its corresponding probability weight and applying the appropriate redshift and dilution factors in order to obtain the present-day quantity $h^2\Omega_{\rm GW}(t_0)$. In this way, the final signal is obtained directly from the fully nonlinear N-body evolution, rather than from an analytical approximation to the collapse dynamics. This is important because, as shown in the previous subsection, the GW emission is highly sensitive to the detailed time dependence of the quadrupole during shell crossing, secondary collapse, and virialization. Accordingly, the statistical average must be performed on the numerical spectra themselves, rather than on simplified fitting procedures that could bias the resulting ensemble average.

\begin{figure}[!htbp]
\centering
\includegraphics[width=0.5\linewidth]{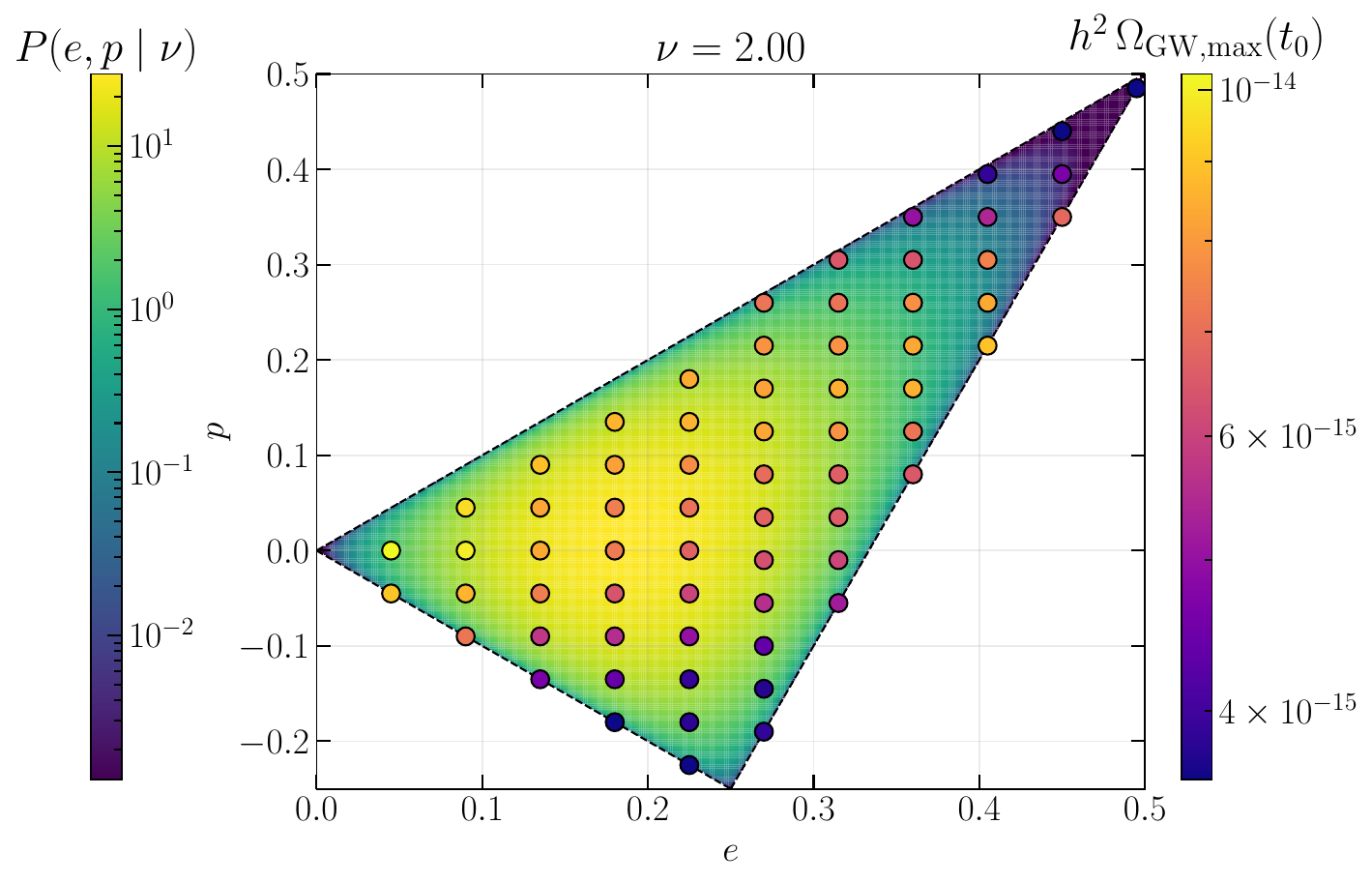}\hfill
\includegraphics[width=0.5\linewidth]{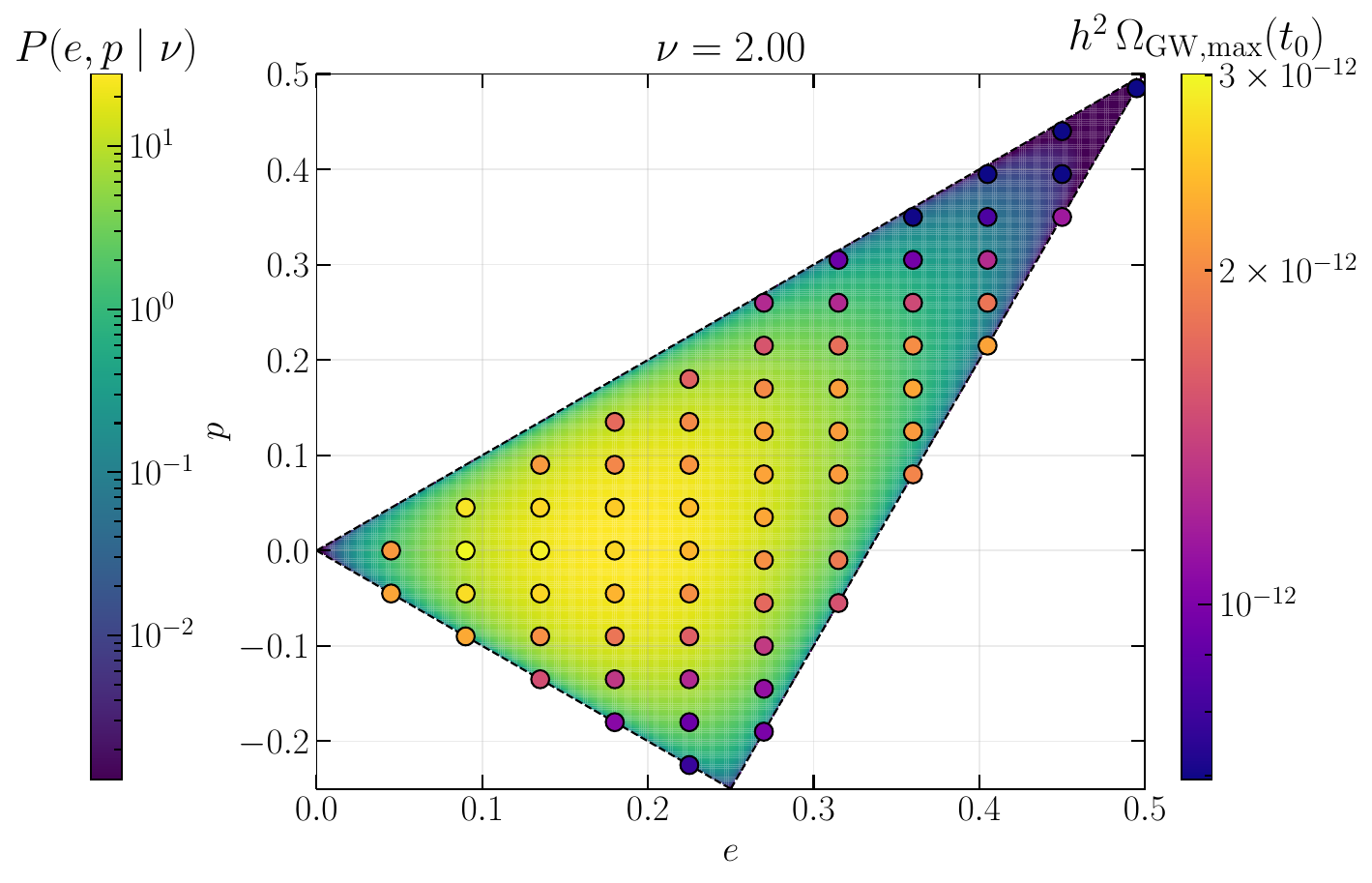}\hfill
\includegraphics[width=0.5\linewidth]{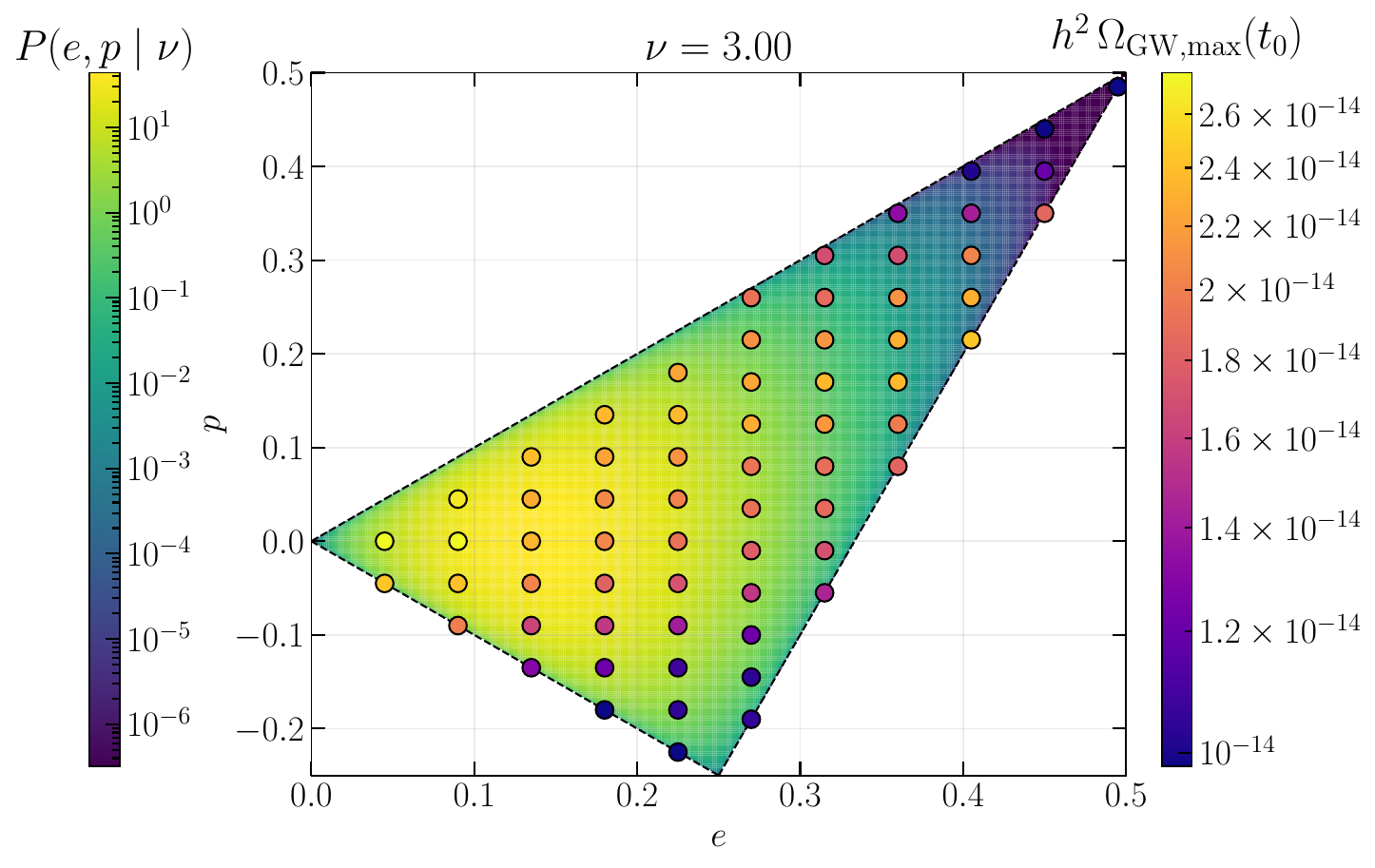}\hfill
\includegraphics[width=0.5\linewidth]{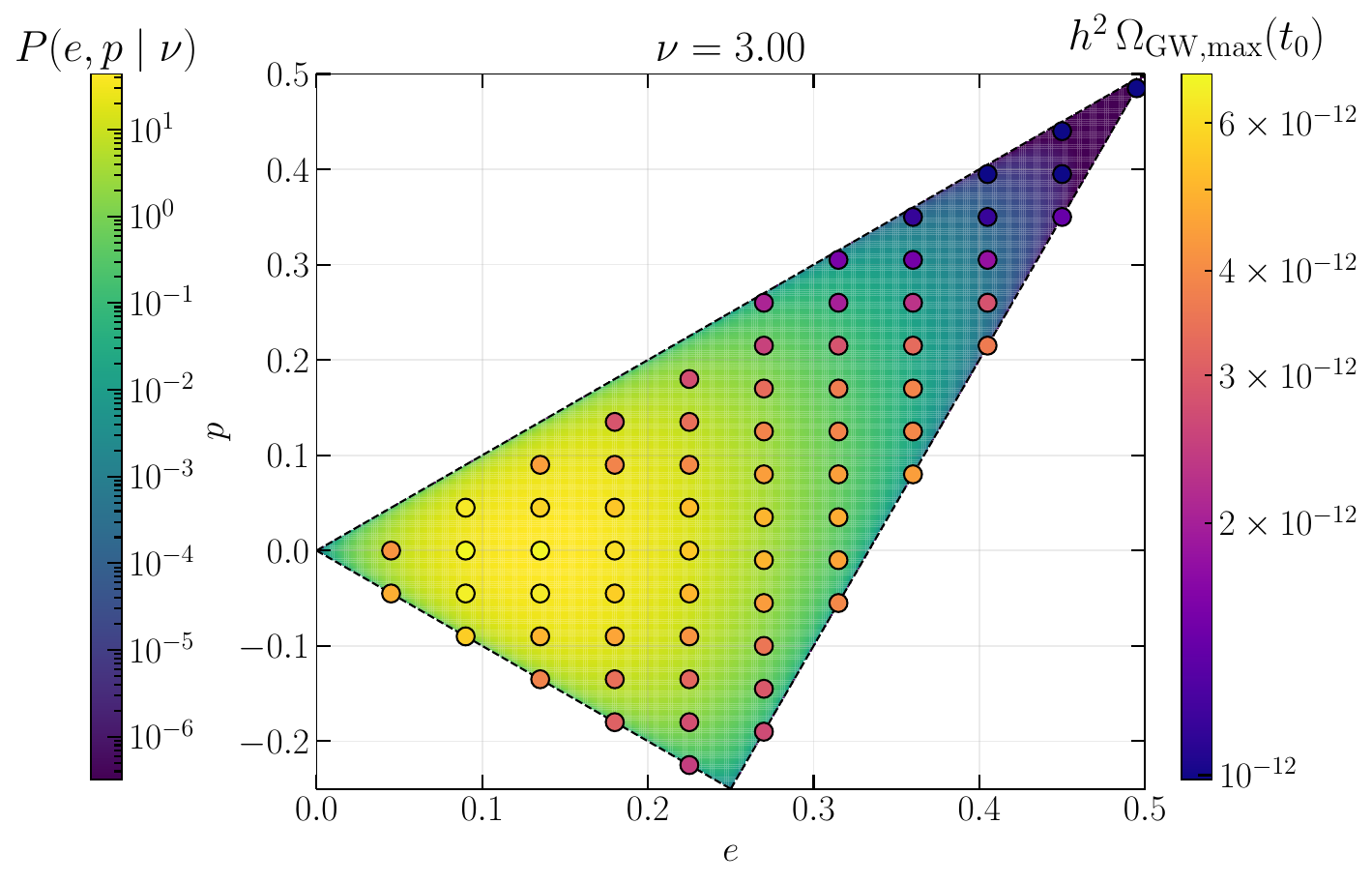}\hfill
\includegraphics[width=0.5\linewidth]{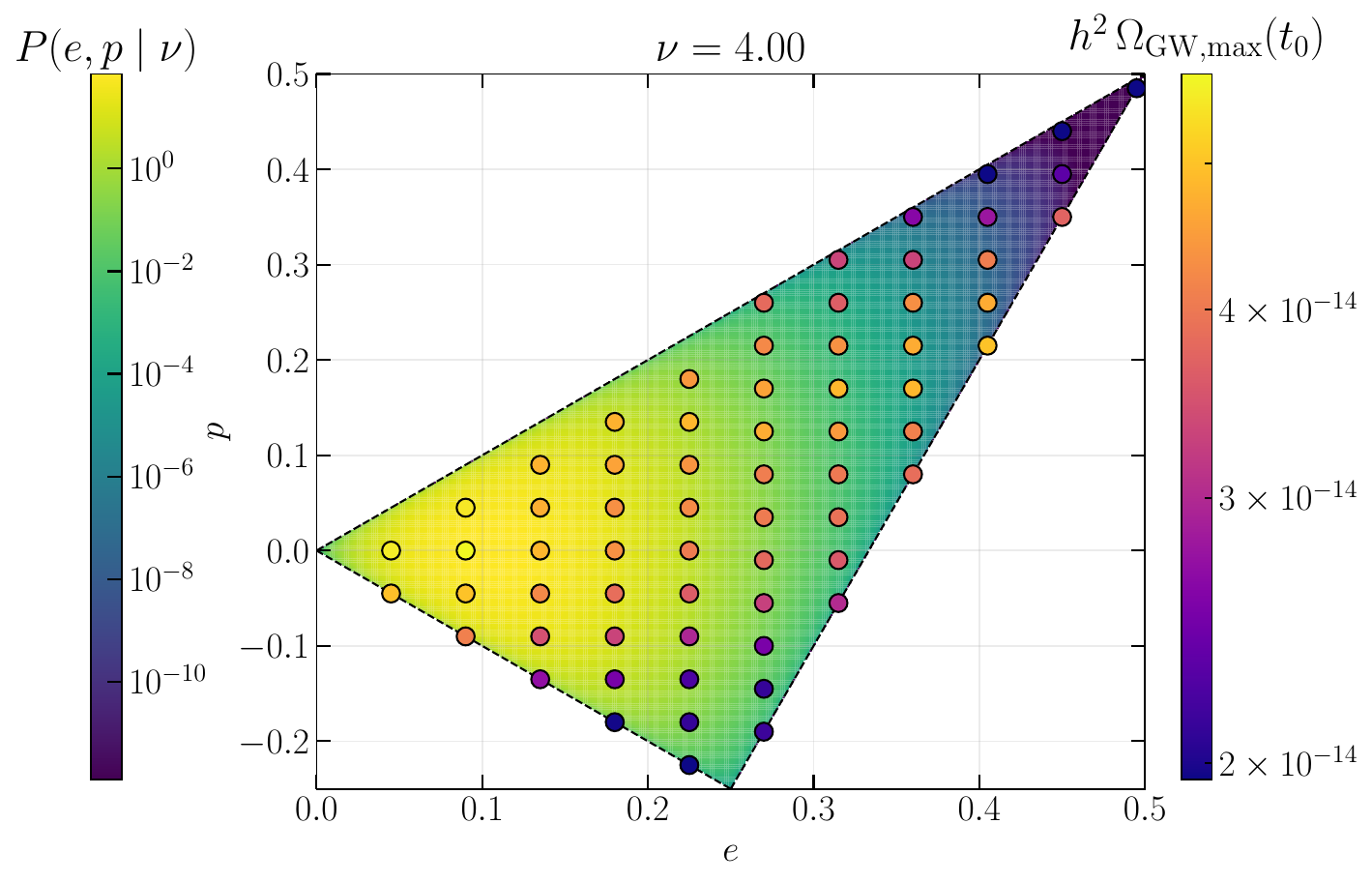}\hfill
\includegraphics[width=0.5\linewidth]{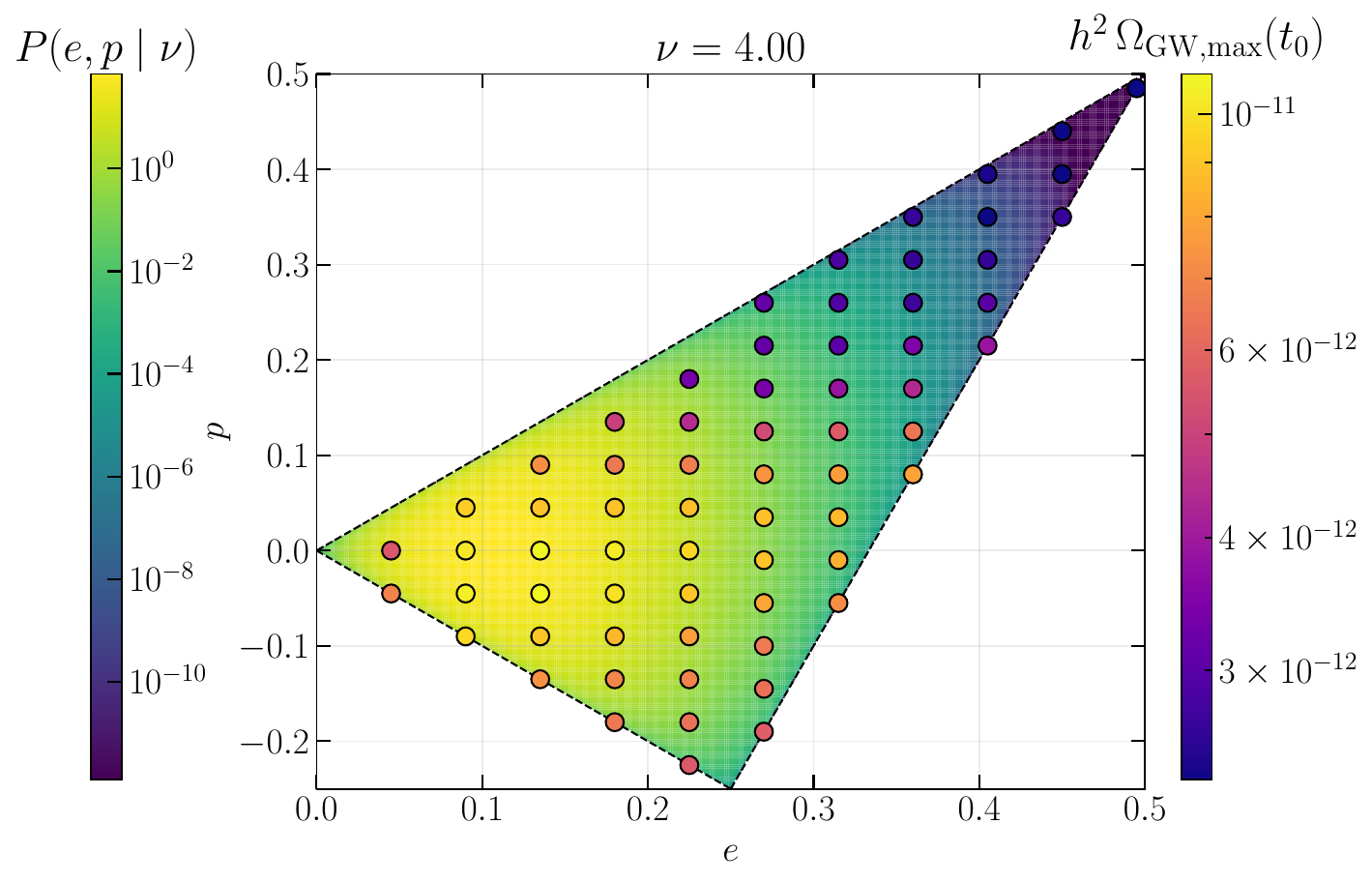}\hfill
\caption{BBKS parameter-space dependence of the GW signal for several fixed values of the peak height $\nu$. 
For each panel, the left colour bar shows the conditional BBKS probability density $P(e,p\,|\,\nu)$ in the allowed $(e,p)$ domain, while the right one shows the corresponding peak amplitude of the present-day GW spectrum, $h^2\Omega_{{\rm GW,max}}(t_0)$, obtained from the N-body simulations. The coloured markers denote the sampled configurations. The top, middle, and bottom row corresponds to $\nu = 2$, $3$, and $4$, respectively.  The left column corresponds to $\sigma=0.01$ and the right one to $\sigma=0.1$ with $M_{k}=10^{-10}M_{\odot}$ and $T_{\rm rh}=0.3 \textrm{GeV}$.}
\label{fig:BBKS_ep_maps}
\end{figure}

Having defined the sampling domain, we now examine which regions of the BBKS parameter space contribute most strongly to the signal. In Fig.~\ref{fig:BBKS_ep_maps}, we show, for several representative values of the peak height $\nu$, the BBKS probability density in the $(e,p)$ plane together with the corresponding peak value of the present-day GW amplitude, $h^2\Omega_{{\rm GW,max}}(t_0)$, obtained from the numerical simulations. For fixed $\nu$, the figure does not show a simple monotonic dependence of the GW amplitude on either $e$ or $p$ separately. Instead, the signal varies across the allowed BBKS domain in a mild and nontrivial way, indicating that the detailed shape parameters induce a secondary modulation of the signal. In practice, the spread in $h^2\Omega_{{\rm GW,max}}(t_0)$ over the $(e,p)$ plane is much smaller than the overall variation produced by changing $\nu$.

This behavior is physically reasonable, since the parameters $(e,p)$ do not determine the GW emission independently, but only through their mapping to the full set of deformation eigenvalues $(\alpha,\beta,\gamma)$, which control the detailed triaxial collapse history. The resulting signal therefore depends on the combined morphology of the collapsing patch, rather than on a simple trend with ellipticity or prolateness alone. By contrast, increasing the peak height $\nu$ produces a clear overall enhancement of the GW amplitude across the whole $(e,p)$ domain. This follows from the relation $\delta_L=\nu\sigma$, so that larger $\nu$ corresponds to a larger absolute initial deformation and hence to a more violent nonlinear collapse. For this reason, Fig.~\ref{fig:BBKS_ep_maps} should mainly be interpreted as showing that the BBKS shape parameters modulate the signal, while the dominant scaling of the peak GW amplitude is controlled by the overall peak height $\nu$.

\begin{figure}[!htbp]
\centering
\includegraphics[width=0.5\linewidth]{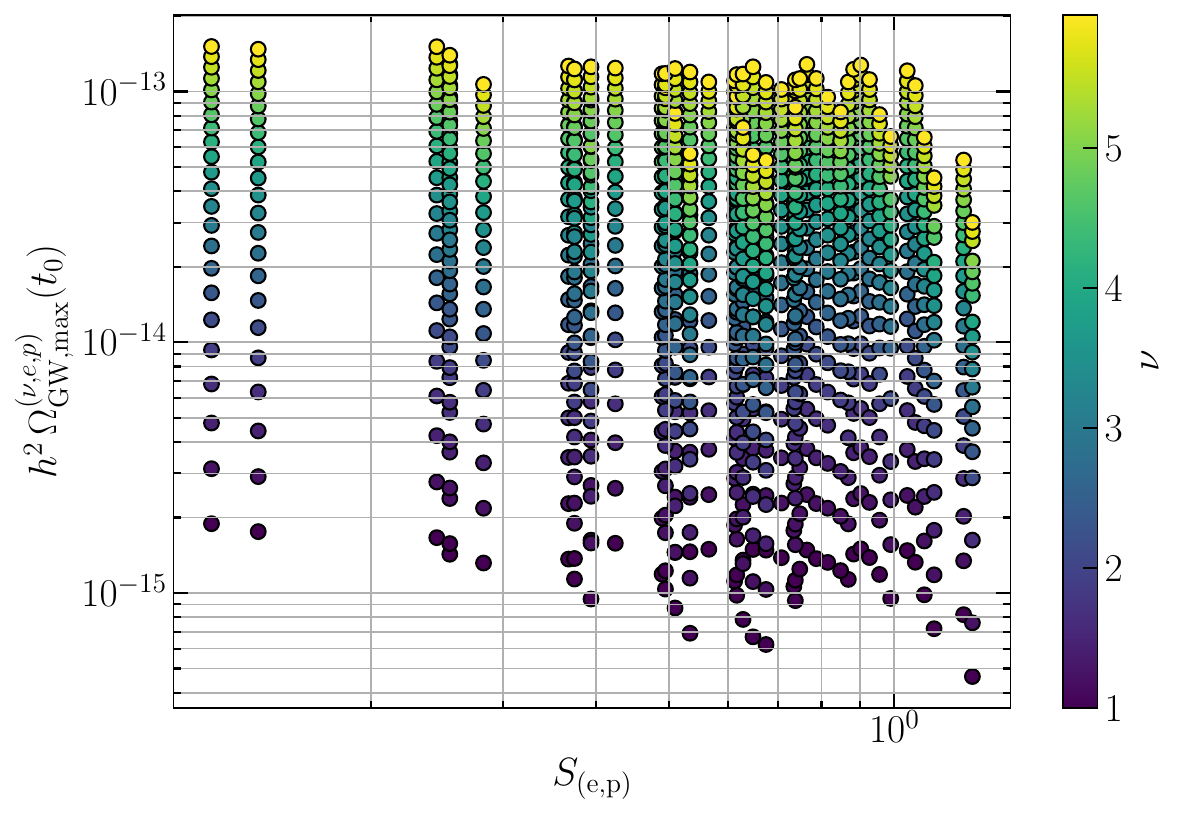}\hfill
\includegraphics[width=0.5\linewidth]{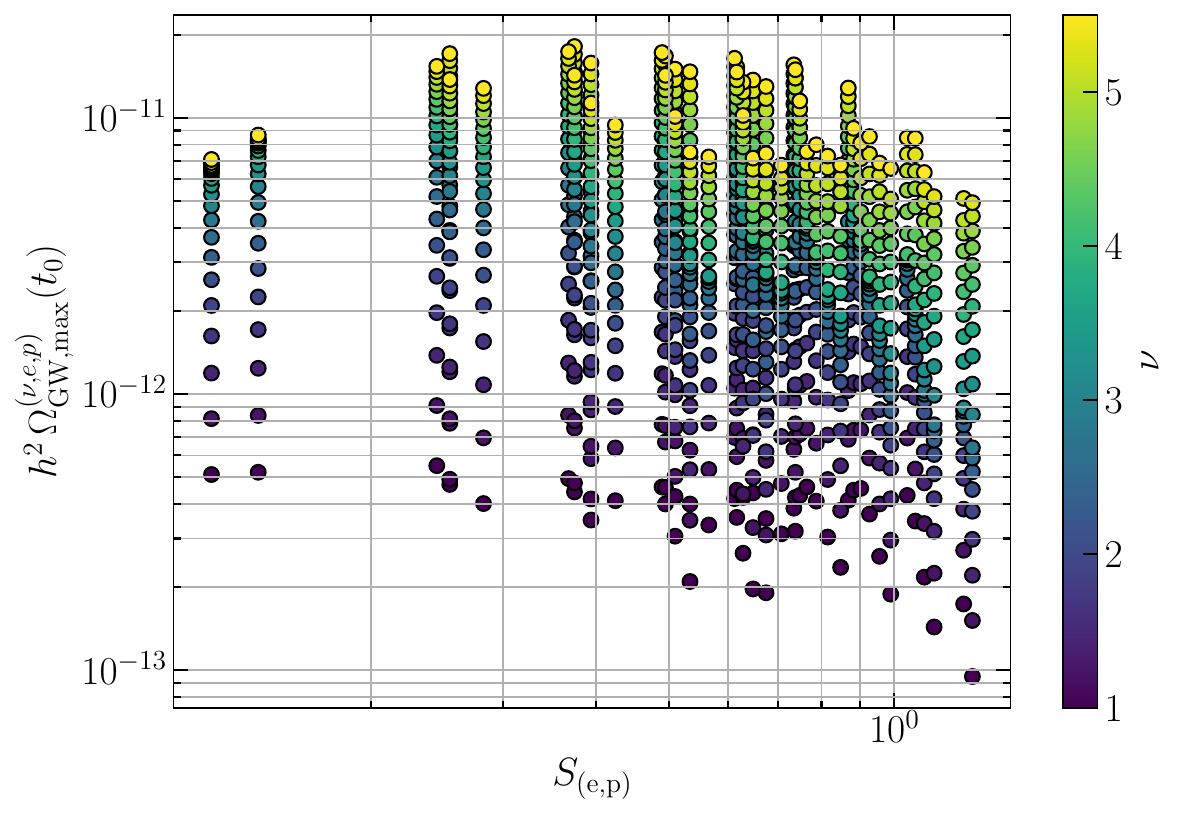}\hfill
\includegraphics[width=0.5\linewidth]{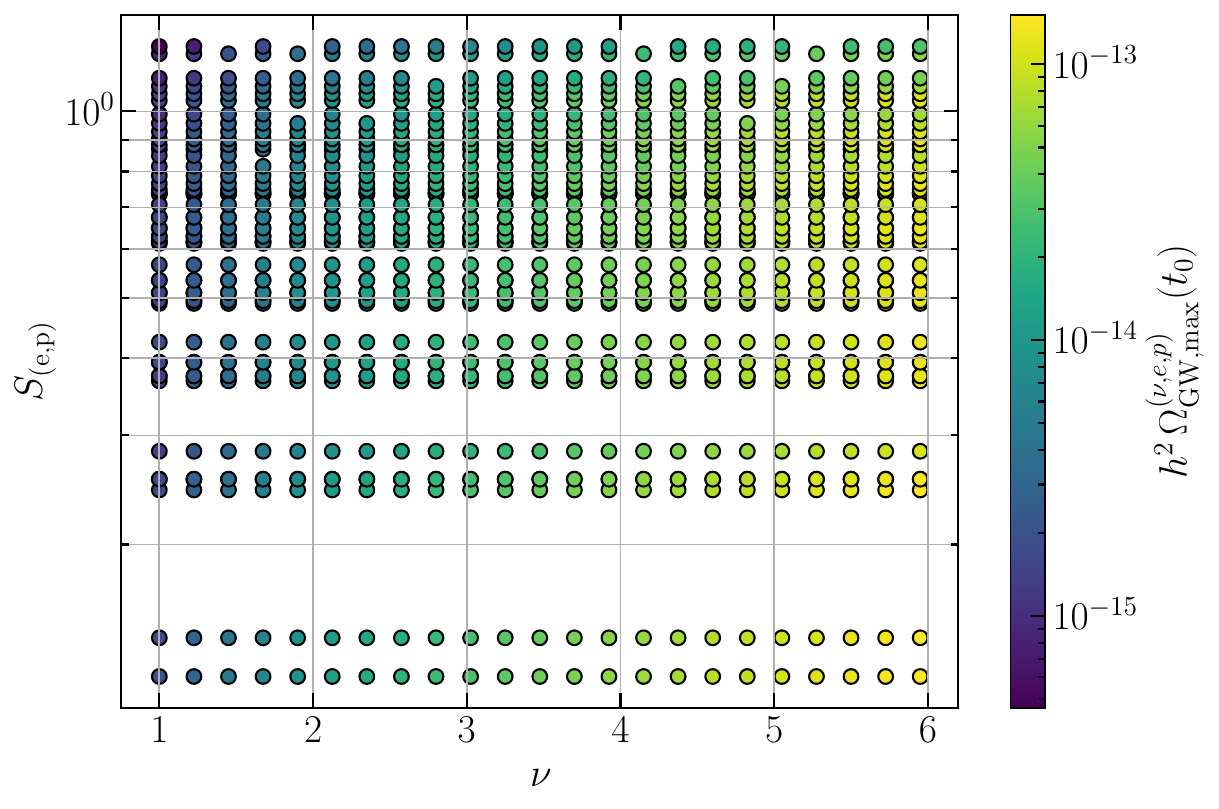}\hfill
\includegraphics[width=0.5\linewidth]{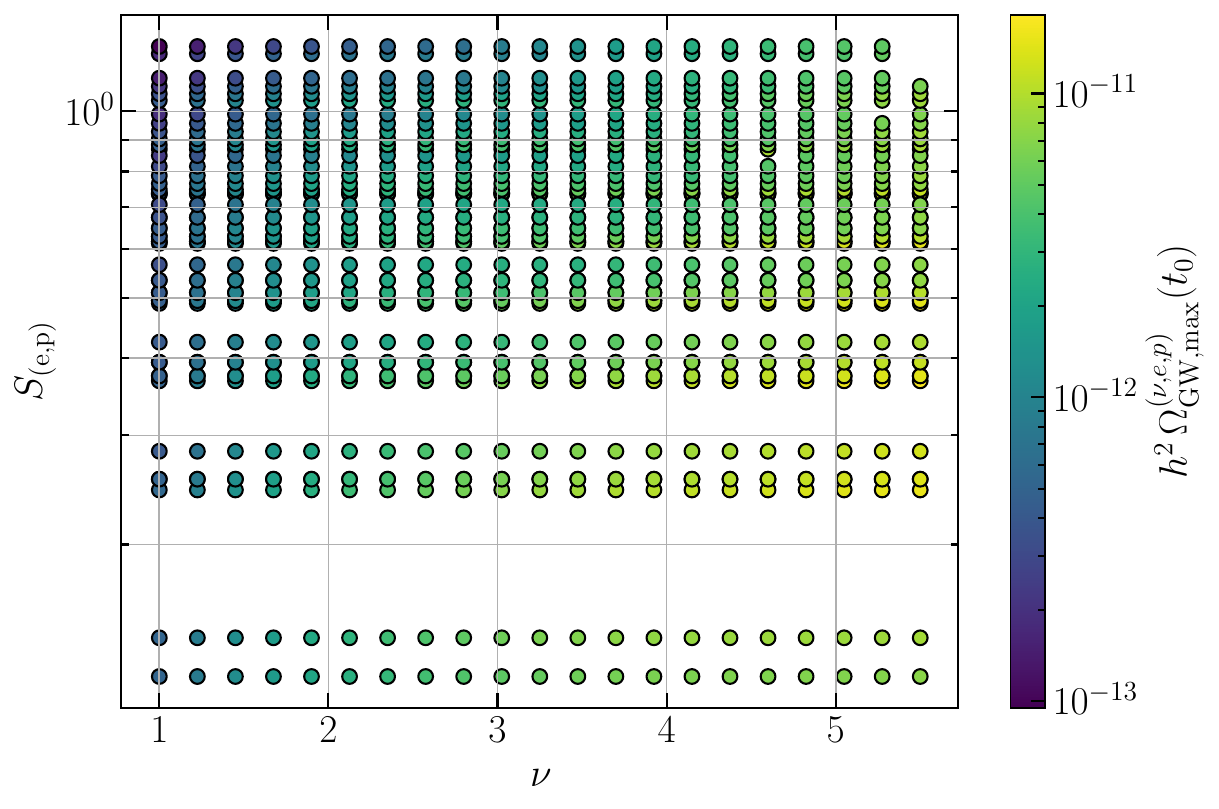}\hfill
\caption{Dependence of the peak GW amplitude on the nonsphericity measure $S(e,p)$ and the peak height $\nu$ for the BBKS realizations. Top panels: Peak value of the present-day GW amplitude, $h^2\Omega_{{\rm GW,max}}(t_0)$, as a function of $S(e,p)$, with the color scale indicating the value of $\nu$. Bottom panels: Distribution of the sampled configurations in the $(\nu,S(e,p))$ plane, with the color scale showing the corresponding value of $h^2\Omega_{{\rm GW,max}}(t_0)$. The left panels correspond to $\sigma=0.01$, while the right panels correspond to $\sigma=0.1$.}
\label{fig:S_parameter}
\end{figure}

To quantify the shape dependence more systematically, we introduce the dimensionless measure of nonsphericity
\begin{equation}
S(e,p)=\frac{1}{\sqrt{3}}
\sqrt{
\left(\frac{\alpha-\bar{\lambda}}{\bar{\lambda}}\right)^2+
\left(\frac{\beta-\bar{\lambda}}{\bar{\lambda}}\right)^2+
\left(\frac{\gamma-\bar{\lambda}}{\bar{\lambda}}\right)^2
}
=
\sqrt{2}\sqrt{3e^2+p^2},
\label{eq:S_ep}
\end{equation}
where $\bar{\lambda}=(\alpha+\beta+\gamma)/3$. This quantity provides a convenient single parameter to characterize the departure from spherical symmetry. The relation between $S(e,p)$ and the peak GW amplitude is shown in Fig.~\ref{fig:S_parameter}. The numerical results indicate that, for both values of $\sigma$ considered here, the largest amplitudes are not produced by the most anisotropic realizations. Rather, the signal tends to be largest for intermediate values of $S$, while it decreases toward the highly anisotropic tail of the distribution. At the same time, the overall scale of the amplitude increases strongly with the peak height $\nu$. This shows that the final GW signal is determined by a nontrivial interplay between the overall amplitude of the peak and the degree of nonsphericity.

\begin{figure}[!htbp]
\centering
\includegraphics[width=0.5\linewidth]{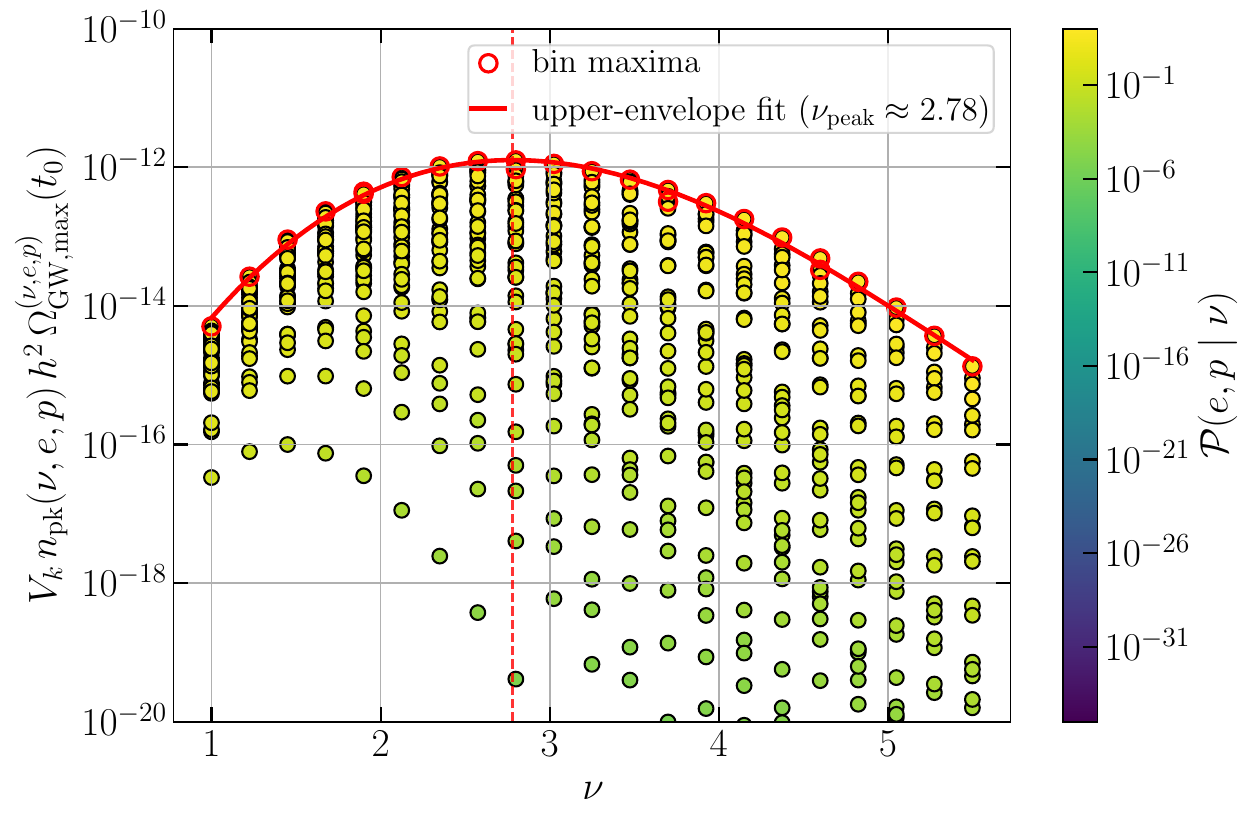}\hfill
\includegraphics[width=0.5\linewidth]{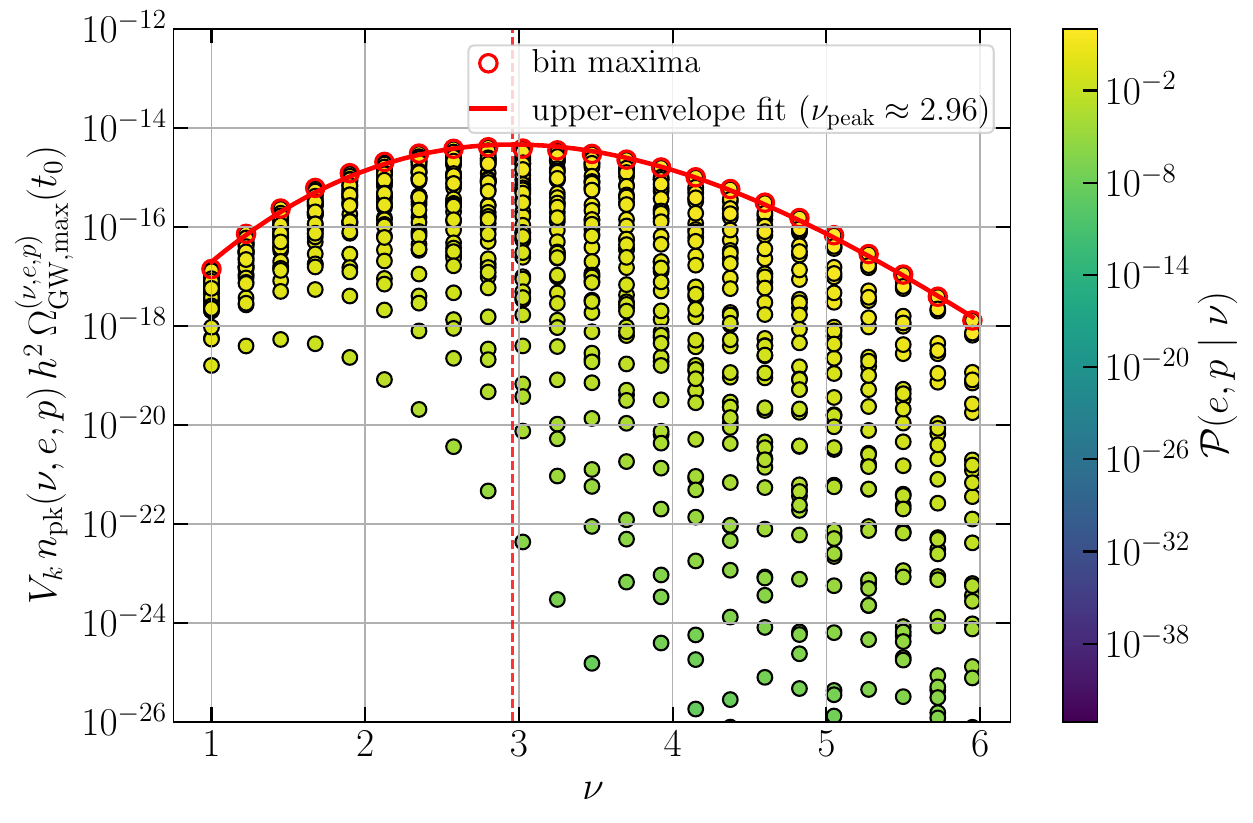}\hfill
\caption{Weighted BBKS contribution to the GW background as a function of the peak height $\nu$. The plotted quantity is $V_k \,n_{\rm pk}(\nu,e,p)\,h^2\Omega_{{\rm GW,max}}^{(\nu,e,p)}(t_0)$ for the sampled realizations. The left and right panels correspond to $\sigma=0.01$ and $\sigma=0.1$, respectively. In both cases, the weighted contribution peaks around $\nu\simeq 3$, showing that the dominant part of the statistically averaged signal arises from an intermediate population of relatively high peaks rather than from either the most common low peaks or the rarest extreme ones.}
\label{fig:weighted_peak_contribution}
\end{figure}

This aspect is made even more explicit in Fig.~\ref{fig:weighted_peak_contribution}, where we plot the quantity $V_k n_{\rm pk}\,h^2\Omega_{{\rm GW,max}}^{(\nu,e,p)}(t_0)$ as a function of $\nu$ for the realizations included in the BBKS analysis. Although the GW amplitude of an individual realization tends to increase with $\nu$, the statistical abundance of such peaks decreases rapidly as the peak height becomes larger. As a consequence, the weighted contribution to the average does not grow monotonically with $\nu$, but instead develops a clear maximum at an intermediate value. From the numerical data, we find that the dominant contribution is concentrated around $\nu\simeq 3$, more precisely near $\nu_{\rm peak}\simeq 2.85$ and $\nu_{\rm peak}\simeq 2.98$ for 
$\sigma = 0.01$ and $\sigma = 0.1$, respectively, 
as shown in Fig.~\ref{fig:weighted_peak_contribution}. Therefore, the dominant contribution to the stochastic background does not come from the most common low peaks, nor from the rarest very large peaks, but rather from an intermediate population of relatively high peaks with intermediate nonsphericity.

\begin{figure}[!htbp]
\centering
\includegraphics[width=0.5\linewidth]{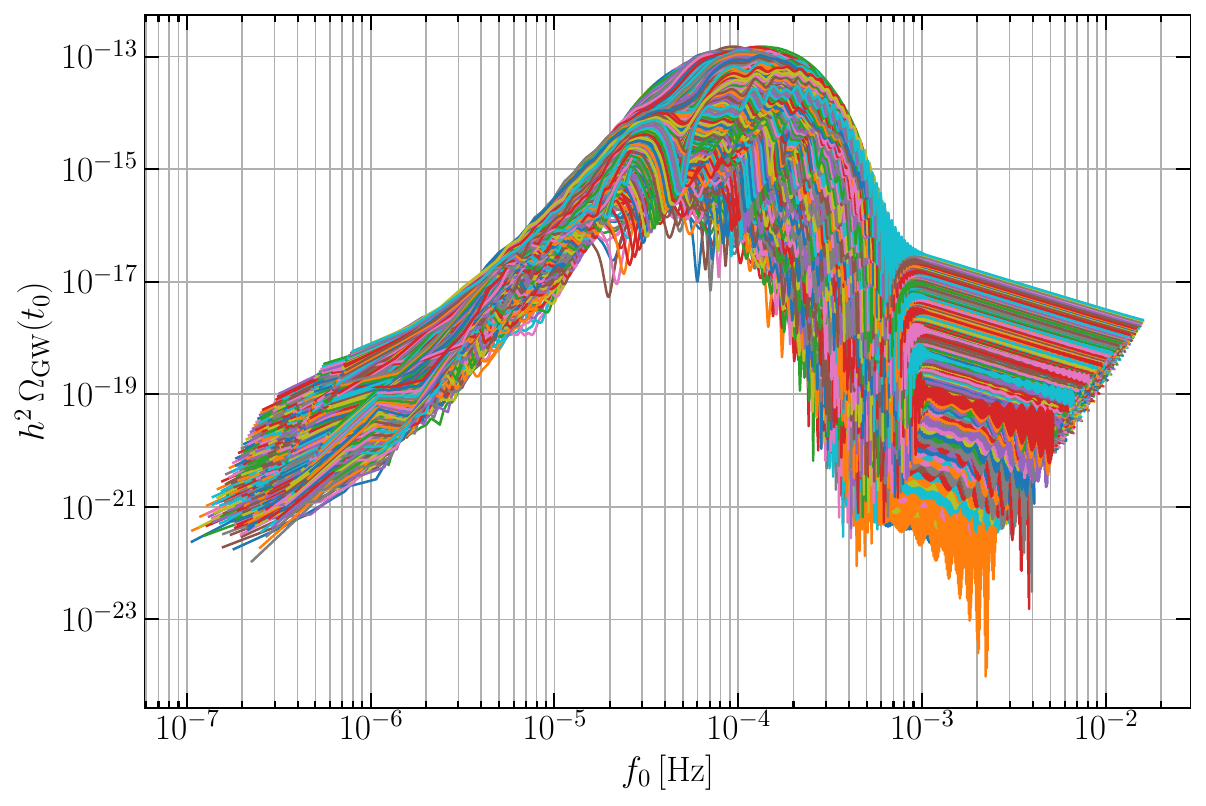}\hfill
\includegraphics[width=0.5\linewidth]{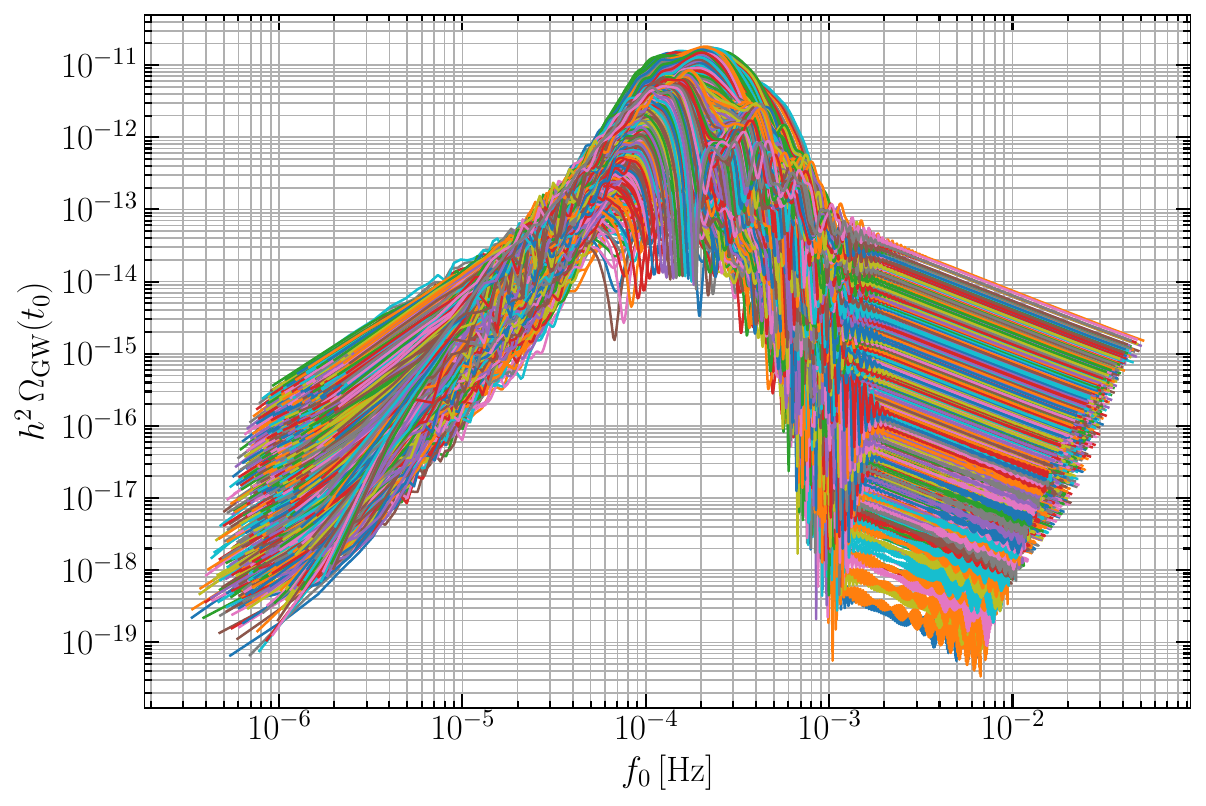}\hfill
\includegraphics[width=0.8\linewidth]{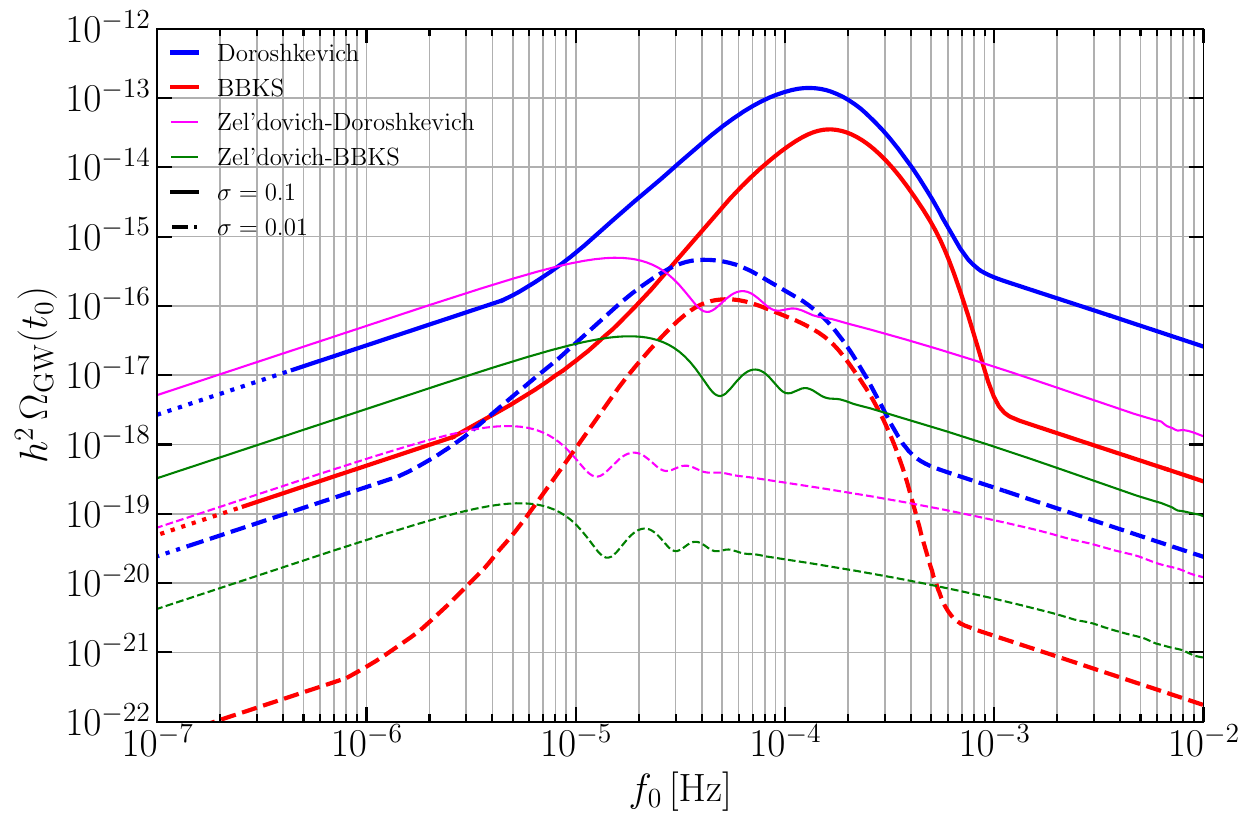}\hfill
\caption{Top panels: Present-day GW spectra, $h^2\Omega_{\rm GW}(t_0)$, for all the configurations included in the statistical analysis, shown for $\sigma=0.01$ (left panel) and $\sigma=0.1$ (right panel) for $M_{k}=10^{-10}M_{\odot}$ and $T_{\rm rh}=0.3\,\textrm{GeV}$. Bottom panel: Statistically weighted average of the spectra. Solid curves correspond to $\sigma=0.1$, while dashed curves correspond to $\sigma=0.01$. The blue and red curves show the fully numerical N-body averages obtained with the Doroshkevich and BBKS weights, respectively, whereas the 
magenta and green curves show the corresponding Zel'dovich-based estimates. The numerical averages are significantly enhanced with respect to the Zel'dovich approximation, demonstrating that the nonlinear post-shell-crossing dynamics and virialization provide an essential contribution to the final GW signal.}
\label{fig:average_spectra}
\end{figure}

We now turn to the final statistically averaged spectra. The full set of individual configurations is shown in the top panels of Fig.~\ref{fig:average_spectra} for the two values of $\sigma$. The statistically weighted averages are shown in the bottom panel of Fig.~\ref{fig:average_spectra}, where we compare the results obtained with the Doroshkevich and BBKS weightings, together with the corresponding estimates based on the Zel'dovich approximation. We refer the reader to Appendix~\ref{subsec:analytic_GW_ZA} for the explicit analytical expression used in the Zel'dovich-based calculation.

Several conclusions follow from this figure. First, the fully numerical N-body averages obtained with the Doroshkevich and BBKS prescriptions are of the same order of magnitude and have a similar spectral shape. This indicates that the overall signal is reasonably robust with respect to the statistical description adopted for the initial conditions, although the BBKS prescription is physically more restrictive since it explicitly counts true peaks of the density field. At the same time, the BBKS result is systematically somewhat smaller than the Doroshkevich one. This is expected, since the peak-theory weighting imposes the additional requirement that the configuration correspond to an actual maximum of the density field, thereby reducing the contribution of some configurations that would otherwise be included in the Doroshkevich approach.

Second, the fully numerical spectra are substantially larger than those obtained from the Zel'dovich-based treatment. In particular, the discrepancy reaches several orders of magnitude in the region around the peak of the signal. This confirms that the nonlinear evolution beyond the Zel'dovich regime plays a crucial role in the generation of gravitational waves. The analytical approximation fails to capture the violent post-shell-crossing dynamics, the subsequent matter redistribution, and the sharp time dependence of the quadrupole during the approach to virialization. As a result, it significantly underestimates the emitted power. This conclusion is fully consistent with the single-realization comparison discussed in the previous subsection and reinforces the need for a direct numerical treatment of the GW source. This differs from the conclusions of previous work \cite{DalianisKouvaris2024}, where the numerical results were found to remain broadly consistent with the Zel'dovich-based estimate, typically within about one order of magnitude. By contrast, within our framework the deviation from the Zel'dovich approximation is significantly larger, reaching up to two-three orders of magnitude in the comparison with the BBKS-based calculation. Relevantly, in our results we don't find a two-peak structure observed in the previous study for the gravitational wave spectrum.

Finally, Fig.~\ref{fig:average_spectra} also shows the dependence on the variance $\sigma$ of the initial perturbations. The case with larger $\sigma$ produces a much stronger GW signal, while the case with smaller $\sigma$ leads to a suppressed spectrum. Thus, although the qualitative features of the averaged spectra are similar in both cases, the overall amplitude depends sensitively on the statistical properties of the primordial perturbations.

It is useful to clarify the distinct roles of the parameters $\nu$ and $\sigma$. By definition, the peak height satisfies $\nu=\delta_L/\sigma$, so that the linear overdensity is $\delta_L=\nu\sigma$. In our implementation, the deformation eigenvalues $(\alpha,\beta,\gamma)$ are proportional to $\delta_L$, and therefore at fixed $(e,p)$ both increasing $\nu$ and increasing $\sigma$ correspond to a larger absolute initial deformation of the collapsing patch. Physically, this implies an earlier turnaround, a more violent nonlinear collapse, and sharper time variations of the quadrupole, which in turn enhance the GW emission. The difference between the two parameters appears at the statistical level: while the amplitude of an individual realization increases with $\nu$, the abundance of such peaks decreases rapidly, so that the weighted contribution to the stochastic background is maximal only at an intermediate value, around $\nu\simeq 3$. This reflects a competition between individual emission strength and statistical abundance: low peaks are relatively common but weak GW emitters, whereas very high peaks produce stronger individual signals but are exponentially rare. Consequently, the stochastic background is dominated by the intermediate population around $\nu\simeq3$. By contrast, increasing $\sigma$ shifts the ensemble toward larger initial overdensities and therefore toward more violent collapse across the sampled configurations, which explains why the spectra for $\sigma=0.1$ are systematically larger than those for $\sigma=0.01$.

Overall, the results of this subsection show that a realistic estimate of the stochastic GW background from nonspherical collapse in an early matter-dominated era requires two ingredients simultaneously: a statistically meaningful sampling of the initial conditions and a fully numerical treatment of the subsequent nonlinear dynamics. The first determines which configurations contribute most strongly to the ensemble average, while the second is essential for obtaining the correct amplitude and shape of the emitted spectrum.

\subsection{Comparison with observational gravitational wave prospects}

Finally, we compare our numerical results with the sensitivities of present and future GW detectors, for which we take the sensitivity curves of different experiments (see for example Ref.~\cite{Schmitz:2020syl}). The purpose of this analysis is to determine whether the signal produced by the collapse and virialization of overdense nonspherical patches during an early matter-dominated era can fall within experimentally relevant frequency windows once it is redshifted to the present epoch.

For the observational comparison presented in this section, we adopt the BBKS-weighted numerical spectrum as our reference result. This choice is motivated by the fact that the BBKS formalism explicitly describes true peaks of the smoothed field and provides their number density, whereas the Doroshkevich distribution describes the deformation eigenvalues at a generic point of the Gaussian field. The BBKS prescription is therefore more directly suited to estimating the contribution from a population of collapsing overdense peaks. The Doroshkevich-weighted result shown in Fig.~\ref{fig:average_spectra} is retained for comparison and illustrates the dependence of the predicted signal on the adopted statistical prescription.

As discussed in Sec.~\ref{sec:theory_gws_calculation}, the present-day frequency and amplitude of the signal depend on both the horizon mass scale $M_k$ and the reheating temperature $T_{\rm rh}$. In particular, changing these parameters modifies the time of collapse relative to reheating and therefore changes the redshift from emission to today. For convenience, the relation between the horizon-entry time and the reheating time may be written as
\begin{equation}
\frac{t_k}{t_{\rm rh}}
\simeq
21.03
\left(\frac{g_{*\rho}(T_{\rm rh})}{g^{\rm SM}_{*\rho}}\right)^{1/2}
\left(\frac{M_k}{M_\odot}\right)
\left(\frac{T_{\rm rh}
}{{\rm GeV}}\right)^2,
\label{eq:tk_trh_obs}
\end{equation}
where we set $k_\text{B}=\hbar=c=1$ and the numerical value $g^{\rm SM}_{*\rho} =106.75$ corresponds to the Standard Model value of the effective energy-density degrees of freedom at high temperature. Requiring the relevant collapse stage to occur before reheating implies an upper bound on the reheating temperature,
\begin{equation}
T_{\rm rh}
\lesssim
0.218\,
\left(\frac{g^{\rm SM}_{*\rho}}{g_{*\rho}(T_{\rm rh})}\right)^{1/4}
\Theta^{-1/2}
\left(\frac{M_k}{M_\odot}\right)^{-1/2}
{\rm GeV},
\label{eq:Trh_bound_obs}
\end{equation}
where $\Theta \equiv t_{\rm e}/t_k$ parametrizes the time of the dominant GW emission in units of the horizon-entry time. These relations are useful for identifying the regions of parameter space in which the signal is generated during the early matter-dominated stage and for understanding how the spectrum is shifted across the observable frequency range. On the other hand, if we assume that these over-dense patches form PBHs and constitute a fraction of the dark matter, this applies to PBHs with masses larger than $M_{\rm PBH} \gtrsim 10^{-16} M_{\odot}$, which are expected not to have evaporated yet. Then, we use the estimation of Ref.~\cite{Ye:2025wif} which infers that the abundance of the PBHs is given by $\beta_{\rm PBH} \sim 1.08 \cdot \sigma^5$ when we consider the anisotropy of the density fluctuation in the early matter-dominated epoch~\cite{Khlopov:1980mg,Harada:2016mhb,Harada:2017fjm}.~\footnote{There are also the other mechanisms which can affect the formations of the PBHs in early matter-dominated epoch, e.g., by the inhomogeneity~\cite{Khlopov:1980mg,Kokubu:2018fxy} or the velocity dispersion~\cite{Harada:2022xjp}.} It is important to stress, however, that this PBH interpretation is not an unavoidable outcome of the collapse. Depending on the subsequent nonlinear dynamics and on the details of reheating, the overdense patches may instead form other compact or virialized structures rather than PBHs. In that case, their contribution to the present dark-matter
abundance is model-dependent and need not be described by the PBH abundance parameter $\beta_{\rm PBH}$. Even within the PBH interpretation, the estimate of $\beta_{\rm PBH}$ carries intrinsic theoretical uncertainties associated with the statistical description of the initial perturbations and with additional physical effects affecting collapse. Therefore, in the following comparison with observational GW prospects, the PBH abundance should be regarded as a reference normalization rather than as a definitive prediction.

\begin{figure}[!htbp]
\centering
\includegraphics[width=0.5\linewidth]{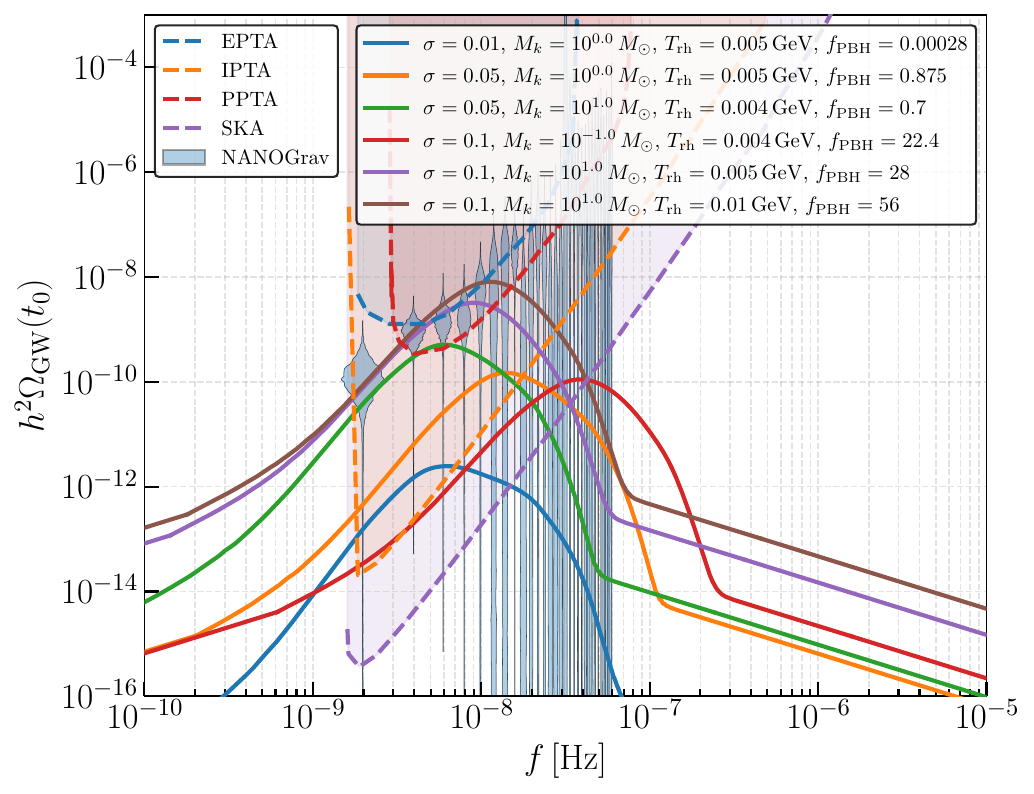}\hfill
\includegraphics[width=0.5\linewidth]{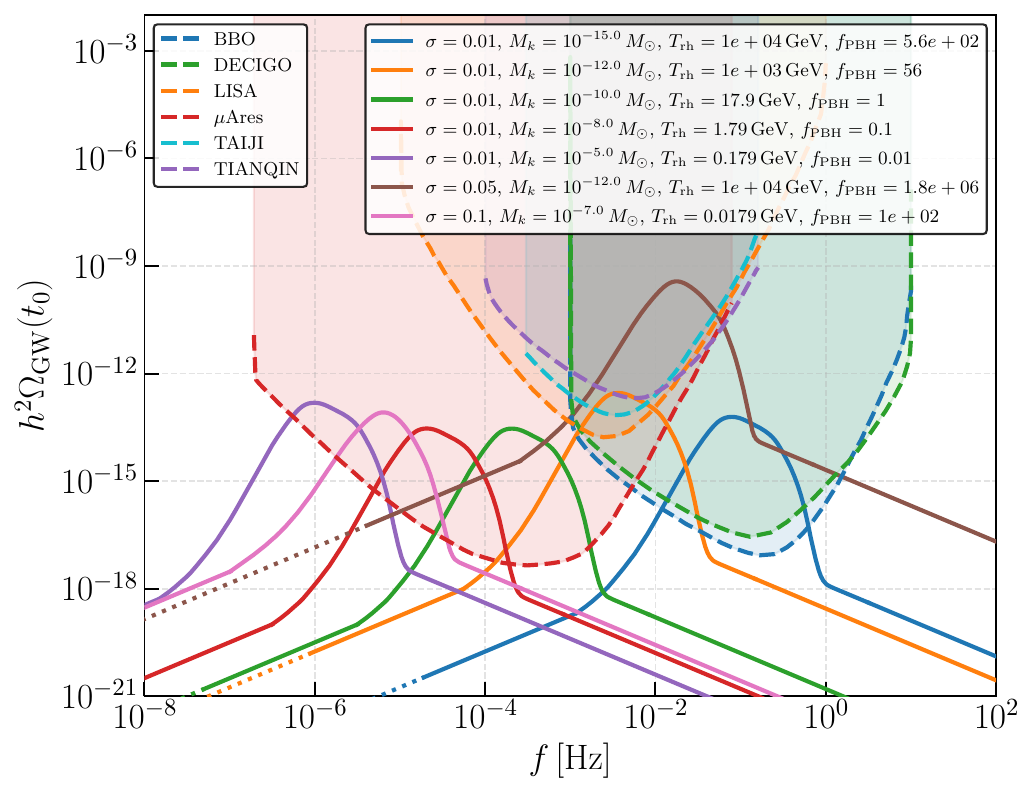}\hfill
\includegraphics[width=0.5\linewidth]{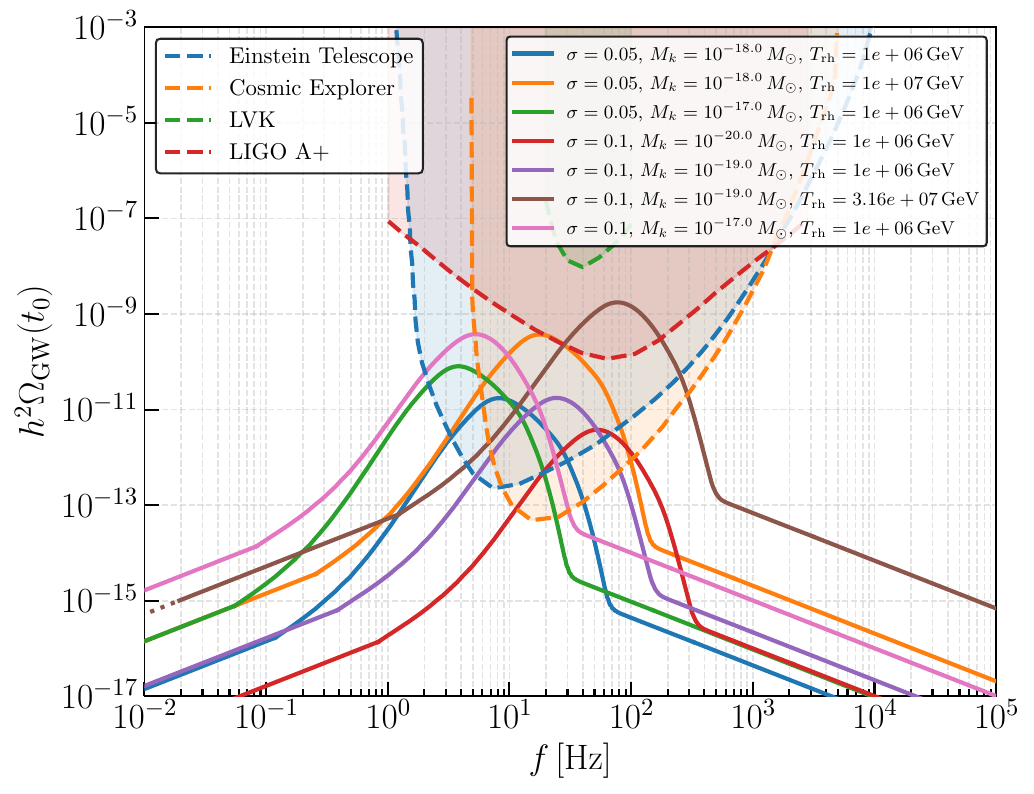}\hfill
\includegraphics[width=0.5\linewidth]{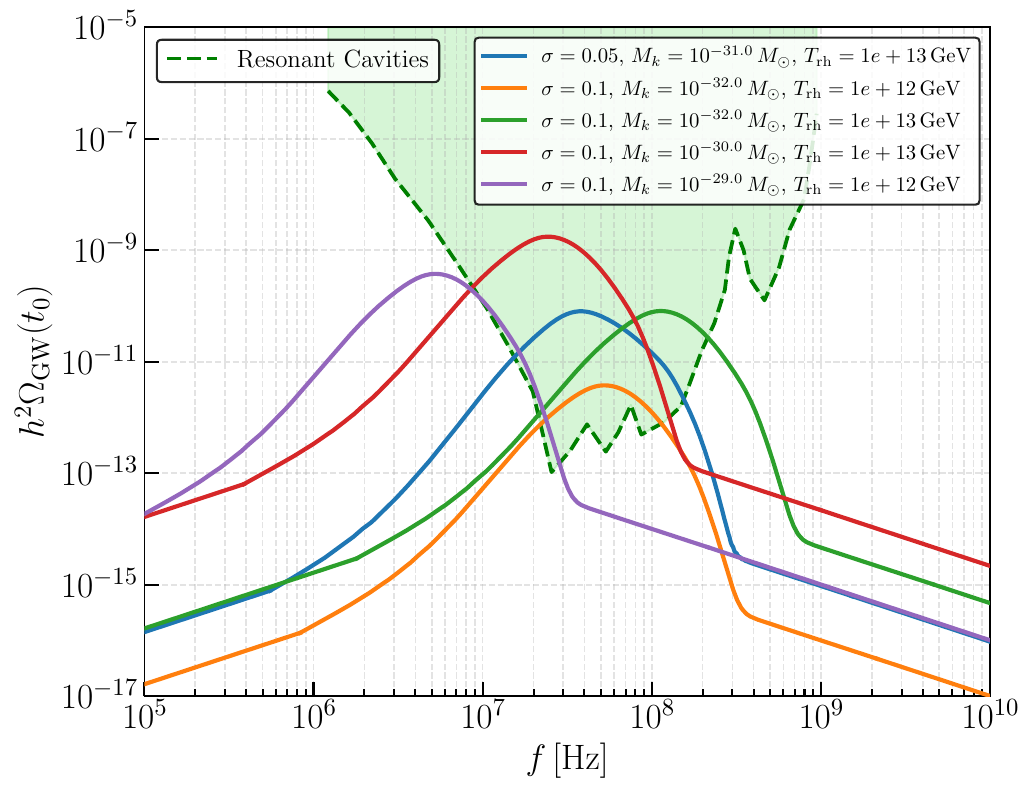}\hfill
\caption{Comparison between the present-day numerical GW spectra and the sensitivities of different detector classes. The panels show representative configurations selected so that the spectral peak falls within the frequency windows of pulsar timing arrays (top left), space-based interferometers (top right), ground-based interferometers (bottom left), and resonant-cavity searches (bottom right). The spectra shown in this figure are obtained using the BBKS statistical weighting. The quoted $f_{\rm PBH}$ values are indicative estimates subject to the theoretical uncertainties of the abundance calculation; values above unity signal formal PBH overproduction within the adopted prescription.}
\label{fig:average_spectra_graw_prospects}
\end{figure}

Then, the 
abundance of PBHs in the form of dark matter can be related with the temperature of reheating $T_{\rm rh}$ 
as (see for instance Refs.~\cite{Dalianis:2018ymb,Allahverdi2020}) $f_{\rm PBH} \approx 5.6 \cdot 10^{8} \cdot (T_{\rm rh}/\textrm{GeV}) \cdot \sigma^5$. Accordingly, the values of $f_{\rm PBH}$ displayed in Fig.~\ref{fig:average_spectra_graw_prospects} should be regarded as indicative estimates obtained within this abundance prescription. In particular, values $f_\text{PBH} >1$ indicate formal PBH overproduction within the adopted estimate and should not be interpreted as the definite prediction that physical dark-matter fractions exceed unity. For a fixed 
value of $\sigma$, we can obtain the reheating temperature for a given desired $f_{\rm PBH}$. It is important to remark, however, that we have a lower bound for the reheating temperature $T_{\rm rh} \gtrsim 4 \textrm{MeV}$ to be consistent with the thermal history of the Universe, e.g., the BBN (Big-Bang Nucleosynthesis) and the CMB (Cosmic Microwave Background)~\cite{Kawasaki:1999na,Kawasaki:2000en,Ichikawa:2005vw,deSalas:2015glj,Hasegawa:2019jsa,Barbieri:2025moq} (see also a review article~\cite{Allahverdi2020}). 

The comparison with detector sensitivities is shown in Fig.~\ref{fig:average_spectra_graw_prospects}. The four panels correspond, respectively, to the frequency bands probed by pulsar timing arrays (NANOGrav \cite{NANOGrav:2023gor}, EPTA \cite{EPTA:2023sfo}, IPTA \cite{InternationalPulsarTimingArray:2023mzf}, PPTA \cite{Reardon:2023gzh}, and SKA \cite{Janssen:2014dka})\footnote{Since the evidence of GW signals was found by pulsar timing arrays~\cite{NANOGrav:2023gor, EPTA:2023fyk, Reardon:2023gzh, Xu:2023wog, InternationalPulsarTimingArray:2023mzf}, the found GW signals constitute the background/foreground for our GW signals. Therefore, the sensitivity degrades compared to those shown in the figure~\cite{Babak:2024yhu}.} , space-based interferometers (Big Bang Observatory-BBO \cite{Crowder:2005nr}, LISA \cite{2017arXiv170200786A}, DECIGO \cite{Seto:2001qf}, TAIJI \cite{Luo:2021qji}, TIANQIN \cite{Liang:2021bde} and $\mu$-Ares \cite{Sesana:2019vho}), ground-based interferometers LVK \cite{KAGRA:2021kbb,LIGOScientific:2025bgj}, LIGO A $+$ \cite{Barsotti:2018Aplus}, Einstein Telescope \cite{Sathyaprakash:2012jk}, and Cosmic Explorer \cite{Reitze:2019iox}), and high-frequency resonant-cavity searches \cite{Berlin:2021txa,Herman:2022fau} (we take the sensitivity band from Ref.~\cite{Jiang:2024akb}). The numerical curves shown in each panel correspond to representative configurations selected from the N-body results after varying $M_k$ and $T_{\rm rh}$ so as to place the present-day peak of the spectrum within the relevant observational window.

The figure reveals a clear overall trend. Signals in the pulsar-timing band are obtained for comparatively large horizon masses and very low reheating temperatures. In contrast, moving the peak to higher frequencies requires progressively smaller values of $M_k$ together with larger reheating temperatures.\footnote{
For fixed dimensionless collapse parameters and emission time relative to horizon entry, the approximate dependence of the present-day peak frequency and peak amplitude on the horizon mass $M_k$ and reheating temperature $T_{\rm rh}$ is
\begin{equation}
f_{0,\mathrm{peak}}
\propto M_k^{-1/3}T_{\rm rh}^{1/3},
\qquad
\Omega_{\rm GW}\left(f_{0,\mathrm{peak}}\right)
\propto M_k^{2/3}T_{\rm rh}^{4/3},
\end{equation}
in agreement with Ref.~\cite{DalianisKouvaris2024}. Thus, decreasing $M_k$ or increasing $T_{\rm rh}$ shifts the peak toward higher present-day frequencies. These scaling relations describe how a given dimensionless source spectrum is mapped to the present epoch, but they do not determine its intrinsic amplitude and spectral shape. A numerical treatment remains essential to calculate these quantities, as well as the characteristic emission time, because they depend on the fully nonlinear post-shell-crossing collapse, matter redistribution, and subsequent virialization of the overdense patch.}

In this way, the characteristic frequency of the signal can be shifted from the nHz range relevant for PTA searches, to the mHz--Hz range of space-based detectors, then to the Hz--kHz range of ground-based interferometers, and finally to the MHz--GHz regime relevant for resonant-cavity experiments. Therefore, within this framework, the pair $(M_k,T_{\rm rh})$ acts as the main control of the present-day location of the GW peak.

More specifically, the PTA window is populated by examples with $T_{\rm rh}\sim 10^{-3}$--$10^{-2}\,{\rm GeV}$ and comparatively large horizon masses, typically around $M_k\sim 10^{-1}$--a few $M_\odot$. The space-based window relevant for LISA, DECIGO, BBO, TAIJI, TIANQIN and $\mu$-Ares is reached for intermediate masses and reheating temperatures, approximately $M_k\sim 10^{-15}$--$10^{-7}M_\odot$ and $T_{\rm rh}\sim 10^{-2}$--$10^4\,{\rm GeV}$. The ground-based window of Einstein Telescope, Cosmic Explorer, LVK and LIGO $A^{+}$ requires still smaller masses, around $M_k\sim 10^{-20}$--$10^{-17}M_\odot$, together with $T_{\rm rh}\sim 10^6$--$10^7\,{\rm GeV}$. Finally, the resonant-cavity regime is reached only for extremely small horizon masses, $M_k\sim 10^{-32}$--$10^{-29}M_\odot$, and very large reheating temperatures, $T_{\rm rh}\sim 10^{12}$--$10^{13}\,{\rm GeV}$.

Another feature visible in Fig.~\ref{fig:average_spectra_graw_prospects} is the dependence on the variance $\sigma$ of the primordial perturbations. In agreement with the results of Sec.~\ref{sec:statistics_weighted}, the configurations with $\sigma=0.1$ systematically produce stronger GW signals than those with $\sigma=0.01$. This reflects the fact that, at fixed shape parameters, increasing $\sigma$ increases the linear overdensity $\delta_L=\nu\sigma$, leading to a more violent nonlinear collapse and therefore to a larger quadrupole variation. As a consequence, the observational prospects are considerably more favorable for the larger-$\sigma$ realizations.

Overall, Fig.~\ref{fig:average_spectra_graw_prospects} shows that the numerical signal obtained in this work can populate a very broad range of observable frequencies, from the PTA band up to very high-frequency searches, depending on the values of $M_k$ and $T_{\rm rh}$. In this sense, the signal carries direct information about the expansion history prior to reheating. At the same time, the figure should be interpreted as identifying illustrative regions of parameter space in which the signal can enter the sensitivity windows of different detector classes, rather than as a definitive forecast of detectability. A complete assessment would require a dedicated exploration of the full model parameter space together with all relevant cosmological and primordial-black-hole constraints.

Nevertheless, the comparison shows that the substantially enhanced amplitudes obtained from the fully nonlinear N-body treatment improve the observational prospects relative to Zel'dovich-based estimates. This provides further evidence that resolving the nonlinear collapse and virialization dynamics is essential not only for a reliable theoretical prediction of the spectrum, but also for a realistic assessment of its possible observability.

\section{Conclusions}
\label{sec:conclusions}

In this work, we have investigated the production of gravitational waves from the collapse of nonspherical overdense patches during an early matter-dominated era by means of semirelativistic N-body simulations.

To this end, we have developed a new numerical procedure and methodology from scratch for performing the N-body simulations accounting for the cosmological expansion (see the GitHub repository \cite{escriva_github} for details). We constructed the collapsing region from a Zel'dovich deformation of a homogeneous Lagrangian sphere and evolved it in an Einstein--de Sitter background with a spline-softened particle--particle force law and semirelativistic particle kinematics. We also formulated the appropriate virial diagnostics for this framework and used them to identify the onset of virialization. This allowed us to analyze the collapse dynamics in detail and to quantify the role of numerical ingredients such as the softening scale and particle number $N$. We found that the choice of softening has a nontrivial impact on both the virialization history and the GW emission: overly small values enhance
close-encounter effects and numerical roughness, whereas overly large values oversmooth the inner collapse and suppress the emitted signal. Therefore, the predicted GW signal retains some dependence on the adopted softening value. Within our setup, $\epsilon_{\rm com}=0.05$ and $N=3000$ provide a satisfactory compromise between numerical stability, convergence, and computational cost. Nevertheless, the detailed high-frequency tail shows a comparatively stronger dependence on the adopted softening scale, reflecting its sensitivity to the shortest-scale post-shell-crossing dynamics. In addition, we studied the particle velocity distribution during the collapse. As a general trend, we find that the later the collapse begins relative to horizon crossing, the milder the particle velocities become. In particular, when the system is already well inside the horizon at the onset of collapse, which is the case for small $\sigma$, the particle velocities remain only mildly relativistic. This contrasts with cases of larger $\sigma$ in which the collapse starts closer to the horizon-crossing time, where moderately relativistic velocities can be reached during the peak of the collapse dynamics.

A central result of our analysis is that the GW signal is highly sensitive to the detailed time dependence of the quadrupole during the strongly nonlinear stage. In particular, the fitting-based prescription adopted in previous work does not accurately reproduce the direct numerical quadrupole evolution once the system approaches the bottleneck and virialization stages. Since the GW power depends on the third time derivative of the quadrupole, even moderate mismatches in the fitted time-domain signal are strongly amplified in the final spectrum. More generally, we showed that Zel'dovich-based estimates fail to capture the violent post-shell-crossing dynamics and therefore substantially underestimate the emitted power.

We then performed a statistically weighted average over realizations using both the Doroshkevich and BBKS distributions. The resulting averaged spectra have similar shapes and remain within the same overall order of magnitude, indicating that the signal is reasonably robust with respect to the statistical prescription adopted for the initial conditions. At the same time, the BBKS result is systematically smaller, as expected from the more restrictive requirement that the configurations correspond to true peaks of the density field. The dominant contribution to the gravitational wave spectrum does not arise from either the most common low peaks or the rarest extreme ones, but from an intermediate population of relatively high peaks, around $\nu \simeq 3$, with intermediate nonsphericity. We also found that increasing the variance $\sigma$ enhances the signal significantly, since it increases the absolute deformation amplitude and leads to a more violent collapse.

On the other hand, we have also made a comparison between the fully numerical results and the Zel'dovich-based treatment. Around the peak of the spectrum, the discrepancy reaches several orders of magnitude, especially in the BBKS-based comparison with $\sim 3$ orders of magnitude. This differs qualitatively from previous conclusions in the literature and shows that a precision prediction of the GW signal requires resolving the full nonlinear dynamics numerically. In this sense, one of the main messages of the present paper is that the post-shell-crossing and virialization stages are not a minor correction to the signal, but rather a dominant part of the source. In addition, the two-peak structure reported previously is not found within our setup and full numerical methodology.

Finally, we translated the spectra to the present epoch and compared them with the sensitivity windows of current and future experiments. By varying the horizon mass scale $M_k$ and the reheating temperature $T_{\rm rh}$, we found that the present-day peak can be shifted over a very broad range of frequencies, from the nHz PTA band to the MHz--GHz regime. Within our framework, representative configurations can populate the sensitivity windows of pulsar timing arrays, space-based interferometers, ground-based interferometers, and resonant-cavity searches. The observational prospects are considerably more favorable for the larger-$\sigma$ realizations, reflecting the enhanced collapse strength in those cases. 

The present analysis can be extended in several directions. First, it would be interesting to study how the GW spectrum changes for primordial power spectra beyond the monochromatic case considered here, as well as to investigate the signal produced during a standard matter-dominated epoch rather than an early one. Second, it would be valuable to incorporate additional relativistic corrections and, ultimately, extend the numerical framework developed in this work to a fully general relativistic treatment, in order to compare the resulting GW emission and collapse dynamics with those obtained in the present semirelativistic approach. On the phenomenological side, a more complete assessment of detectability will require a dedicated scan over the relevant cosmological parameters, together with all associated constraints, using Bayesian analysis techniques, including those related to primordial black hole formation and reheating. We leave all these aspects for future work.

\acknowledgments
A.E thanks the support from the Young Leader Cultivation program (YLC) at Nagoya University and the Junior Research Group Leader program at APCTP. This work was supported by JSPS KAKENHI Grant Numbers 26K17141 (A.E), 25K07281 (C.Y.), 24K07027  (TH, KK, CY) and 26H00403 (TT). TT's work was partly supported by the 34th (FY 2024) Academic research grant (Natural Science) No.~9284 from DAIKO FOUNDATION.

\appendix
\section{Appendix}

\subsection{Additional numerical results}
\label{sec:extra_simulations}

\begin{figure}[!htbp]
\centering
\includegraphics[width=0.3\linewidth]{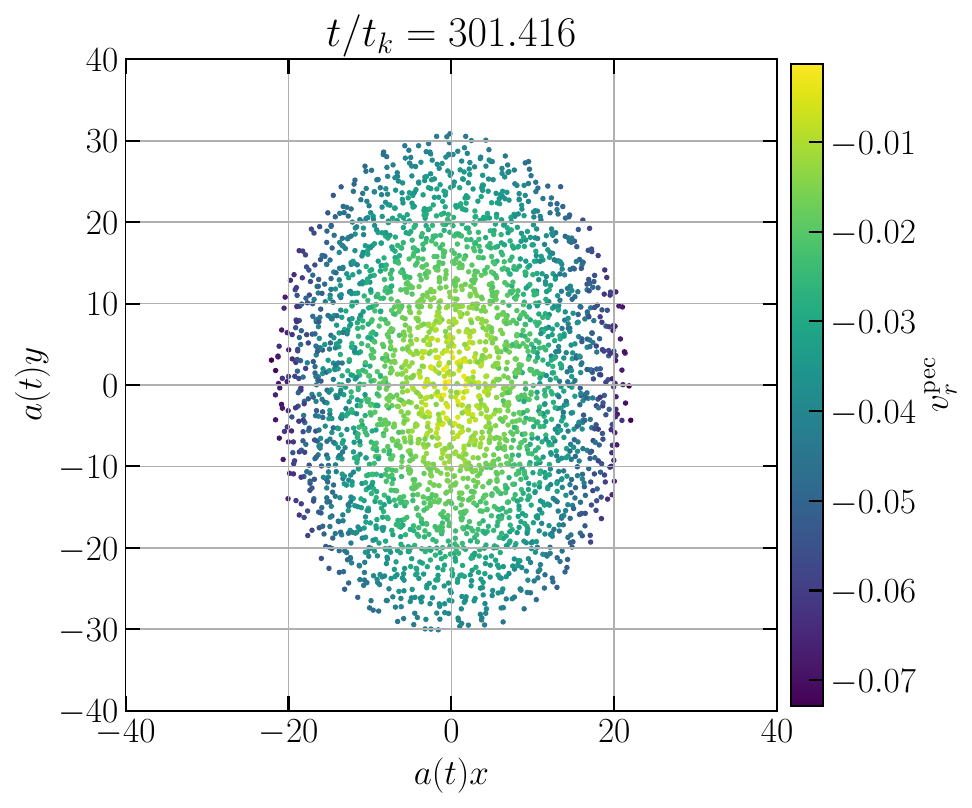}\hfill
\includegraphics[width=0.3\linewidth]{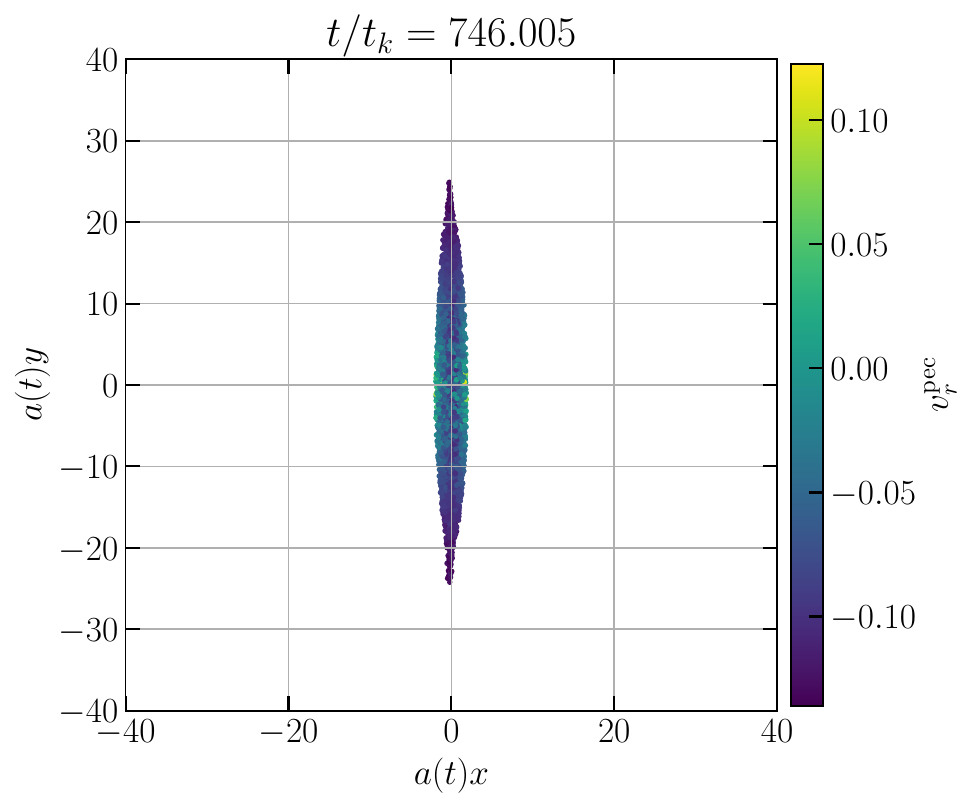}\hfill
\includegraphics[width=0.3\linewidth]{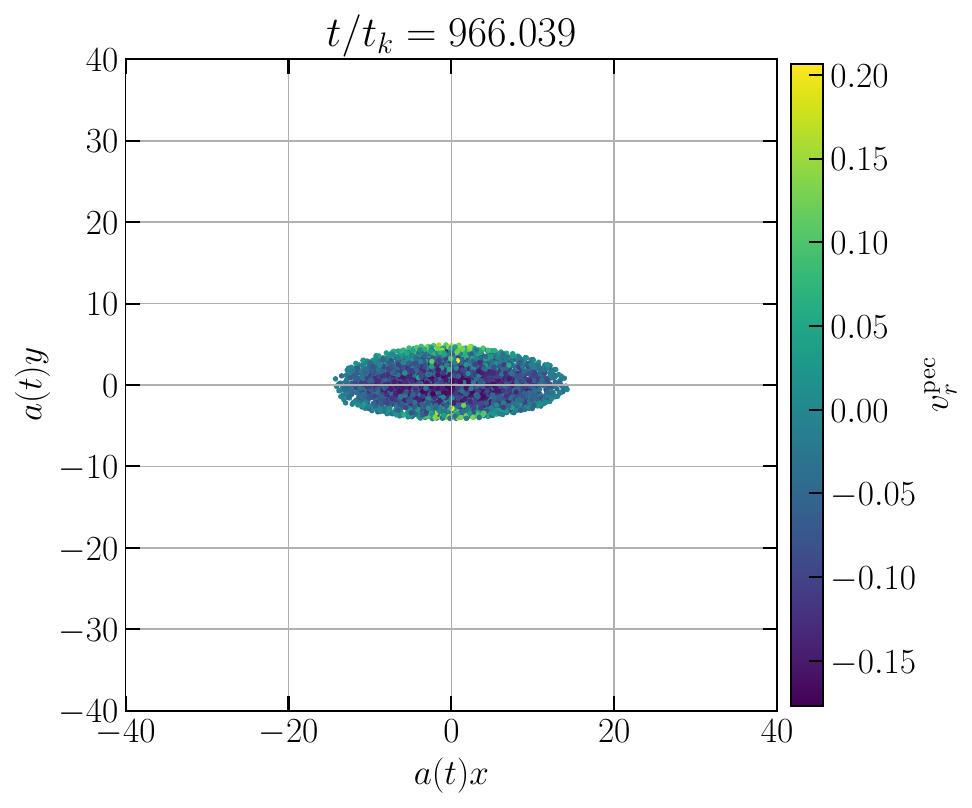}\hfill
\includegraphics[width=0.3\linewidth]{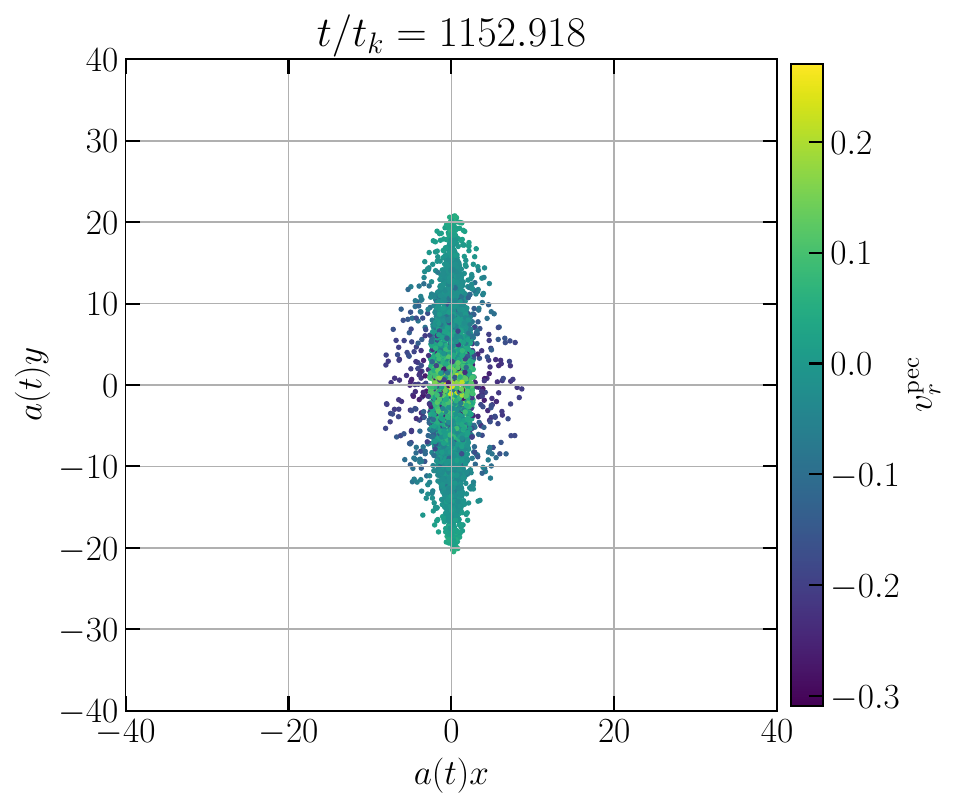}\hfill
\includegraphics[width=0.3\linewidth]{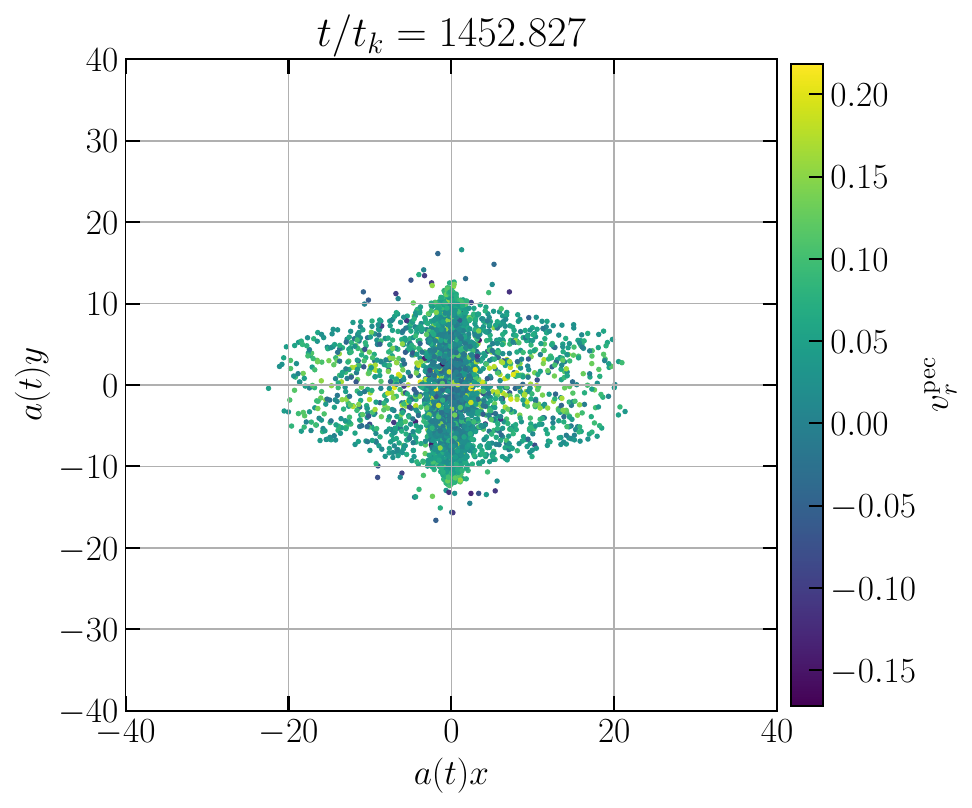}\hfill
\includegraphics[width=0.3\linewidth]{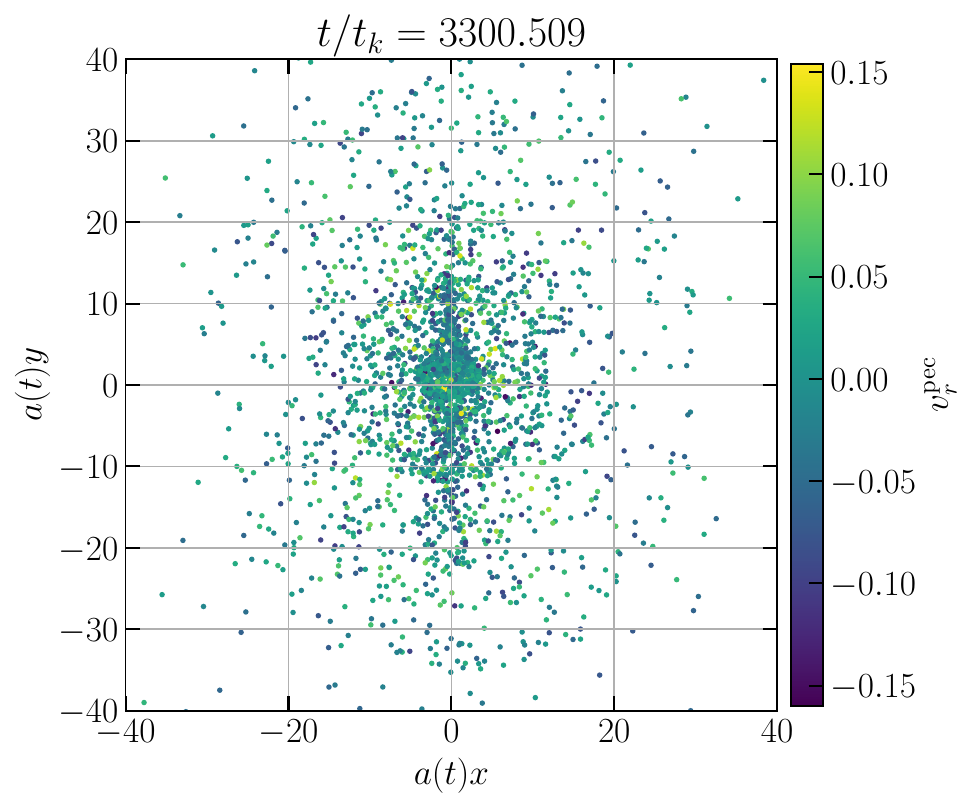}\hfill
\includegraphics[width=0.5\linewidth]{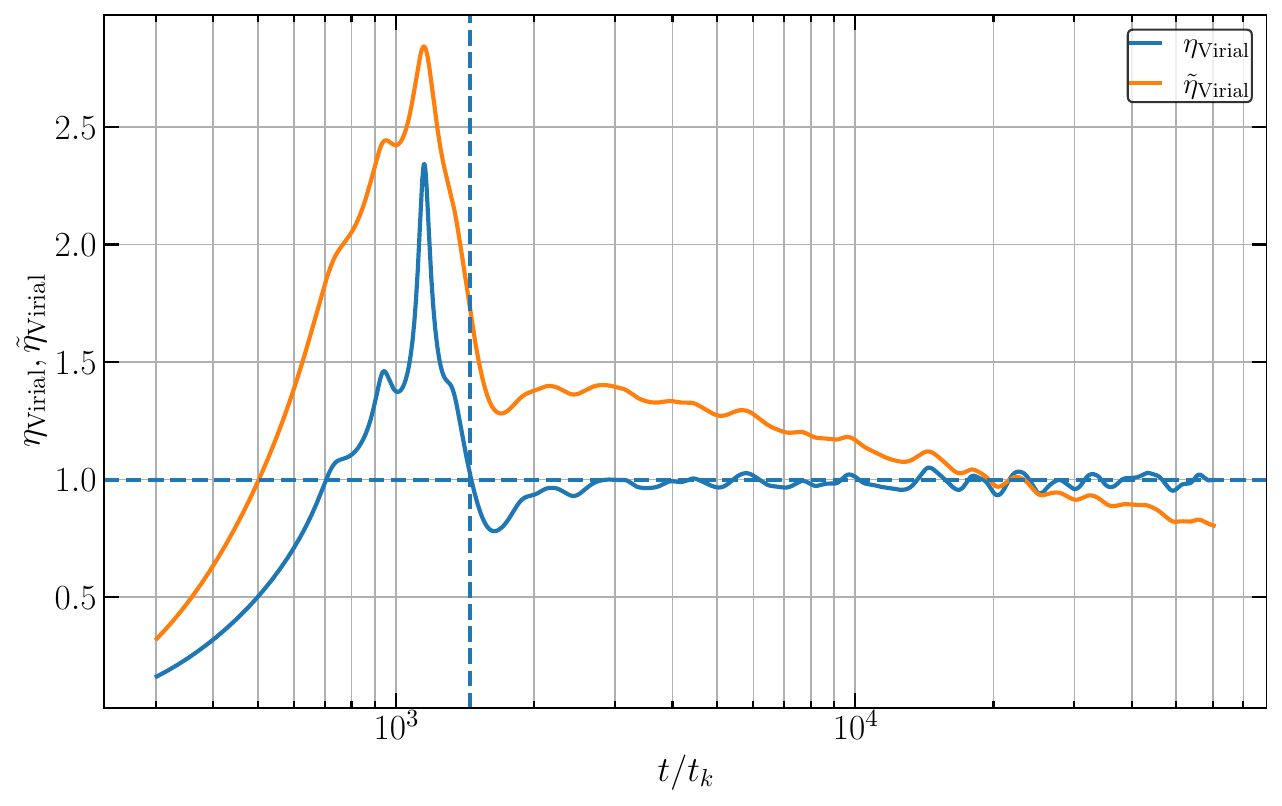}\hfill
\includegraphics[width=0.5\linewidth]{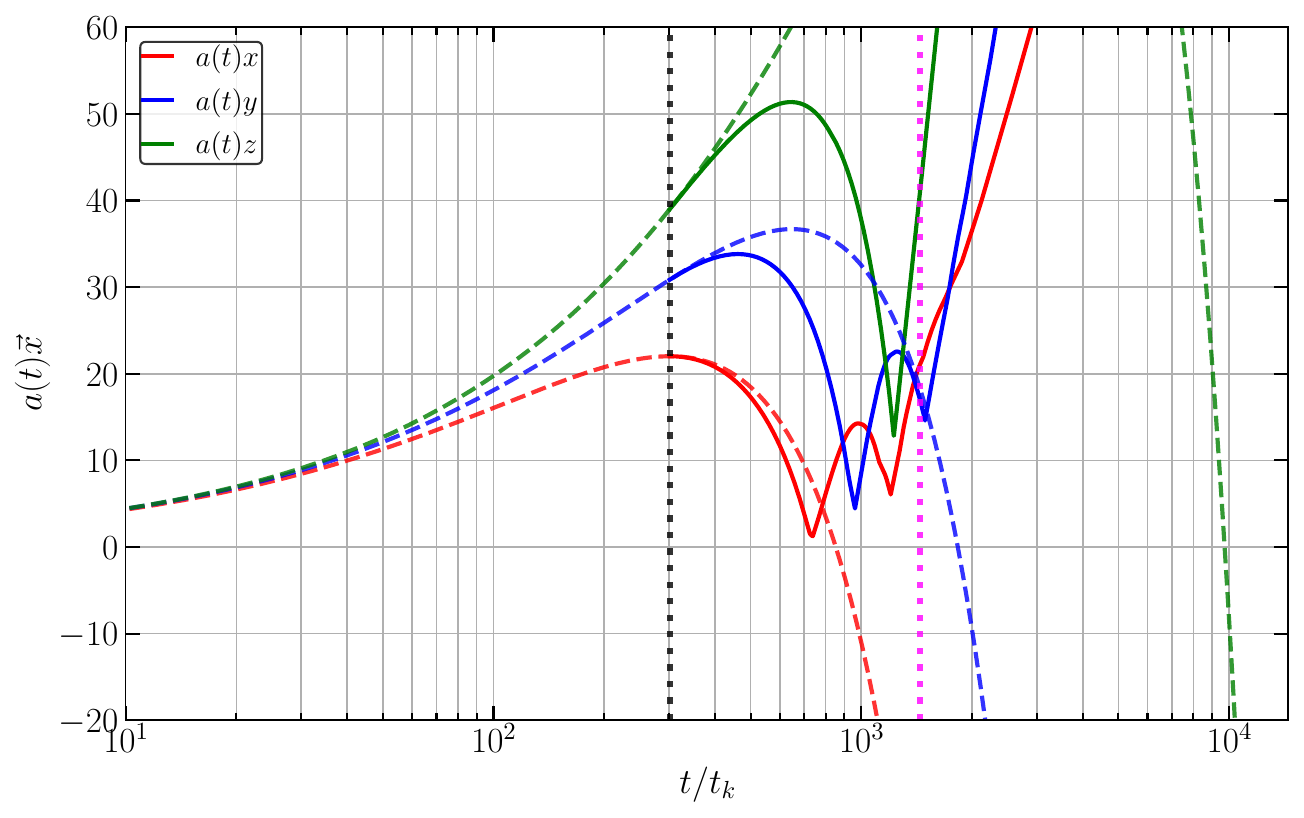}\hfill
\caption{Top panels: Snapshots of the particle distribution at different times, expressed in units of the horizon-crossing time of the overdense patch. Bottom panels: Time evolution of the virial theorem (left) and comparison between the Zel’dovich approximation and the numerical result (right). Case with $\nu=2, e=0.223196, p=-0.00125251$, $\alpha \approx 0.0111, \beta \approx 0.00668, \gamma \approx 0.00219$, and $\sigma=0.01$.}
\label{snapshots_particles_Nbody2}
\end{figure}

\begin{figure}[!htbp]
\centering
\includegraphics[width=0.3\linewidth]{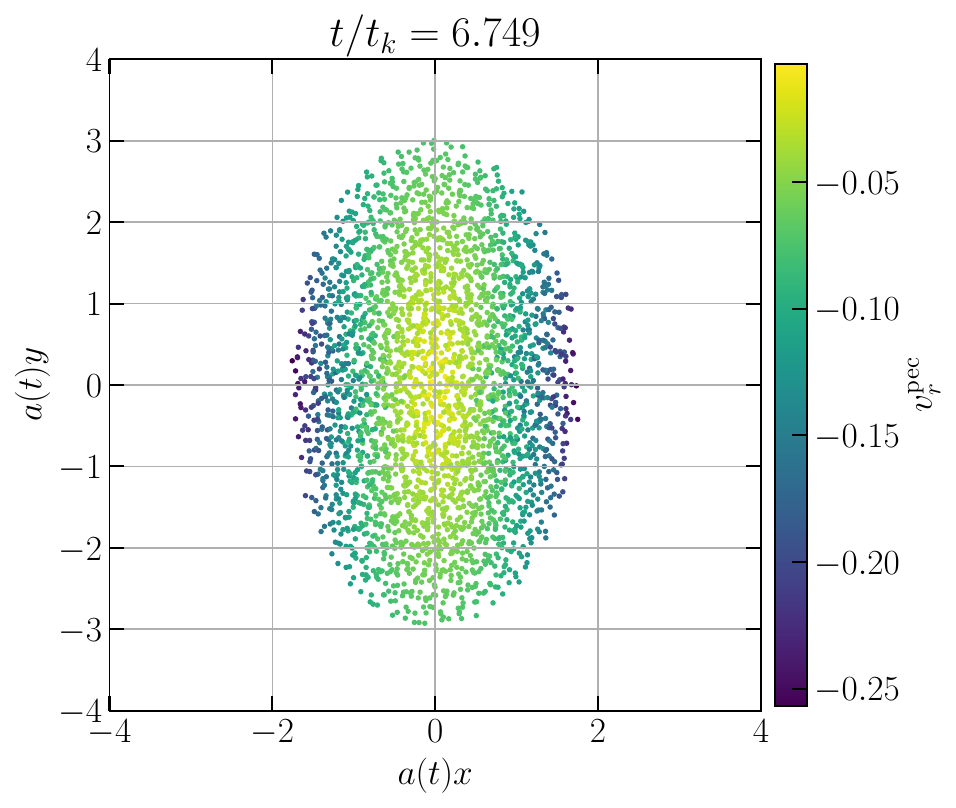}\hfill
\includegraphics[width=0.3\linewidth]
{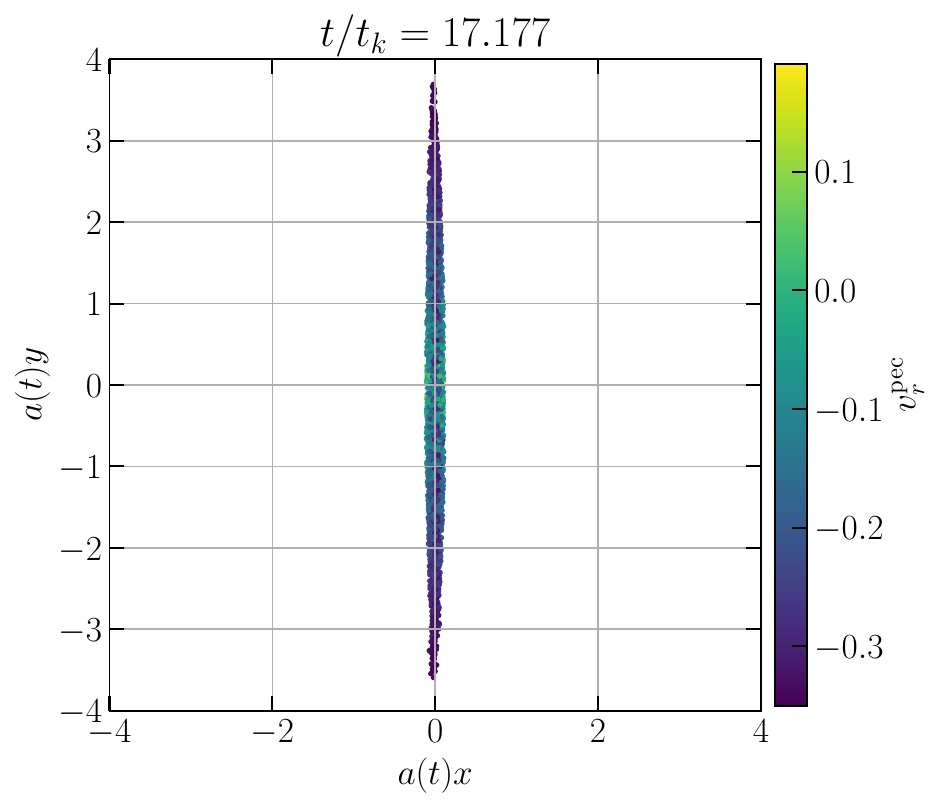}\hfill
\includegraphics[width=0.3\linewidth]
{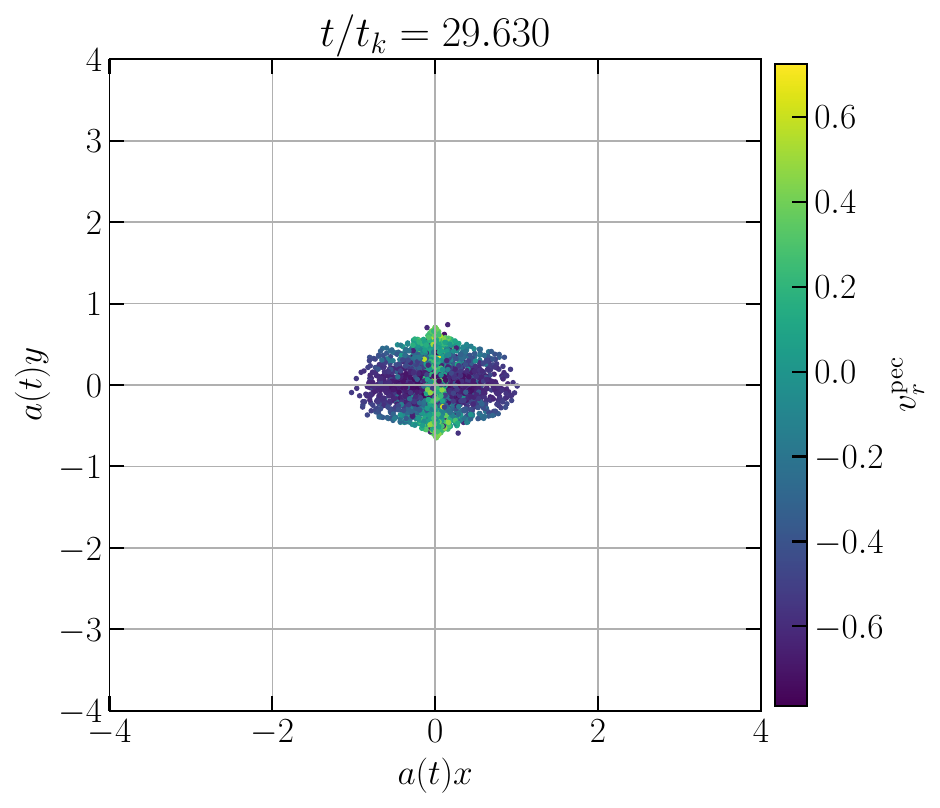}\hfill
\includegraphics[width=0.3\linewidth]{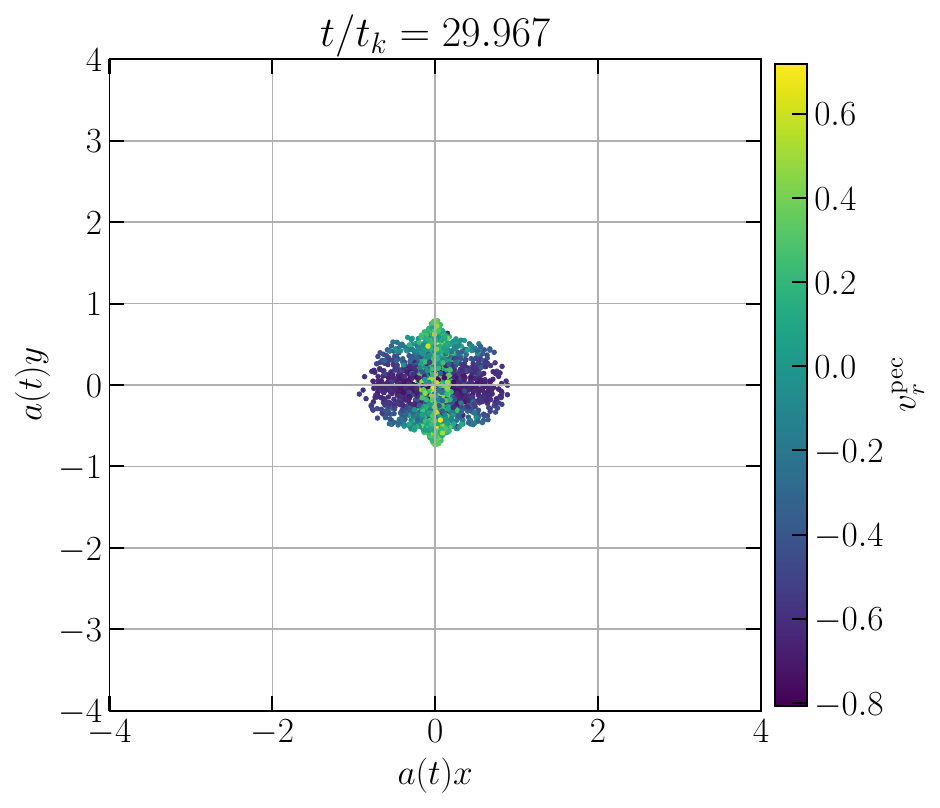}\hfill
\includegraphics[width=0.3\linewidth]{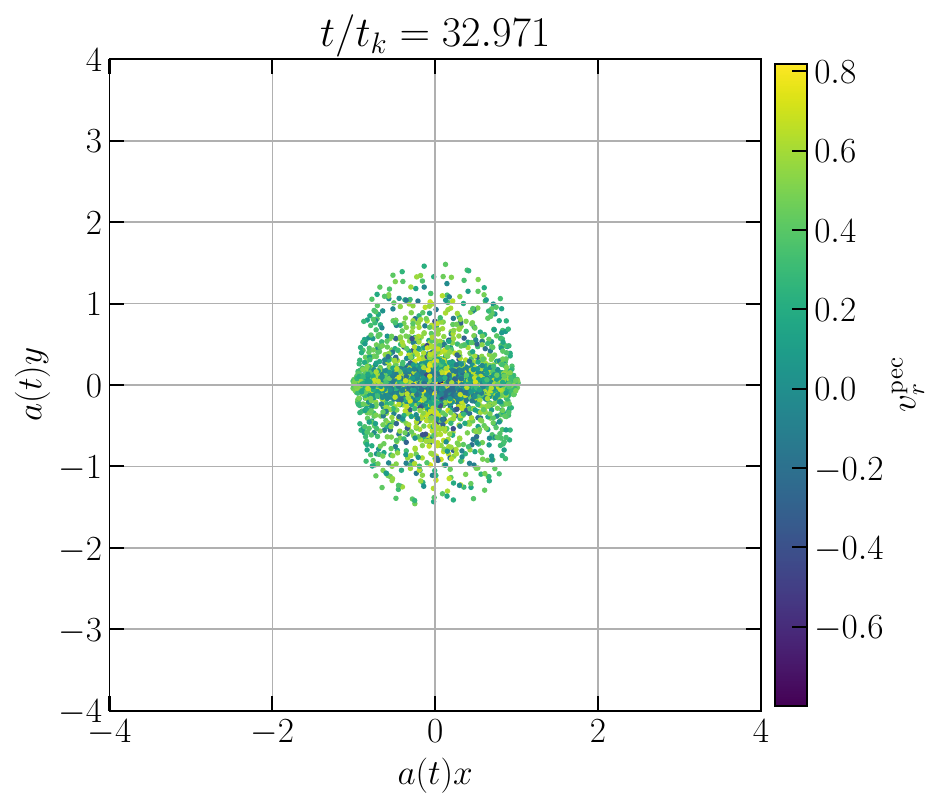}\hfill
\includegraphics[width=0.3\linewidth]{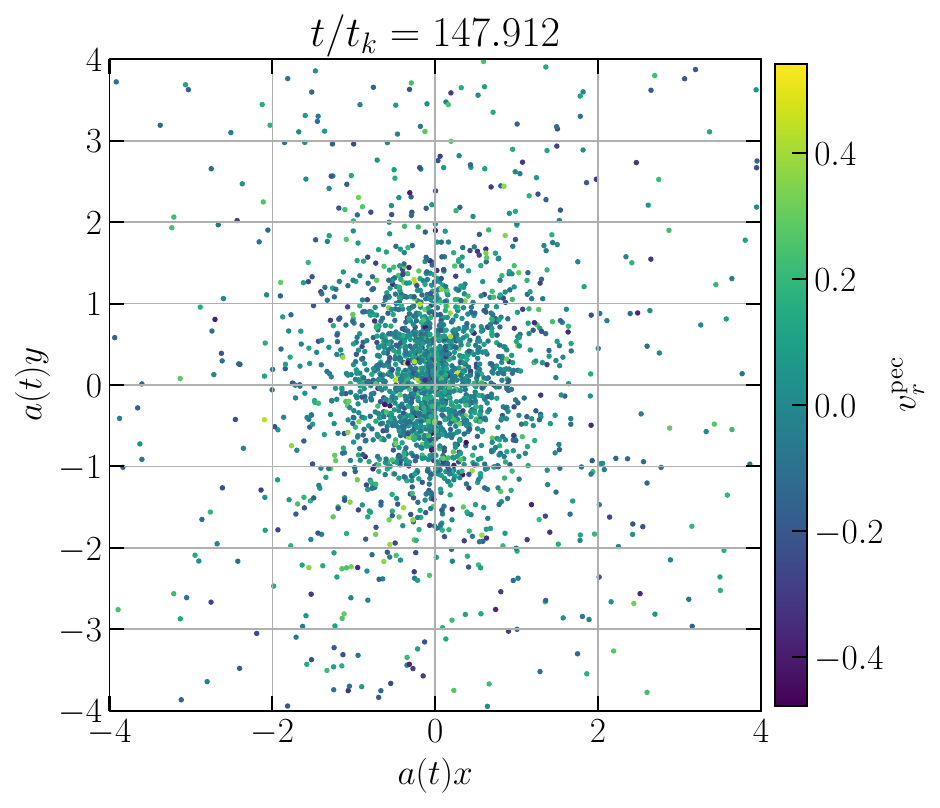}\hfill
\includegraphics[width=0.5\linewidth]{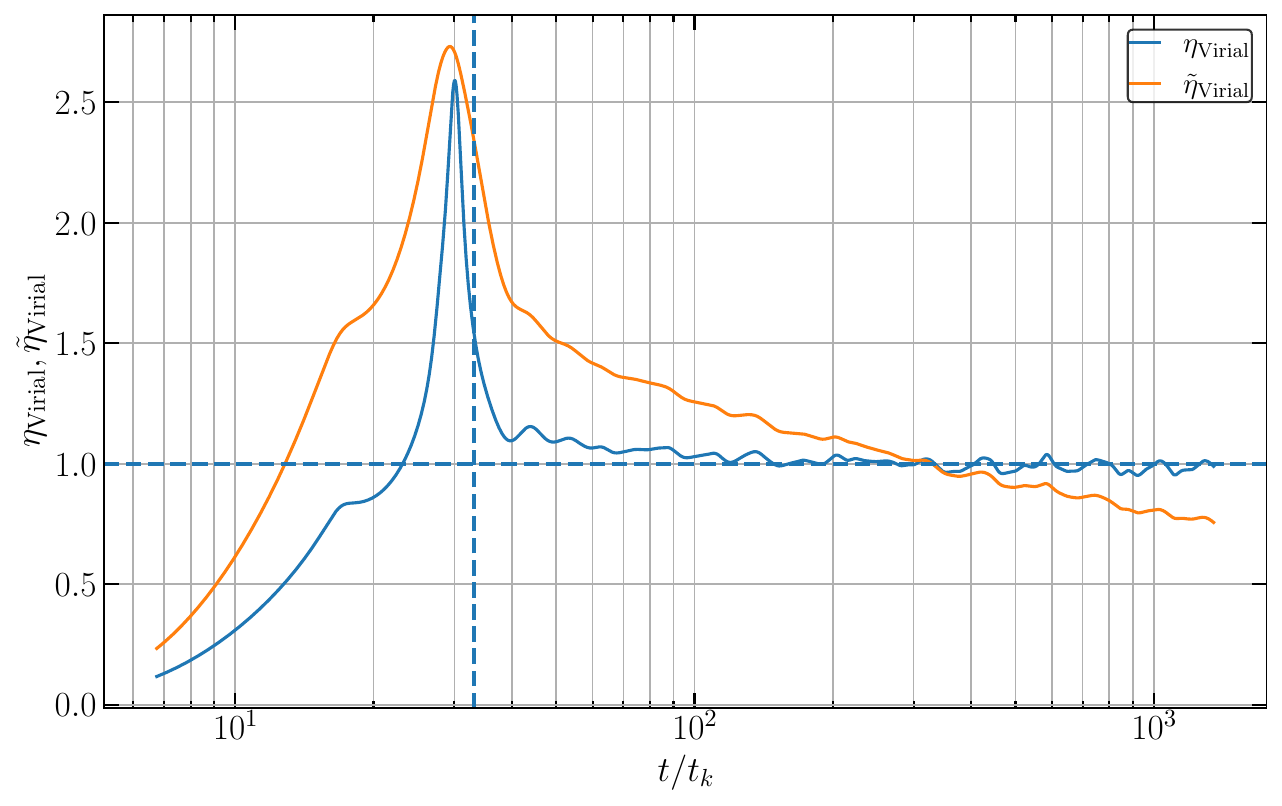}\hfill
\includegraphics[width=0.5\linewidth]{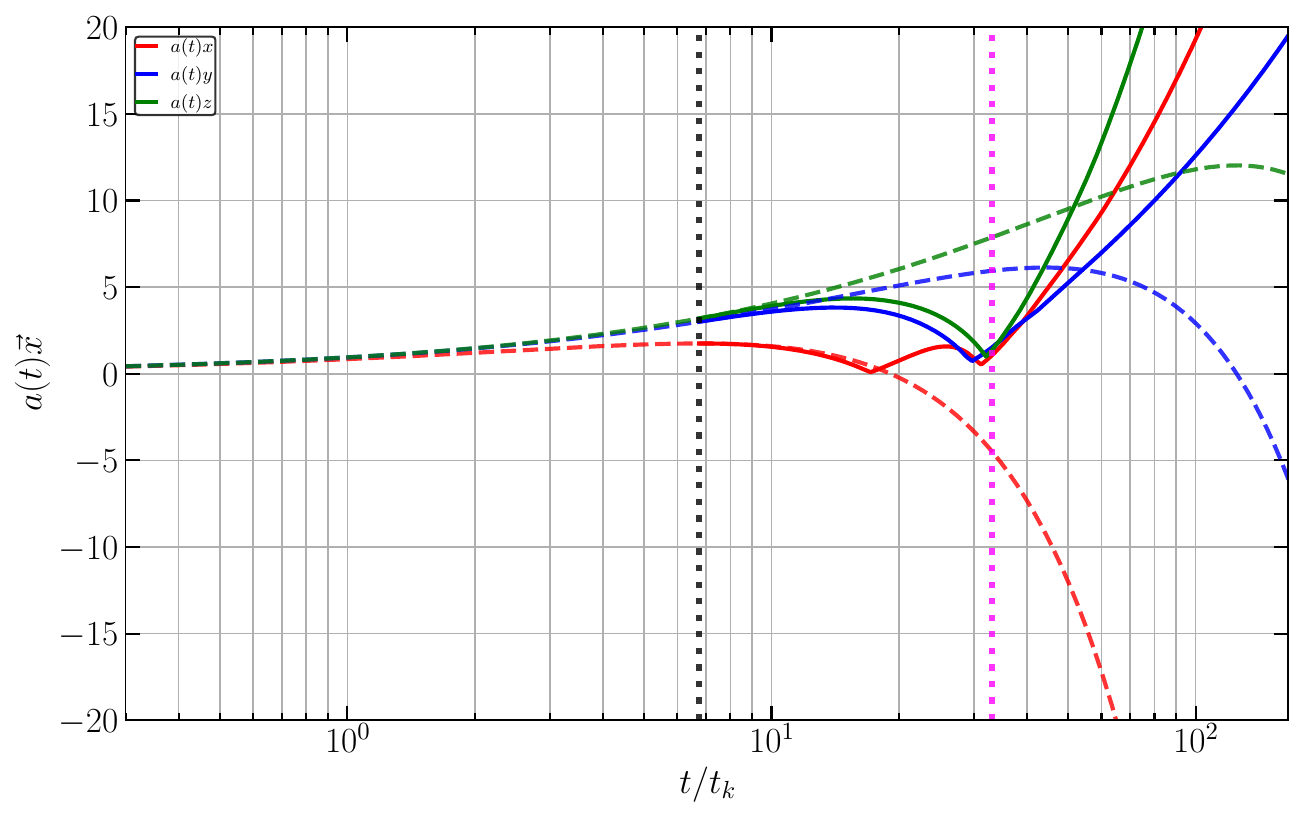}\hfill
\caption{Same caption as in Fig.\ref{snapshots_particles_Nbody2} with parameters $\sigma=0.1$, $\nu=2, e=0.3, p=0.2$, $\alpha \approx 0.14$, $\beta \approx 0.04$, and $\gamma \approx 0.02$.}
\label{snapshots_particles_Nbody3}
\end{figure}

\begin{figure}[!htbp]
\centering
\includegraphics[width=0.3\linewidth]{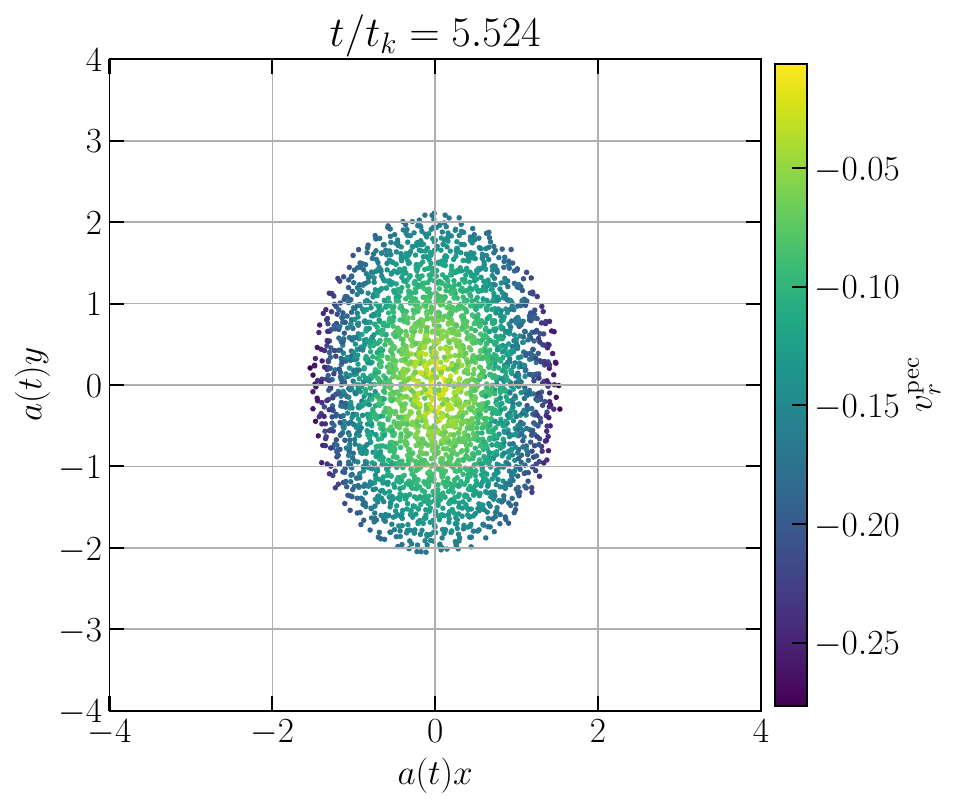}\hfill
\includegraphics[width=0.3\linewidth]{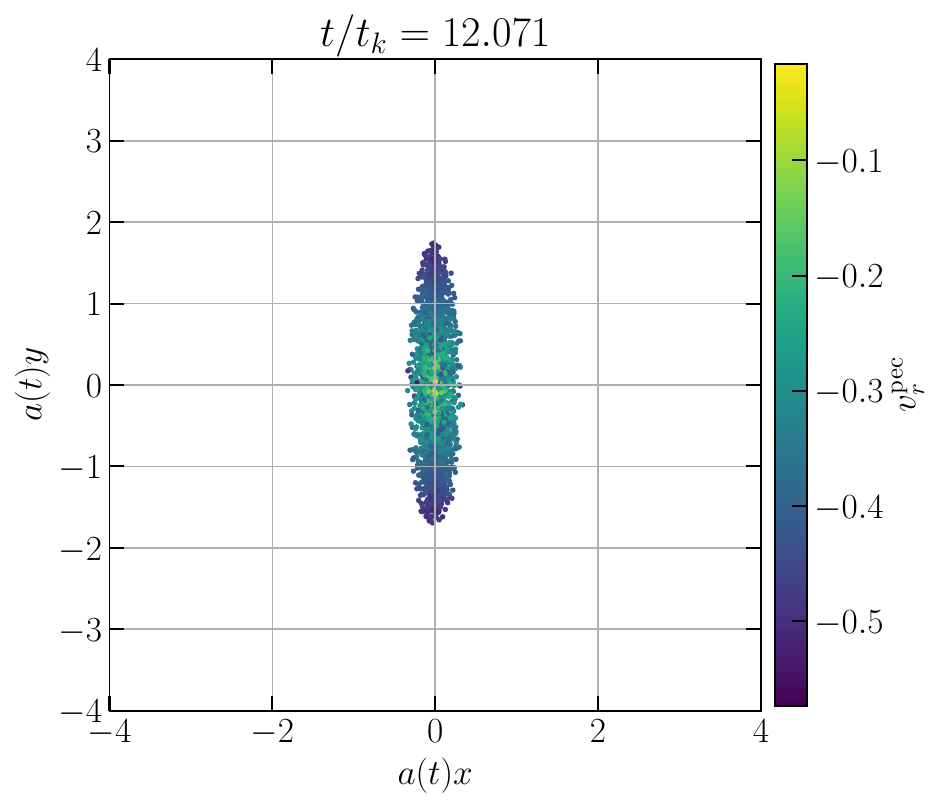}\hfill
\includegraphics[width=0.3\linewidth]{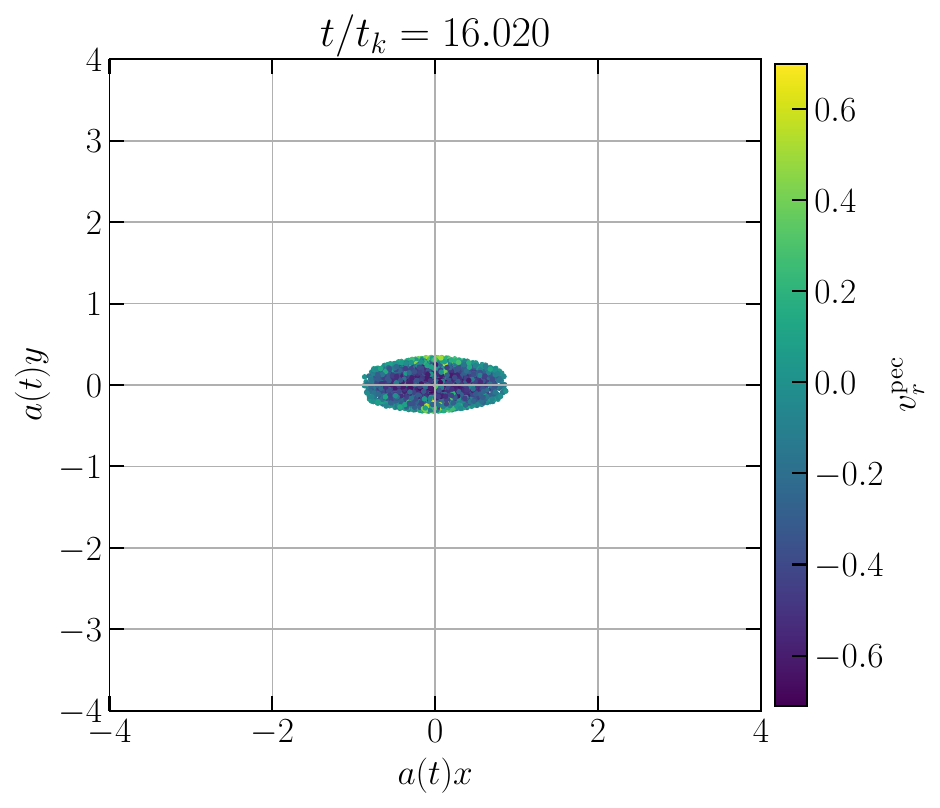}\hfill
\includegraphics[width=0.3\linewidth]{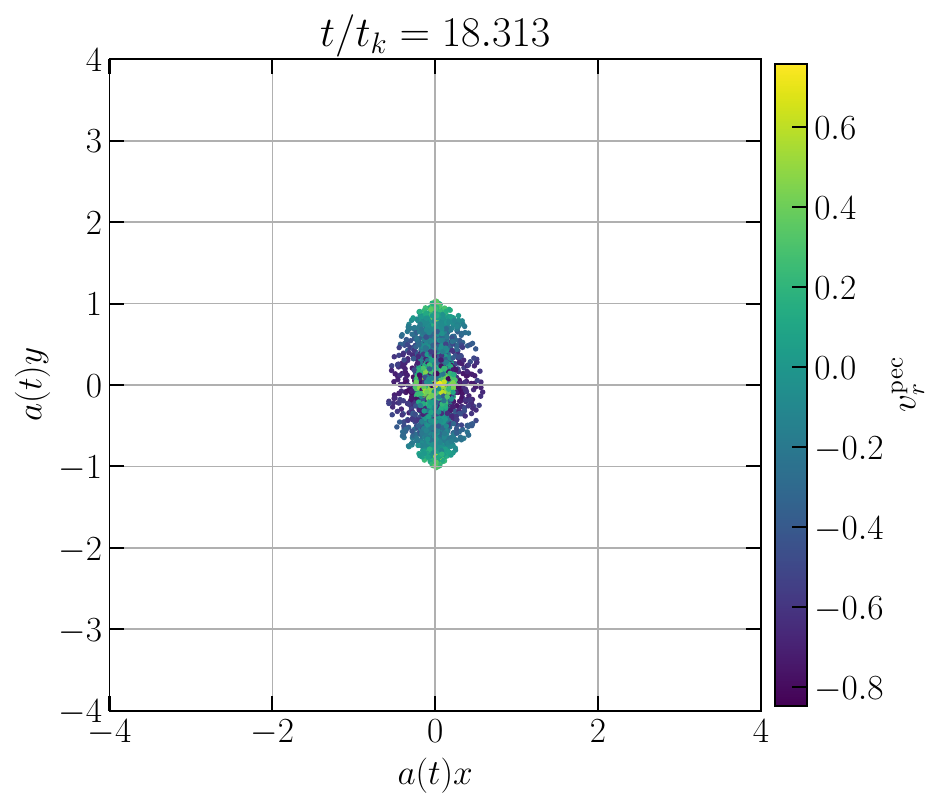}\hfill
\includegraphics[width=0.3\linewidth]{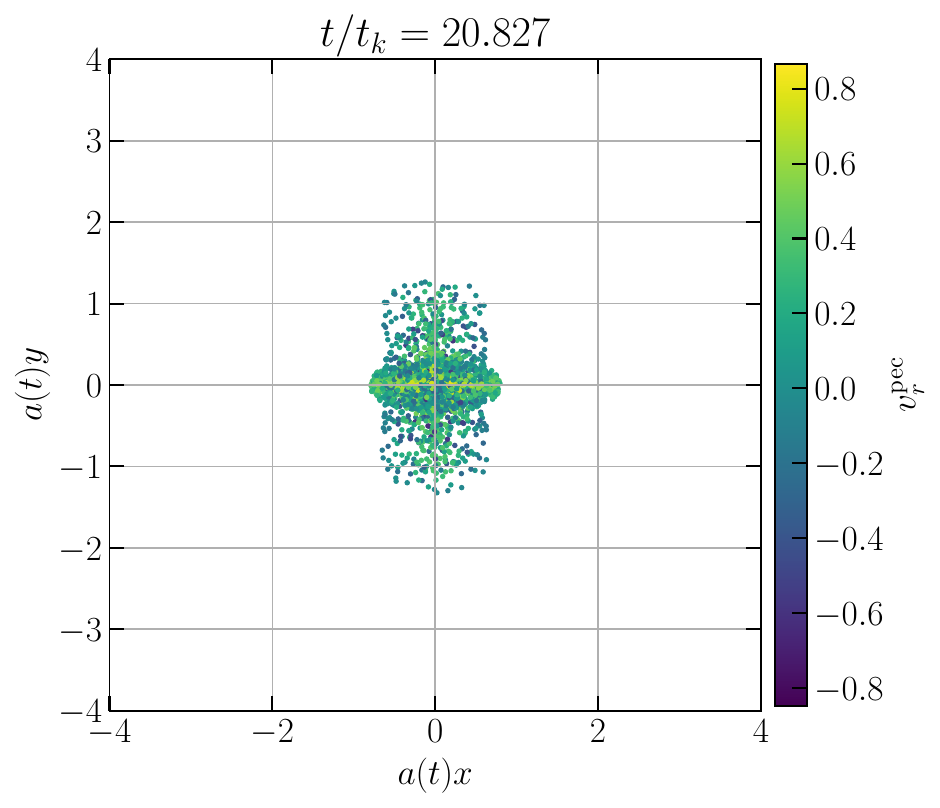}\hfill
\includegraphics[width=0.3\linewidth]{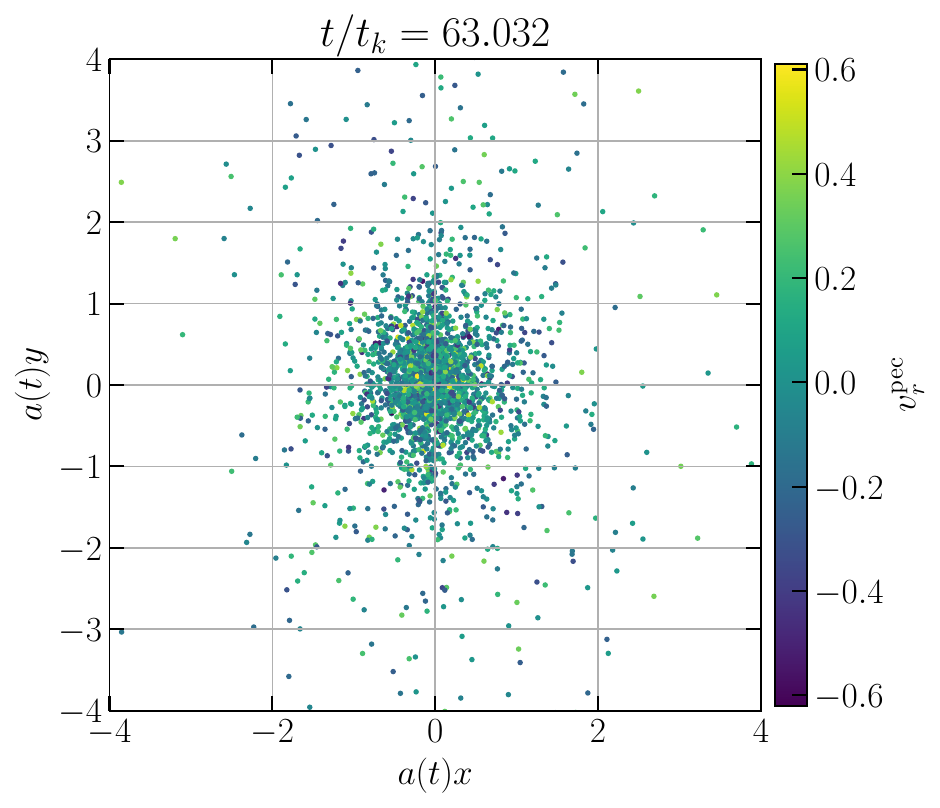}\hfill
\includegraphics[width=0.5\linewidth]{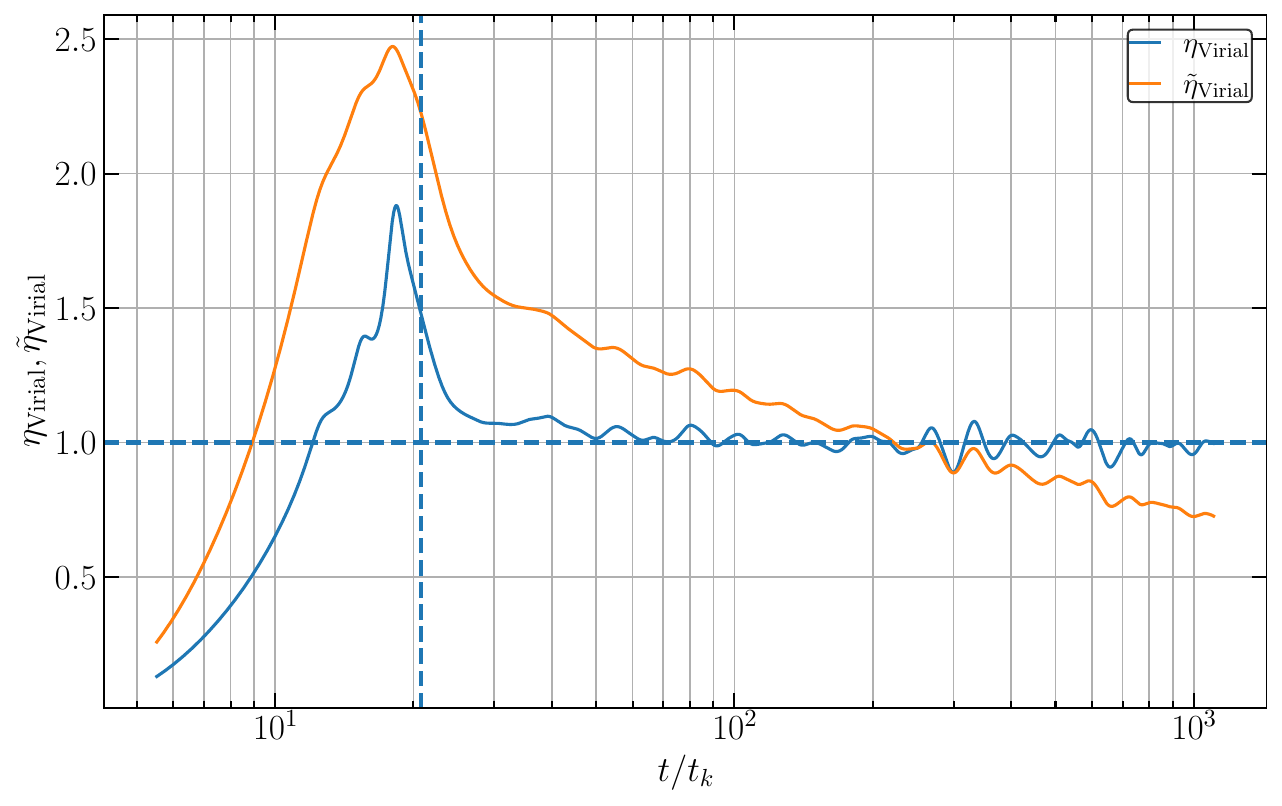}\hfill
\includegraphics[width=0.5\linewidth]{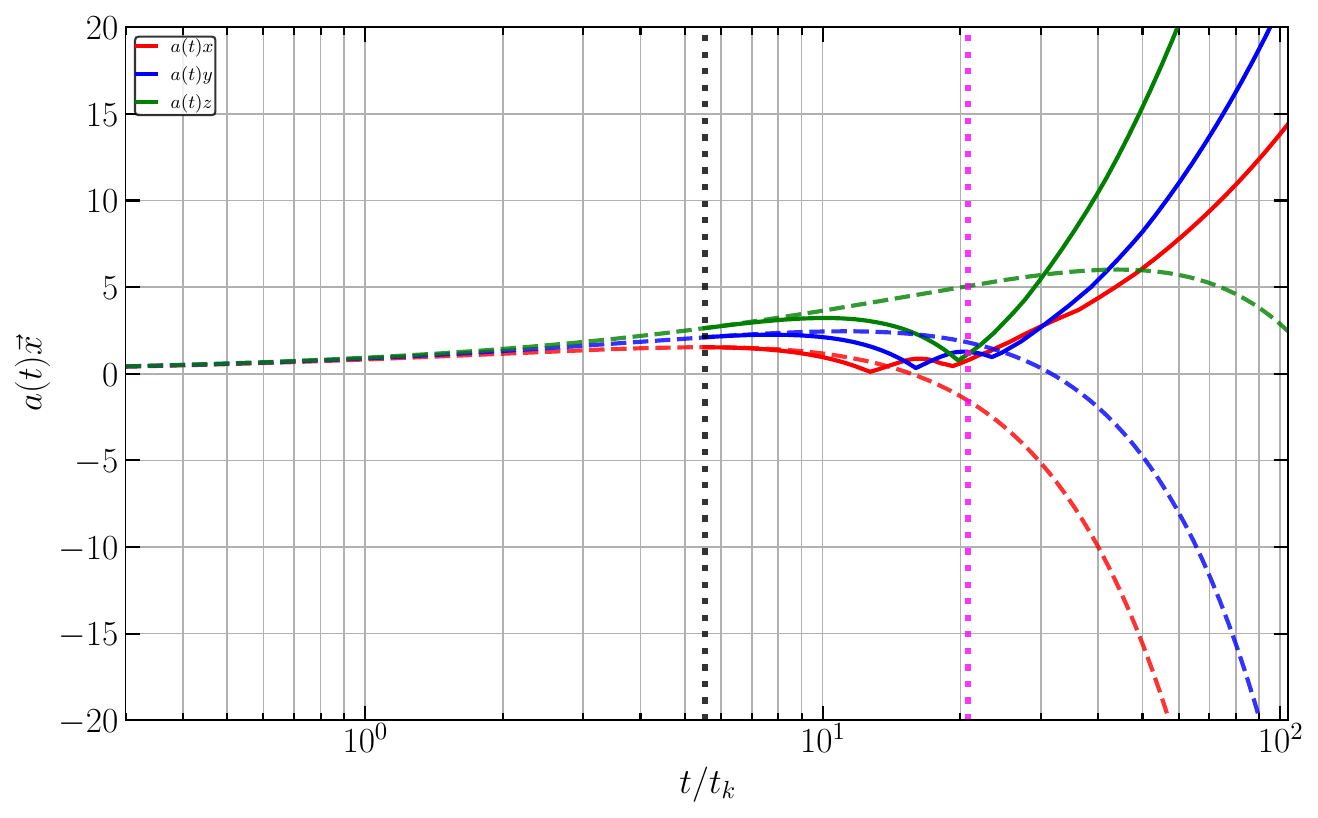}\hfill
\caption{Same caption as in Fig.~\ref{snapshots_particles_Nbody2} with parameters $\sigma=0.1$, $\nu=3, e=0.2, p=0.0$, $\alpha \approx 0.16$, $\beta \approx 0.10$, and $\gamma \approx 0.04$.}
\label{snapshots_particles_Nbody4}
\end{figure}

In this section, we present additional numerical results, similar to those shown in Section~\ref{sec:numerical_results_all}, for different non-spherical configurations. The qualitative behaviour of the results shown in Figs.~\ref{snapshots_particles_Nbody2}, \ref{snapshots_particles_Nbody3}, and \ref{snapshots_particles_Nbody4} is similar to that of the case presented in the main text. The case shown in Fig.~\ref{snapshots_particles_Nbody2} corresponds to the same non-spherical configuration studied in the main text in section \ref{sec:collapse_dynamics}, but with $\sigma=0.01$.

\subsection{Comparison with previous analytical fitting procedure}
\label{subsec:fitting_comparison}

\begin{figure}[!htbp]
\centering
\includegraphics[width=0.4\linewidth]{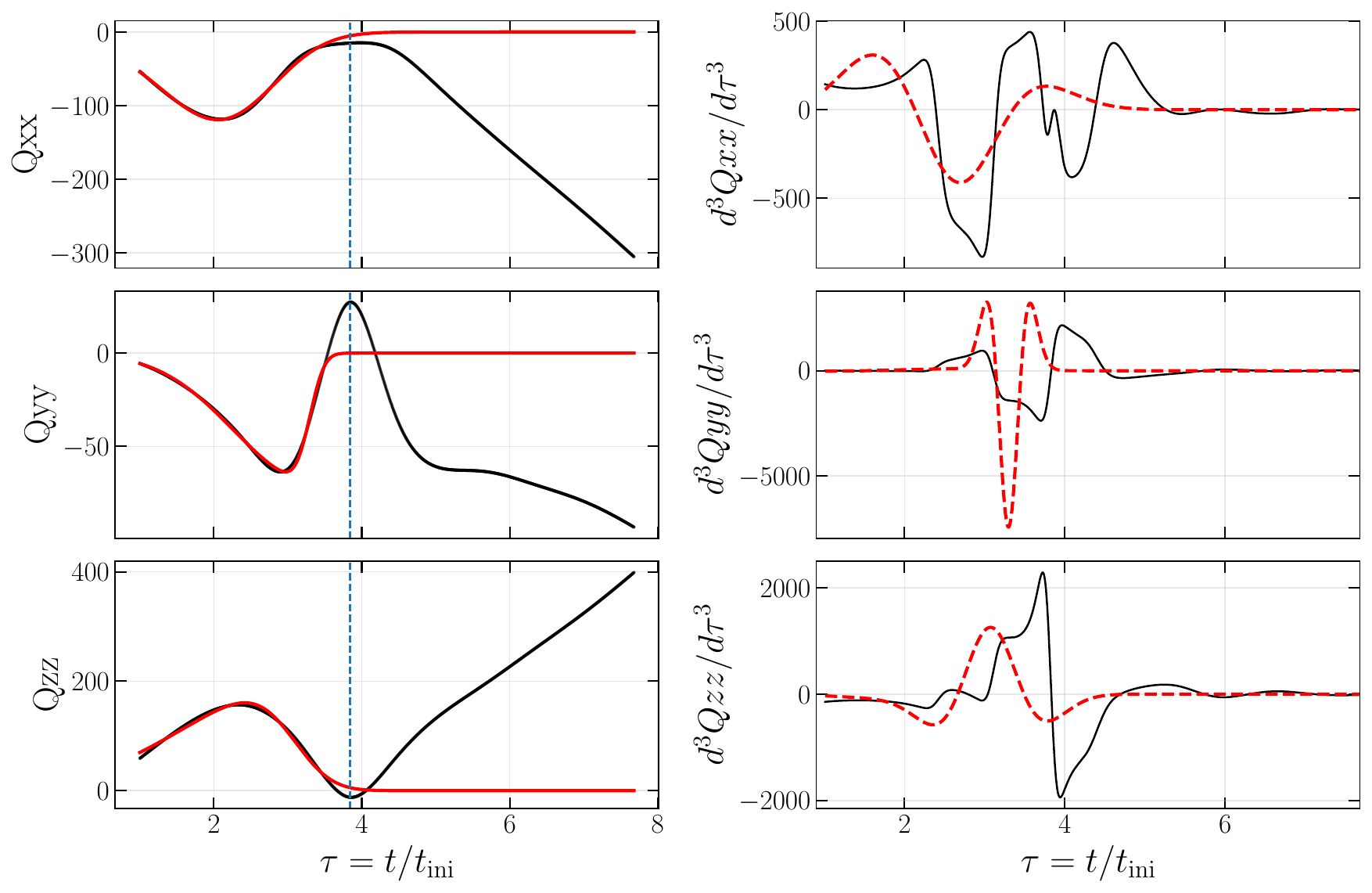}\hfill
\includegraphics[width=0.5\linewidth]{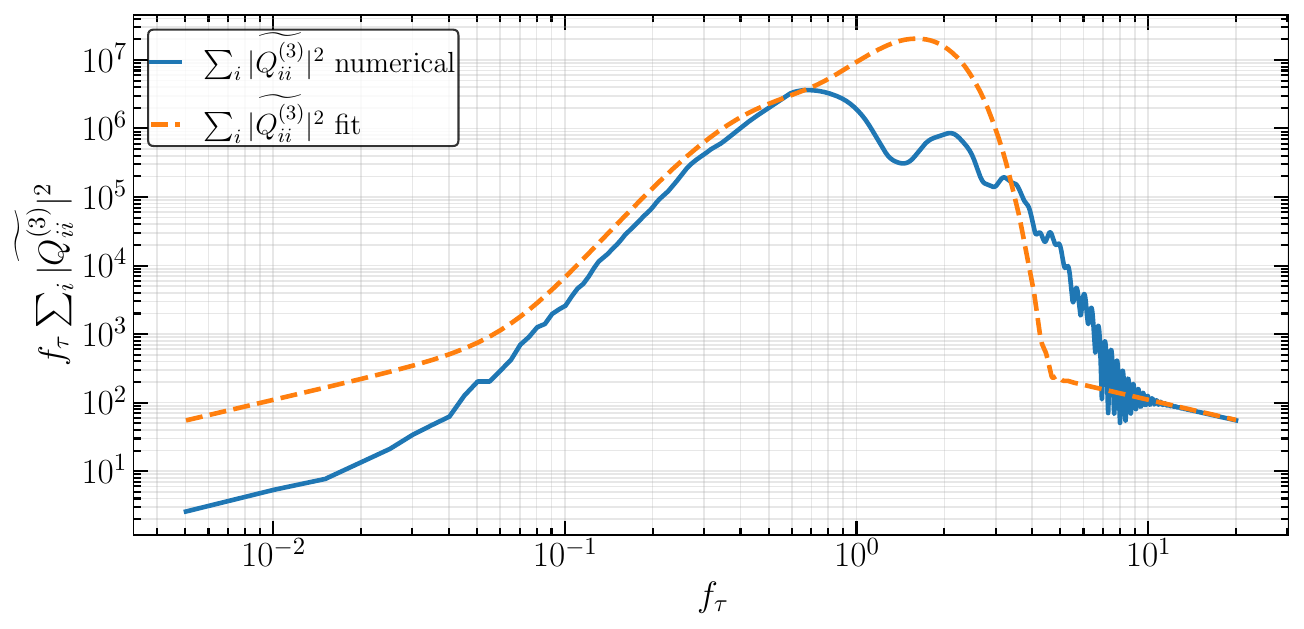}\hfill
\caption{Left panels: Time evolution of the diagonal quadrupole components (left column) and their third time derivatives (right column). The red lines correspond to the numerical fit, while the black lines correspond to the direct numerical results. Right panel: The corresponding Fourier transform of the quadrupole components in code units comparing the numerical result (blue line) with the one inferred from the fit (orange dashed line). Case with $\sigma=0.01$, $\nu=2, e=0.223, p \approx -0.00125$, $\alpha \approx 0.0111, \beta \approx 0.00668$, and $\gamma \approx 0.00219$.}
\label{fig:quadrupole_components_comparison}
\end{figure}

\begin{figure}[!htbp]
\centering
\includegraphics[width=0.4\linewidth]{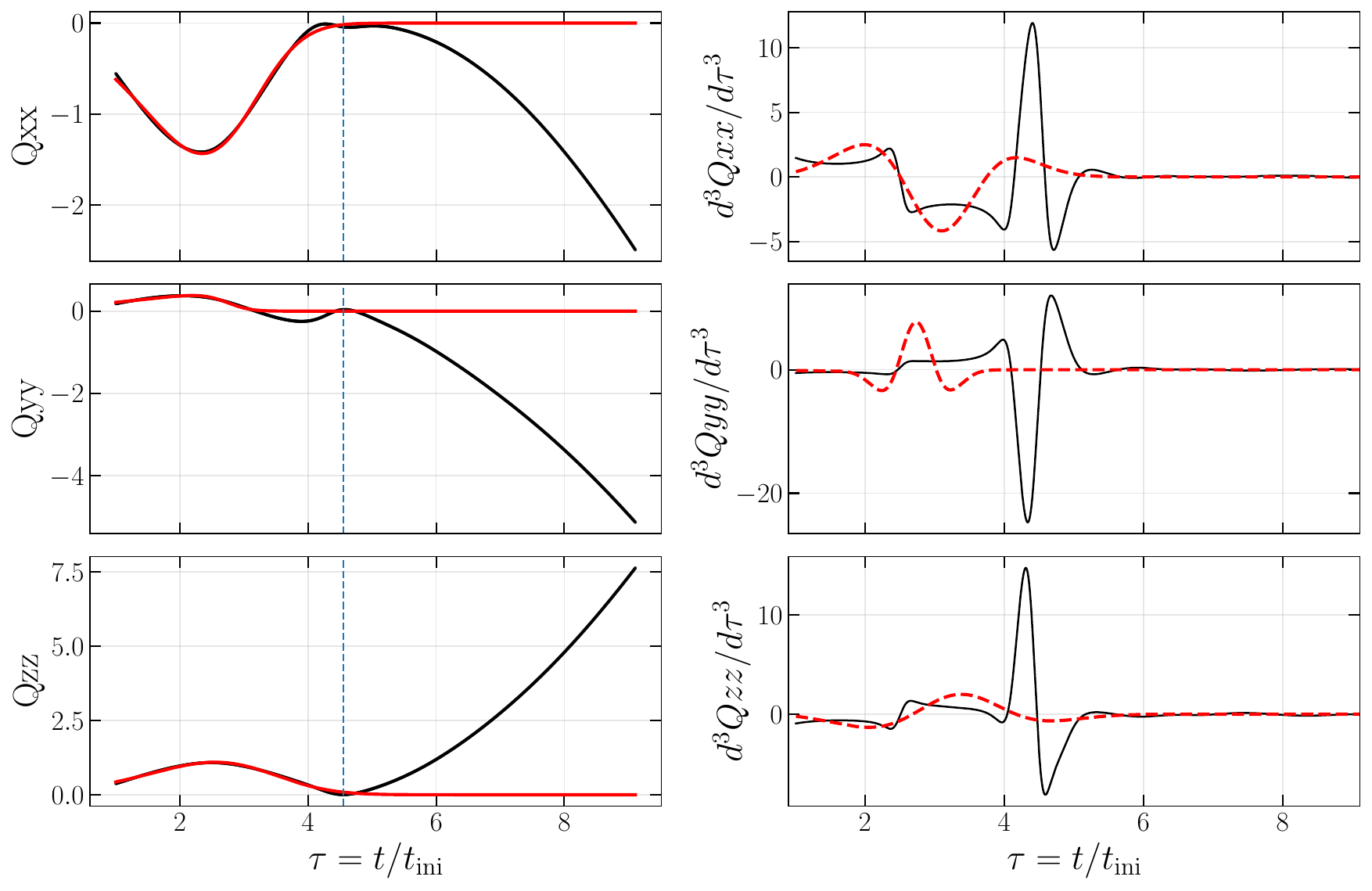}\hfill
\includegraphics[width=0.5\linewidth]{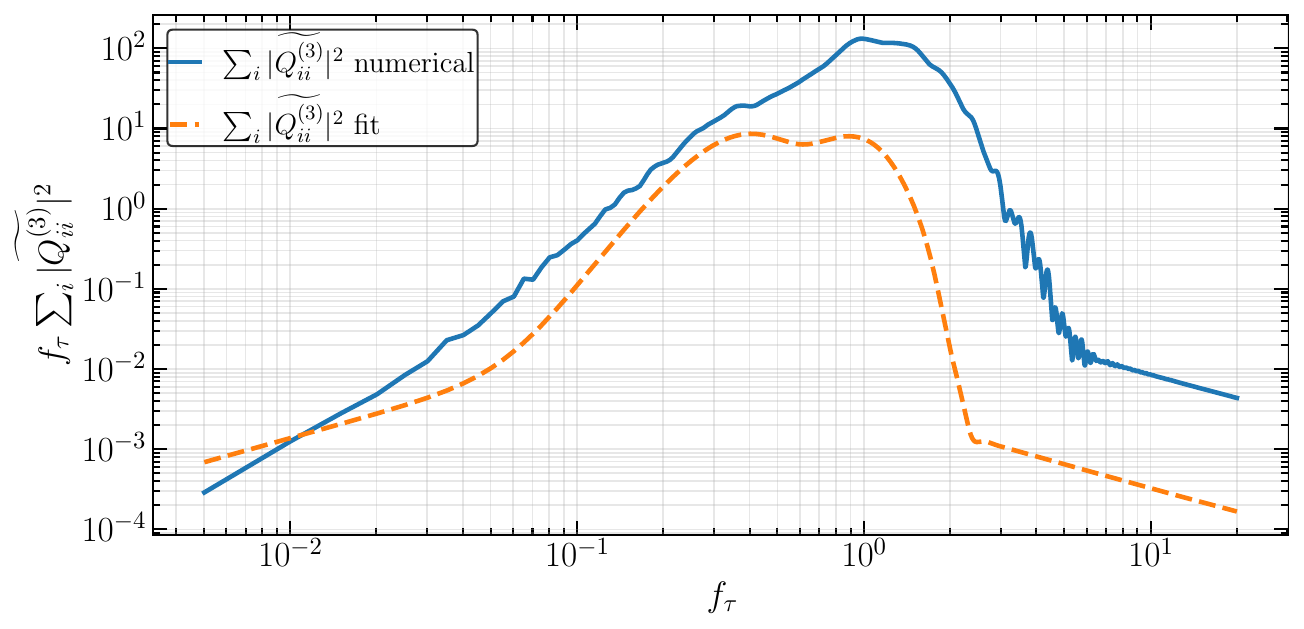}\hfill
\caption{Same caption as in Fig.\ref{fig:quadrupole_components_comparison} but for the case with $\sigma=0.1, \nu=2, e=0.3,p=0.2$, $\alpha = 0.14, \beta = 0.04$, and $\gamma = 0.02$.}
\label{fig:quadrupole_components_comparison3}
\end{figure}

\begin{figure}[!htbp]
\centering
\includegraphics[width=0.4\linewidth]{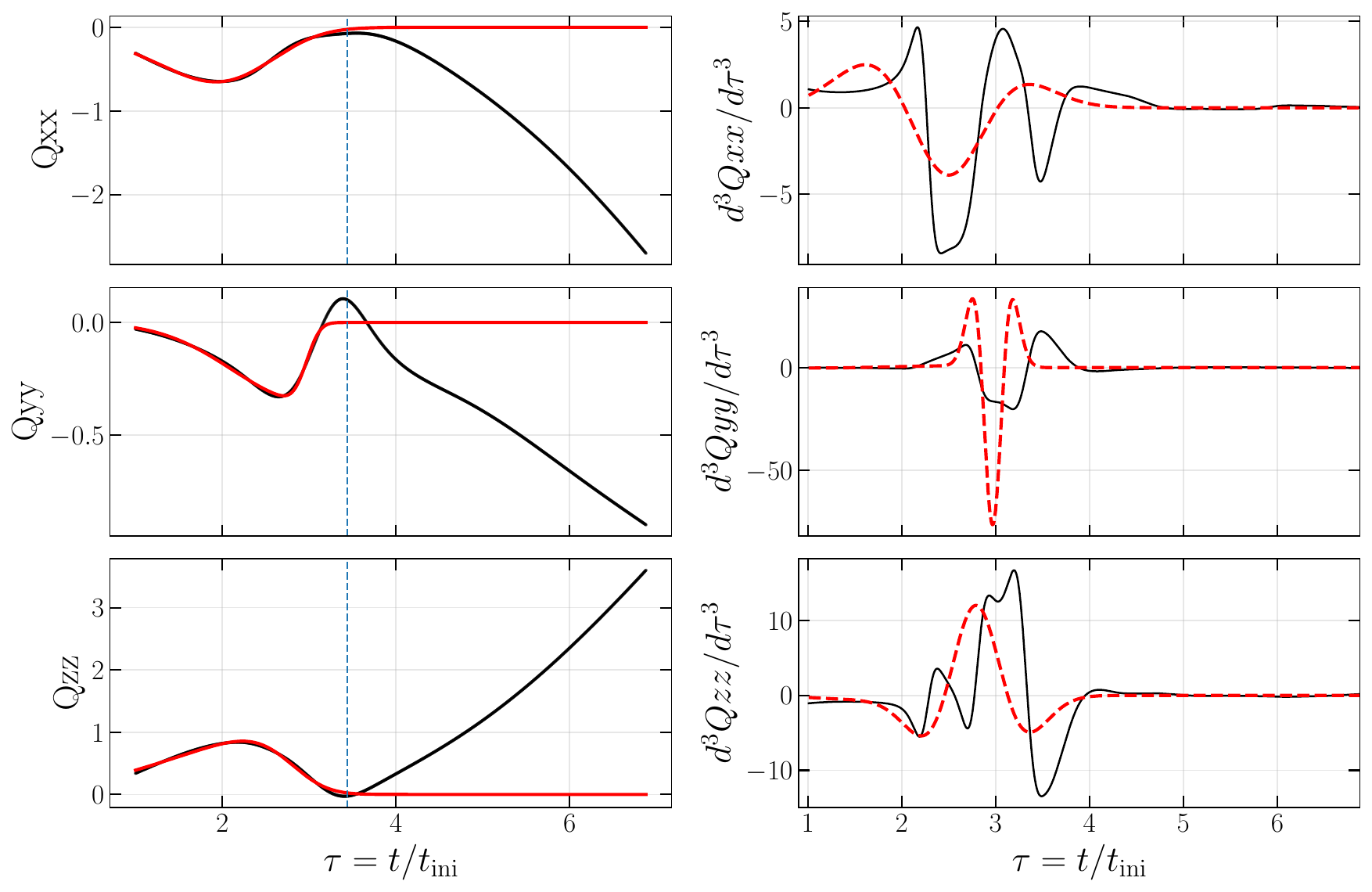}\hfill
\includegraphics[width=0.5\linewidth]{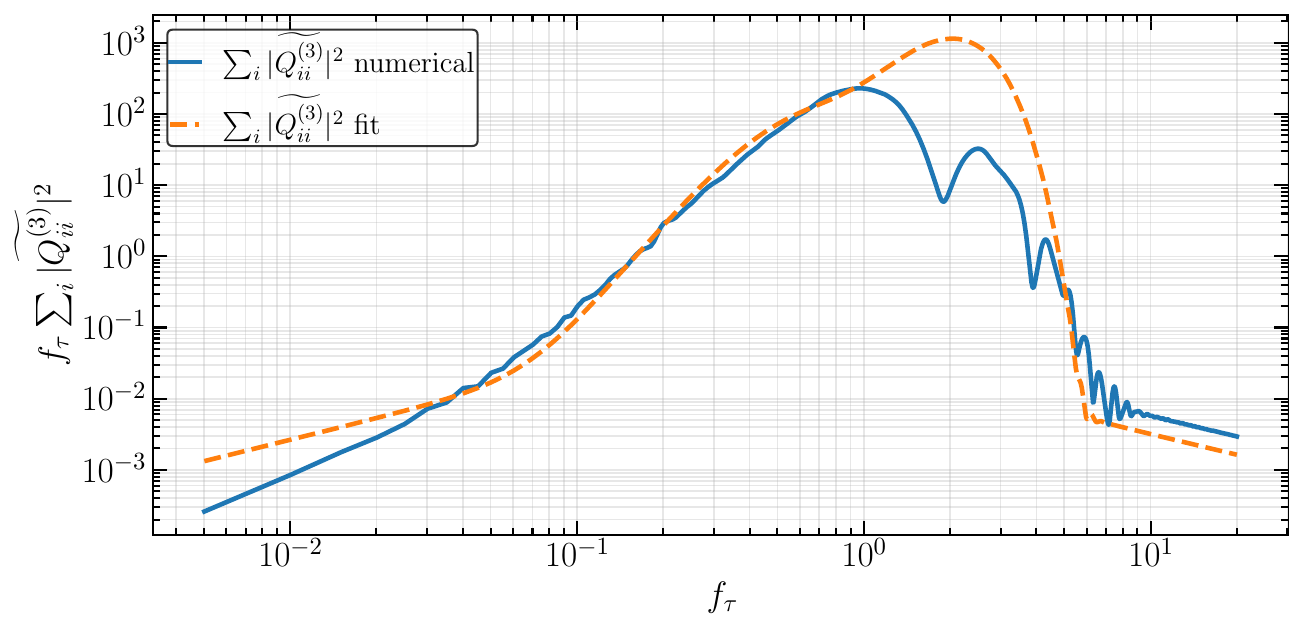}\hfill
\caption{Same caption as in Fig.\ref{fig:quadrupole_components_comparison} but for the case with $\sigma=0.1, \nu=3, e=0.2, p=0.0$, $\alpha =0.16, \beta = 0.1$, and $\gamma = 0.04$.}
\label{fig:quadrupole_components_comparison4}
\end{figure}

In this appendix, primarily for comparison purposes, we compare our numerical results for the quadrupole components with the phenomenological fitting-based procedure proposed in Ref.~\cite{DalianisKouvaris2024}. To assess the reliability of the fitting-based procedure adopted in Ref.~\cite{DalianisKouvaris2024}, we compare in Figs.~\ref{fig:quadrupole_components_comparison},~\ref{fig:quadrupole_components_comparison3} and~\ref{fig:quadrupole_components_comparison4} the direct numerical evolution of the quadrupole for different configurations with the fitting ansatz used there. In that approach, the diagonal quadrupole components are modeled by a skew-normal profile,
\begin{equation}
Q_{ii}^{\rm fit}(t)
=
A_i \exp\!\left[-\frac{(t-\mu_i)^2}{2\sigma_i^2}\right]
\left[
1+\operatorname{erf}\!\left(
\frac{\lambda_i (t-\mu_i)}{\sqrt{2}\,\sigma_i}
\right)
\right],
\label{eq:skewnormal_fit_quadrupole}
\end{equation}
with fitting parameters $(A_i,\mu_i,\sigma_i,\lambda_i)$ determined from the numerical data. This ansatz is intended to provide a smooth approximation to the quadrupole evolution and to reduce the numerical noise that would otherwise be amplified when taking time derivatives.

However, the comparison shown in the following figures demonstrates that this fitting prescription is not sufficiently accurate to describe the GW source term in the regime of interest. Although the fit may reproduce the broad envelope of the diagonal quadrupole components during the early stage of collapse, it fails to capture the detailed time dependence that develops near the bottleneck and during the subsequent violent relaxation. In particular, the fitted quadrupole is systematically smoother than the direct numerical result and does not reproduce the sharp features associated with shell crossing, secondary contraction, and the redistribution of matter in the post-collapse phase.

This is therefore not merely a cosmetic difference associated with smoothing the data, but a physically relevant discrepancy that directly affects the predicted GW spectrum. In particular, the dominant GW contribution is generated during the strongly nonlinear stage of the collapse, precisely where the dynamics departs most significantly from any simple fitting form. Moreover, our simulations consistently include the cosmological background expansion, which can also modify the functional behavior of the quadrupole components relative to the treatment of Ref.~\cite{DalianisKouvaris2024}.

For this reason, in the present work we compute the quadrupole evolution and its derivatives directly from the N-body simulation, without imposing any {\it a priori} functional ansatz. This fully numerical treatment preserves the sharp time structure generated by the nonlinear collapse and provides a more faithful description of the virialization process. Our results therefore indicate that a robust prediction of the GW signal requires a direct numerical treatment of the full nonlinear dynamics.

\subsection{Comparison of the GW signal between the semirelativistic and Newtonian treatments}
\label{subsec:rel_non_rel}

\begin{figure}[!htbp]
\centering
\includegraphics[width=0.8\linewidth]{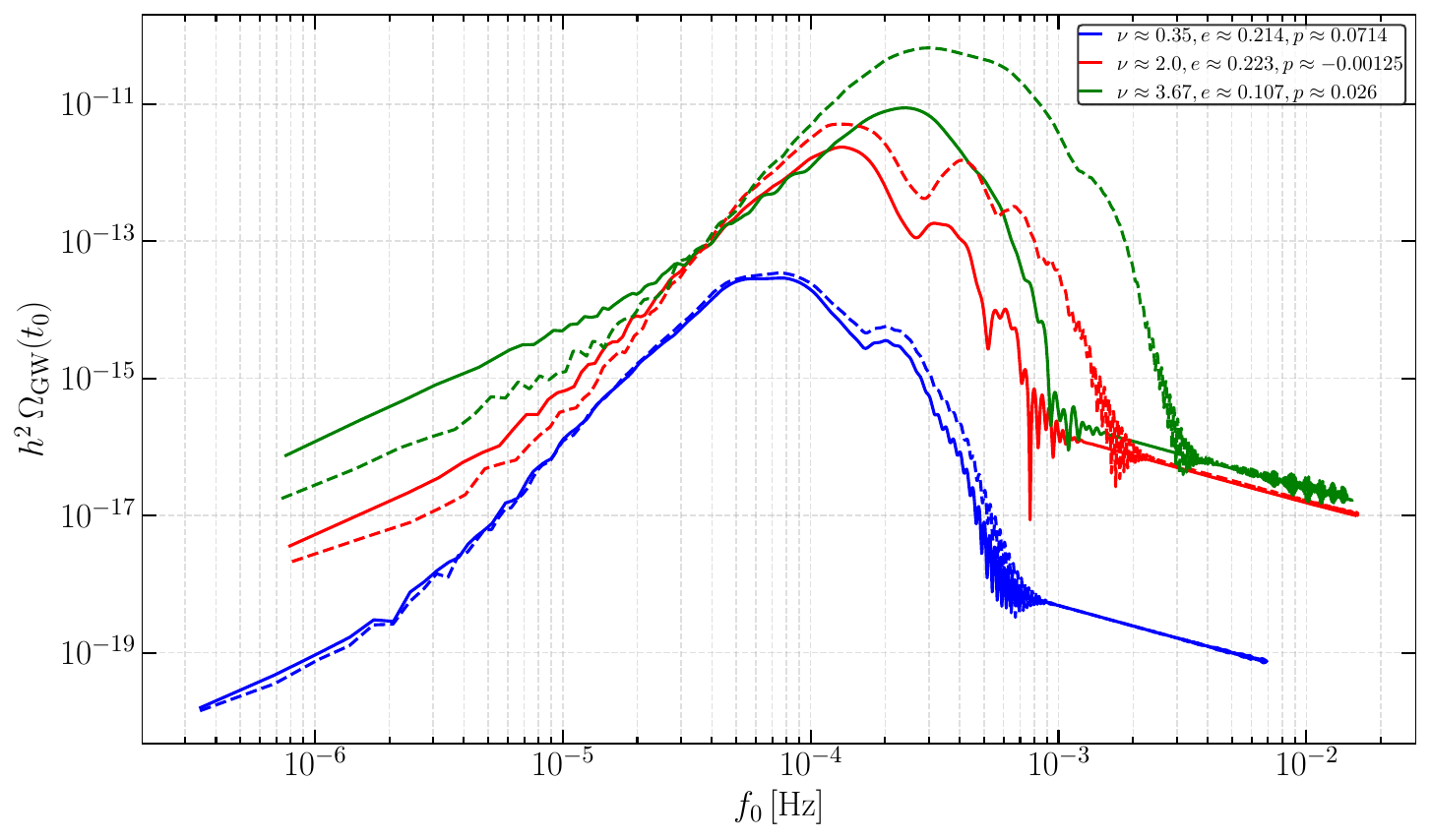}\hfill
\caption{The present-day GW spectra for specific configurations $(\nu,e,p)$. The solid lines correspond to the semi-relativistic approach, while the dashed lines represent the case without relativistic corrections. For all cases $N=3000$, $\epsilon=0.05$, $\sigma=0.1$.}
\label{fig:compare_sr_newt_specific}
\end{figure}

Figure~\ref{fig:compare_sr_newt_specific} compares the GW spectra obtained with the purely Newtonian treatment (dashed lines) and with the semirelativistic calculation (solid lines) for three representative configurations. Overall, the inclusion of semirelativistic corrections does not strongly modify the characteristic frequency scale of the signal, but it can noticeably reduce the amplitude of the spectrum, especially around the main peak.

The impact of the semirelativistic correction is configuration dependent. For the weakest-collapse case, shown in blue, the difference between the two calculations remains relatively small over most of the spectrum. In contrast, for the intermediate and strongest-collapse configurations, shown in red and green, respectively, the semirelativistic treatment leads to a visible suppression of the GW amplitude, reaching factors of a few and approaching nearly one order of magnitude around the peak in the most extreme case.

This behaviour is physically consistent with the expectation that semirelativistic kinematical effects can modify the nonlinear collapse dynamics once the peculiar velocities become sufficiently large. Since the emitted signal is determined by the third time derivative of the quadrupole tensor, even moderate changes in the particle trajectories during the violent post-shell-crossing stage can translate into a significant variation in the resulting GW spectrum.

The comparison therefore shows that the treatment of particle kinematics can noticeably affect the predicted GW signal for the most nonlinear configurations considered here. For the representative cases shown in Fig.~\ref{fig:compare_sr_newt_specific}, neglecting the semirelativistic momentum--velocity relation tends to give a larger estimate for the peak amplitude, while the overall spectral shape and peak location are only moderately affected. Thus, within the leading-quadrupole prescription adopted in this work, semirelativistic kinematical effects provide a relevant correction to the modelling of the violent post-shell-crossing stage. Further work based on a more complete weak-field, post-Newtonian, or general-relativistic treatment could help quantify the remaining corrections.

\subsection{Convergence with the time-step criterion}
\label{sec:time_step_convergence}

\begin{figure}[!htbp]
\centering
\includegraphics[width=0.5\linewidth]{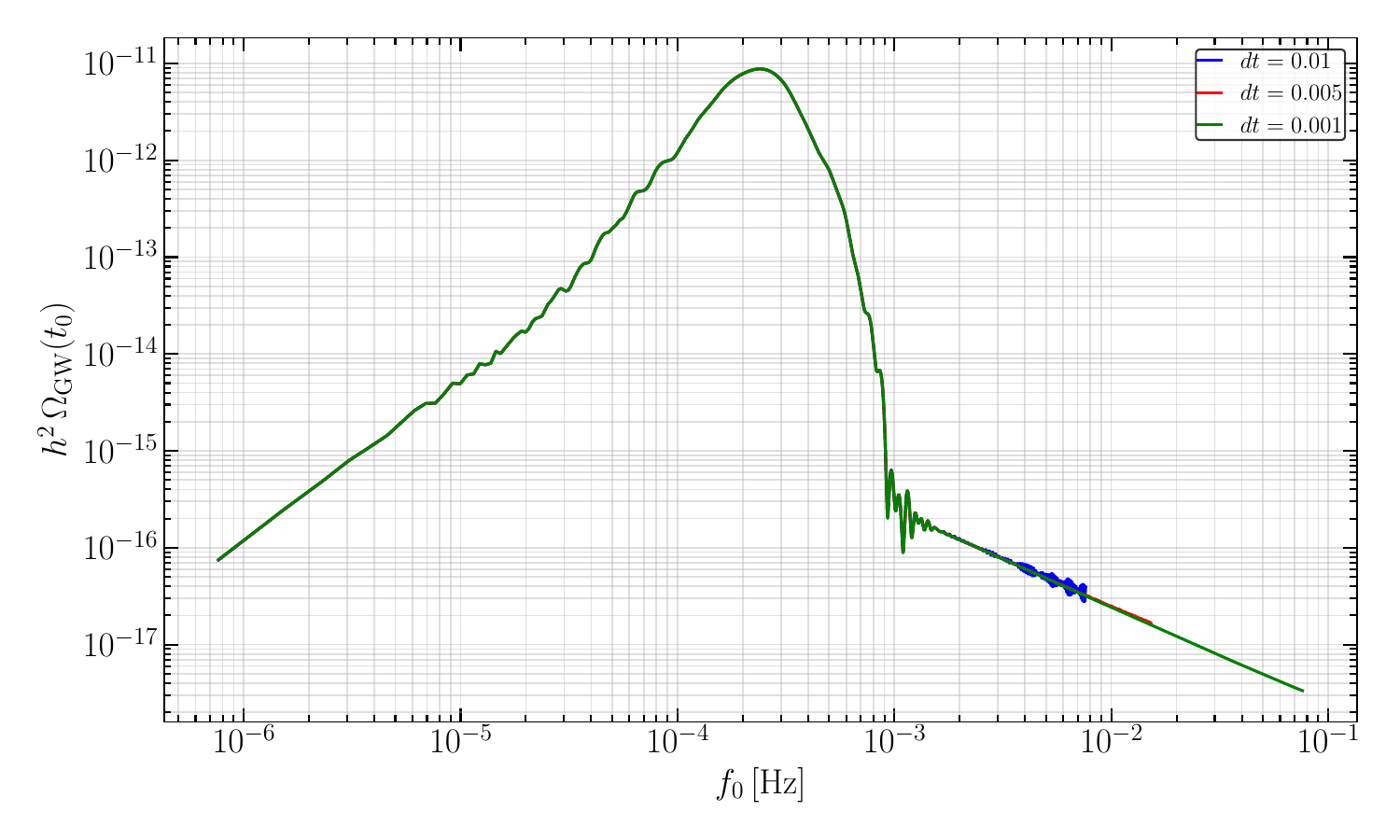}\hfill
\includegraphics[width=0.5\linewidth]{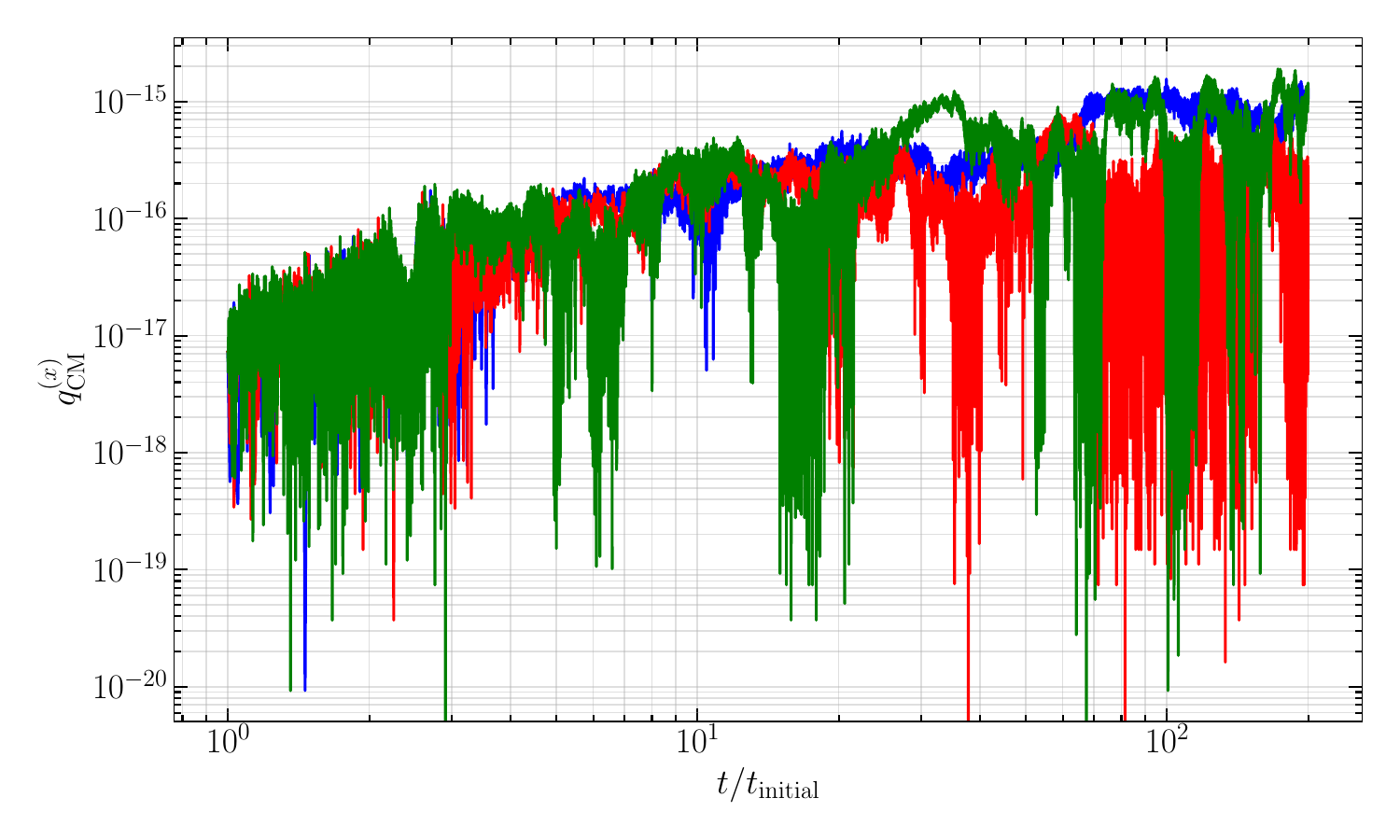}\hfill
\caption{Left panel: present-day GW spectrum obtained using different constant time-steps. Right panel: time evolution of the \(x\)-component of the center-of-momentum variable, \(q_{\rm CM}^{(x)}\), for the same set of runs, similar result is found for the other components $(y,z)$ of $\bf q_{\rm CM}$. The parameters chosen are $N=3000,\epsilon_{\rm com}=0.05,\sigma=0.1$ with shape configuration $\nu=3.67, e=0.107, p=0.026$.}
\label{fig:compare_time_step}
\end{figure}

In this appendix we assess the convergence of our results with respect to the choice of the constant time-step used in the integration. Since the GW signal is obtained from the time evolution of the quadrupole tensor and, in particular, depends on its third time derivative, a natural quantity to monitor is the jerk-based characteristic timescale. For this reason, we consider the criterion
\begin{equation}
d t_{\rm jerk} \sim \min_n \left( \frac{|\ddot{\bm{x}}_n|}{|\dddot{\bm{x}}_n|} \right),
\end{equation}
which estimates the characteristic timescale over which the particle acceleration varies significantly. This criterion is particularly relevant in the present case, since the dominant contribution to the GW emission arises during the highly non-linear stage of collapse, where both the force and its time variation can increase rapidly. In typical runs with the example of Fig.~\ref{fig:compare_time_step}, this timescale is found to lie in the range $\mathcal{O}(10^{-2}-10^{-1})$ the code unit.

To test the robustness of the calculation, we repeated representative simulations using different constant values of $dt$, and compared the resulting present-day GW spectra. As shown in the left panel of Fig.~\ref{fig:compare_time_step}, the spectra exhibit good convergence for sufficiently small time-steps, indicating that the fiducial choice adopted in the main analysis is adequate to resolve the relevant collapse dynamics and the associated GW production. In addition, in the right panel we show the time evolution of the \(x\)-component of the center-of-momentum variable, \(q_{\rm CM}^{(x)}\), which provides a useful numerical consistency check. The fact that this quantity remains very small during the evolution confirms that spurious violations of momentum conservation remain under control.

Overall, these results support the numerical stability of the integration scheme and show that the fiducial constant time-step used in this work is sufficiently small compared with the characteristic jerk timescale in the relevant stages of the evolution. We note that resolving the high-frequency part of the spectrum requires a sufficiently small time step $dt$ as discussed in section \ref{sec:theory_gws_calculation}, even in cases where the chosen time step is already adequate for accurately capturing the dynamics.

\subsection{Analytical estimate of the GWs from the Zel'dovich approximation}
\label{subsec:analytic_GW_ZA}

In addition to the direct GW extraction from the N-body evolution, it is
useful to construct an analytical estimate based on the Zel'dovich approximation. The purpose
of this appendix is precisely to provide such a benchmark. Since the definitions of the
shape variables $(\nu,e,p)$, the BBKS support $\chi(e,p)$, and the differential peak number
density have already been introduced in Sec.~\ref{subsec:statistics_initial_conditions}, we do
not repeat them here. Instead, we use those statistical ingredients together with the
analytical evolution of a triaxial overdense patch in order to estimate the stochastic GW
background. We follow the methodology of Ref.~\cite{Dalianis:2020gup} to obtain the expression for $\Omega_{\rm GW}$ using the Zel'dovich approximation with the Doroshkevich-based distribution, and we extend it to account for the number density of peaks.

As in previous sections, the triaxiality of the configuration is characterized by the
ordered deformation eigenvalues $(\alpha,\beta,\gamma)$, which are related to $(\nu,e,p)$
through Eq.~\eqref{eq:abg_inverse_nuep}. In the monochromatic limit, which is the case of
interest here, this identification is exact at the level of the shape variables, so the BBKS
weight can be applied directly to the same parameters entering the Zel'dovich dynamics.

For a homogeneous ellipsoidal patch, the Zel'dovich approximation implies that the physical
principal axes evolve as
\begin{equation}
R_i(t)=a(t)\,R_{L}\,\bigl[1-a(t)\lambda_i\bigr],
\qquad
(\lambda_1,\lambda_2,\lambda_3)=(\alpha,\beta,\gamma),
\label{eq:Ri_ZA}
\end{equation}
where $R_L$ denotes the undeformed Lagrangian radius of the patch. With the normalization
$a(t_k)=1$, the shortest axis reaches turnaround first. Since $\alpha$ is the largest eigenvalue,
its turnaround occurs at
\begin{equation}
a_{\rm max}=\frac{1}{2\alpha},
\qquad
t_{\rm max}=t_k\left(\frac{1}{2\alpha}\right)^{3/2}.
\label{eq:tmax_analytic}
\end{equation}
Following the standard Zel'dovich estimate, the formal collapse time is then
\begin{equation}
t_{\rm col}=2\sqrt{2}\,t_{\rm max}
=\alpha^{-3/2}t_k.
\label{eq:tcol_analytic}
\end{equation}
A finite early matter-dominated era restricts the allowed configurations through
\begin{equation}
t_{\rm col}<t_{\rm rh},
\label{eq:tcol_trh_cond}
\end{equation}
In practice this
translates into a lower cutoff on $\alpha$, while the requirement $t_{\rm max}\ge t_k$ imposes the upper bound $\alpha\le 1/2$.

Within this analytical treatment, the collapsing patch is modeled as a homogeneous ellipsoid
of total mass $M$. In the principal-axis frame, the trace-free quadrupole tensor is diagonal
and can be written as
\begin{equation}
Q_{ij}(t)
=
\frac{M}{15}
\begin{pmatrix}
2R_1^2-R_2^2-R_3^2 & 0 & 0\\
0 & 2R_2^2-R_1^2-R_3^2 & 0\\
0 & 0 & 2R_3^2-R_1^2-R_2^2
\end{pmatrix}.
\label{eq:Qij_ZA_diag}
\end{equation}
Substituting Eq.~\eqref{eq:Ri_ZA}, one finds that the time dependence of each component is
fully determined by the three deformation parameters.

The emitted GW spectrum then follows
from the quadrupole formula in frequency space,
\begin{equation}
\frac{dE_{\rm GW}}{d\ln\omega}
=
\frac{4\pi G}{5c^5}\,
\omega^7
\sum_{i,j}
\bigl|\widetilde{Q}_{ij}(\omega)\bigr|^2.
\label{eq:dEdlnw_quad_general}
\end{equation}
Because the quadrupole is diagonal in this approximation, only the diagonal components contribute. After inserting the Zel'dovich evolution of the axes and performing the time integral between
$t_k$ and $t_{\rm col}$, the emitted spectrum of a single triaxial configuration can be written
as
\begin{equation}
\frac{dE_{\rm GW}^{\rm Zel}}{d\ln\omega}
=
\frac{32}{135\pi}\,
\frac{c^5}{G}\,
\omega
\left(\frac{GM}{c^3}\right)^2
\mathcal{S}(\alpha,\beta,\gamma)\,
\left|
E_{1/3}(i\omega t_k)
-
\alpha^{-1}E_{1/3}(i\omega t_{\rm col})
\right|^2,
\label{eq:dEdlnw_ZA_final}
\end{equation}
where
\begin{equation}
\mathcal{S}(\alpha,\beta,\gamma)
=
\alpha^4+\beta^4+\gamma^4
-\alpha^2\beta^2-\alpha^2\gamma^2-\beta^2\gamma^2,
\label{eq:Sshape_ZA}
\end{equation}
and
\begin{equation}
E_{1/3}(z)\equiv \int_1^\infty du\,u^{-1/3}e^{-zu}
\label{eq:E13_def}
\end{equation}
is the generalized exponential integral. Equation~\eqref{eq:dEdlnw_ZA_final} gives the
emitted GW spectrum of a single configuration labeled by $(\alpha,\beta,\gamma)$, or
equivalently by $(\nu,e,p)$ through Eq.~\eqref{eq:abg_inverse_nuep}.

It is worth emphasizing that Eq.~\eqref{eq:dEdlnw_ZA_final} has the correct physical
dimensions. Indeed,
\begin{equation}
\left[
\frac{c^5}{G}
\left(\frac{GM}{c^3}\right)^2
\omega
\right]
=
{\rm energy},
\end{equation}
so that $dE_{\rm GW}^{\rm Zel}/d\ln\omega$ is an energy, as required.

To relate the emitted spectrum to the present-day observable frequency, we assume an abrupt transition from the early matter-dominated era to radiation domination at reheating as in the main text using the conversion factors described in section \ref{sec:theory_gws_calculation}.

Then, for the Doroshkevich-based Zel'dovich estimate we obtain,

\begin{equation}
h^2\Omega_{\rm GW}^{\rm Dor}(f_0)
=
\frac{h^2}{\rho_{c,0}}
\int d\alpha\,d\beta\,d\gamma\;
\frac{P_{\rm D}(\alpha,\beta,\gamma)}{V_k}\,
\frac{1}{1+z_e(\nu,e,p)}
\left.
\frac{dE_{\rm GW}^{\rm Zel}}{d\ln\omega}
\right|_{\omega=2\pi f_0(1+z_e)},
\end{equation}
where
\begin{equation}
V_k=\frac{4\pi}{3k_*^3},
\end{equation}
the integration is performed over the Doroshkevich domain $e\ge0$, $-e\le p\le e$, together with the same ordering and collapse constraints on $(\alpha,\beta,\gamma)$, and $P_{\rm D}(\nu,e,p)$ is given by Eq.~\eqref{eq:Doroshkevich_abg}.

Let us now obtain the expresion in the BBKS peak theory approach. The statistical weight of the ensemble is provided by the BBKS differential number density
already introduced in Eq.~\eqref{eq:npk_mono_nuep}. Since this quantity is a number density
of peaks, rather than a normalized probability distribution, it can be used directly in the
volume average of the GW background.

Combining the single-source Zel'dovich spectrum with the BBKS peak abundance, the present-day
stochastic background is simply given by
\begin{equation}
h^2\Omega_{\rm GW}^{\rm BBKS}(f_0)
=
\frac{h^2}{\rho_{c,0}}
\int_{\nu_{\min}}^{\nu_{\max}} d\nu
\int_0^{1/2} de
\int_{p_{\min}(e)}^{p_{\max}(e)} dp\;
\frac{d^3 n_{\rm pk}}{d\nu\,de\,dp}\,
\frac{1}{1+z_e(\nu,e,p)}
\left.
\frac{dE_{\rm GW}^{\rm Zel}}{d\ln\omega}
\right|_{\omega=2\pi f_0(1+z_e)} .
\label{eq:h2OmegaGW_ZA_BBKS}
\end{equation}
where the differential number density $n_{\rm pk}$ is given by Eq.~\eqref{eq:d4npk_nuxep} directly assuming the monochromatic power spectrum case. Here the integration limits in the $(e,p)$ plane are those defined by the BBKS support,
namely
\begin{equation}
p_{\min}(e)=
\begin{cases}
-e, & 0<e<\dfrac14,\\[2mm]
3e-1, & \dfrac14<e<\dfrac12,
\end{cases}
\qquad
p_{\max}(e)=e,
\label{eq:pmin_pmax_BBKS}
\end{equation}
while the mapping between $(\nu,e,p)$ and $(\alpha,\beta,\gamma)$ is still given by
Eq.~\eqref{eq:abg_inverse_nuep}. In practice, only configurations satisfying
\begin{equation}
\alpha>0,\qquad
\gamma\le\beta\le\alpha,\qquad
\alpha+\beta+\gamma>0,\qquad
\alpha_{\min}\le \alpha \le \frac12,
\label{eq:alpha_constraints_final}
\end{equation}
are retained in the integral, where $\alpha_{\min}$ is determined by the finite duration of
the early matter-dominated era through Eq.~\eqref{eq:tcol_trh_cond}.

\bibliographystyle{JHEP}
\bibliography{references}

\end{document}